\documentclass[numberedappendix, twocolumn]{emulateapj}
\usepackage{latexsym}
\usepackage{graphicx}
\usepackage{xspace}
\usepackage{amssymb,amsmath}
\usepackage{amsfonts}   % if you want the fonts
\usepackage{natbib}
\usepackage{placeins}
\usepackage{times}
\usepackage[usenames,dvipsnames]{color}
\usepackage{url}
\usepackage{hyperref}

\begin{document}

\defcitealias{2011ApJ...729..106T}{Paper~I}

\newcommand{\ww}{0.48\linewidth}
\newcommand{\valf}{v_\text{Alf}}
\newcommand{\params}{\Theta}
\newcommand{\paramsP}{\Theta_{P}}
\newcommand{\paramsA}{\Theta_{X}}
\newcommand{\nuis}{\vartheta}
\newcommand{\data}{{\bf D}}
\newcommand{\be}{\begin{equation}}
\newcommand{\ee}{\end{equation}}
\newcommand{\like}{{\mathcal L}}
\newcommand{\rt}[1]{{\bf \textcolor{red}{RT: #1}}}
\newcommand{\av}[1]{{\bf \textcolor{blue}{AV: #1}}}
\newcommand{\gj}[1]{{\bf \textcolor{green}{GJ: #1}}}
\newcommand{\pg}[1]{{\bf \textcolor{magenta}{PG: #1}}}
\newcommand{\im}[1]{{\bf \textcolor{YellowOrange}{IM: #1}}}
\newcommand{\prof}{{\mathfrak L}}
\newcommand{\chisq}{\chi^2}
\newcommand{\mdl}{{\mathcal{M}}}
\newcommand{\norm}[2]{${\mathcal N}(#1, #2)$}

\newcommand{\icrc}{Int.\ Cosmic Ray Conf.}

\newcommand{\galprop}{GALPROP}
\newcommand{\Berat}{$^{10}$Be/$^9$Be\xspace}
\newcommand{\gray}{$\gamma$-ray}
\newcommand{\ppbarHe}{$p,\bar p$, He\xspace}
\def\Dpp{D_{pp}}
\def\Dxx{D_{xx}}
\def\ddp{\frac{\partial}{\partial p}}
\def\dfdt{\frac{\partial f(\vec p)}{\partial t}}
\def\dfdp{\frac{\partial f(p)}{\partial p}}
\def\dNpdt{\frac{\partial \psi}{\partial t}}
\def\dNpdp{\frac{\partial \psi}{\partial p}}
\def\Xco{$X_{\rm CO}$}
\newcommand{\hi}{H~{\sc i}}
\newcommand{\hii}{H~{\sc ii}}
\newcommand{\range}[2]{$[#1,#2]$} 

\submitted{}
\title{Bayesian analysis of cosmic-ray propagation: 
evidence against homogeneous diffusion}
\shorttitle{Bayesian analysis of cosmic-ray propagation}
\shortauthors{The GalBayes Project}

\author{G. J\'{o}hannesson\altaffilmark{1}} %\email{gudlaugu@stanford.edu}
\altaffiltext{1}{Science Institute, University of Iceland, Dunhaga 3, IS-107 Reykjavik, Iceland}
\author{R. Ruiz de Austri\altaffilmark{2}}
\altaffiltext{2}{Instituto de F\'isica Corpuscular, IFIC-UV/CSIC, Valencia, Spain}
\author{A.C. Vincent\altaffilmark{3}}
\altaffiltext{3}{Institute for Particle Physics Phenomenology (IPPP),\\ Department of Physics, Durham University, Durham DH1 3LE, UK.}

\author{I.~V. Moskalenko\altaffilmark{4,5}} %\email{imos@stanford.edu}
\altaffiltext{4}
{Hansen Experimental Physics Laboratory, Stanford University, Stanford, CA 94305
%\email{imos@stanford.edu}}
}\altaffiltext{5}
{Kavli Institute for Particle Astrophysics and Cosmology, Stanford University, Stanford, CA 94305}

%\and

\author{E. Orlando\altaffilmark{4,5}}

\author{T.~A. Porter\altaffilmark{4}}   %\email{tporter@stanford.edu}}

\author{A.~W. Strong\altaffilmark{6}} % \email{elena.orlando@mpe.mpg.de}
\altaffiltext{6}{Max-Planck-Institut f\"ur extraterrestrische Physik, Postfach 1312, D-85741 Garching, Germany}

\author{R. Trotta\altaffilmark{7,8}}
\altaffiltext{7}{Astrophysics Group, Imperial Centre for Inference and Cosmology, Imperial College London,
	Blackett Laboratory, Prince Consort Road, London SW7 2AZ, UK }
\altaffiltext{8}{Data Science Institute, William Penney Laboratory, Imperial College London, SW7 2AZ London, UK}

\author{F.~Feroz} %\altaffilmark{8}}
%\altaffiltext{8}{Astrophysics Group, Cavendish Laboratory, J.J. Thomson Avenue, Cambridge CB3 0HE}
\author{P.~Graff\altaffilmark{9}}
\altaffiltext{9}{The Johns Hopkins University Applied Physics Laboratory, 11100 Johns Hopkins Road, Laurel, MD 20723, USA
Department of Physics, University of Maryland, College Park, MD 20742, USA}

\author{M.P.~Hobson\altaffilmark{10}}
\altaffiltext{10}{Astrophysics Group, Cavendish Laboratory, J.J. Thomson Avenue, Cambridge CB3 0HE.}

\begin{abstract}
We present the results of the most complete ever scan of the parameter space for cosmic ray (CR) injection and propagation. We perform a Bayesian search of the main GALPROP parameters, using the MultiNest nested sampling algorithm, augmented by the BAMBI neural network machine learning package. This is the first such study to separate out low-mass isotopes ($p$, $\bar p$ and He) from the usual light elements (Be, B, C, N, O). We find that the propagation parameters that best fit \ppbarHe data are significantly different from those that fit light elements, including the B/C and \Berat secondary-to-primary ratios normally used to calibrate propagation parameters. This suggests each set of species is probing a very different interstellar medium, and that the standard approach of calibrating propagation parameters using B/C can lead to incorrect results. We present posterior distributions and best fit parameters for propagation of both sets of nuclei, as well as for the injection abundances of elements from H to Si. The input galdef-files with these new parameters will be included in an upcoming public GALPROP update. 
\end{abstract}


\maketitle

\section{Introduction}
%#########################################################################

CR physics has entered a data-driven era. 
Until recently CR observations were not accurate enough to warrant 
sophisticated studies of the propagation model parameter space,
although some attempts have been made using mostly analytical propagation codes 
\citep[e.g.,][]{Donato2002,Maurin2001,Maurin2002,Maurin2010,Putze2010}.
The launch of Payload for Antimatter Matter Exploration and Light-nuclei Astrophysics (PAMELA) in 2006 \citep{2007APh....27..296P}, followed by 
the {\it Fermi} Large Area Telescope ({\it Fermi}-LAT) in 2008 \citep{2009ApJ...697.1071A}, 
and finally the Alpha Magnetic Spectrometer -- 02 (AMS-02) in 2011 have changed 
the landscape dramatically. 
The technologies employed by these space missions have enabled measurements 
with unmatched precision and data sets orders of magnitude larger than 
earlier experiments, which allow for searches of subtle signatures of new phenomena in CR and \gray{} data. 
For example, the claimed precision of AMS-02 data reaches 1-3\%. 
This requires propagation models of comparable 
accuracy in order to take full advantage of such high quality data.  
Other high-expectations missions have just launched (e.g., the CALorimetric Electron Telescope -- CALET) or 
are awaiting launch (the Cosmic-Ray Energetics and Mass investigation -- ISS-CREAM).

Our understanding of CR propagation in the Milky Way comes from a combination of substantial observational data and a strong theoretical basis. These include exhaustive maps of the distribution of gas in the Galaxy, interstellar dust, radiation field 
and magnetic field, in addition to the latest data and codes describing  particle and nuclear cross sections.
Incorporation of such information is not possible using analytic methods and 
a fully numerical modeling for the treatment of CR
propagation in the Galaxy is required. 
This was realized about 20 years ago, when some of us started to develop 
the most advanced fully numerical CR propagation code, called \galprop{}\footnote{http://galprop.stanford.edu \label{galprop_link}},
which is also a {\it de facto} standard in astrophysics of CRs \citep{MS1998,SM1998}. \galprop{} makes use of information from astronomy, particle and nuclear physics to predict CRs, \gray{s}, synchrotron and other observables in a self-consistent manner \citep[a review can be found in][]{SMP2007}.  The code's output includes CR spectra and 
intensities in every spatial grid point (in 2D and 3D) in the Galaxy, as well as  the associated diffuse emissions from CR   
interactions with the interstellar gas, radiation, and magnetic fields. 

The first successful attempt to embed such a detailed and fully numerical propagation code within a Bayesian statistical approach was made in 2011 (\citealt{2011ApJ...729..106T}, hereafter \citetalias{2011ApJ...729..106T}). 
This became possible because of an extensive optimization and parallelization of the \galprop{} code (necessary for the fast evaluation of the likelihood
function) combined
with highly efficient sampling algorithms employed in the SuperBayeS~\citep{deAustri:2006jwj} package.
The advantages of such analysis are many-fold. 
Firstly, the Bayesian sampling method used enables a statistical analysis of the entire parameter space, rather than being limited to scanning a reduced number of dimensions at a time. Crucially, this allows  all relevant CR parameters to be fit simultaneously. Second, the parameters that are not of immediate relevance can be marginalized (integrated) over, without sacrificing computing time. Correlations in the global parameter space are thus fully accounted for in the resulting probability distributions. Thirdly, we recover statistically well-defined errors for each parameter in addition to the best fit; this constitutes one of the most important achievements of our earlier work. Finally, a large number of ``nuisance'' parameters can be incorporated, leading to an overall more robust fit. These parameters include the modulation potentials and experimental error rescaling parameters, and allow us to mitigate the effect of potential systematic errors that could arise from the data or the theoretical model. 

This paper is novel in three ways. First, it builds on the framework established in \citetalias{2011ApJ...729..106T} and improves it in several directions. We demonstrate a first application of machine-learning techniques to speed-up the computationally expensive inference from fully numerical codes in an automatically supervised manner. We introduce neural network training in the form of the BAMBI algorithm, which reduces computational effort by 20\%. The ensuing trained neural network can then in principle be used to conduct a (usually more costly) profile likelihood analysis with almost no computational effort.
Second, we now constrain both the CR propagation model parameters and the source abundances, using an iterative scheme to convergence. 
Third, for the first time we split the data sets used into low-mass isotopes ($p$, $\bar p$ and He) and light nuclei (Be--Si). 
Significantly different inelastic cross sections of protons (and antiprotons) and heavier nuclei ($\sim$40 mb for protons vs.\ $\sim$250 mb for carbon), 
result in different CR species propagating from different distances in the Galaxy. Treating them separately allows us to directly probe different diffusion length scales in the Galaxy for the first time. 

This paper is organized as follows: In section~\ref{theory} we give an overview of CR propagation and the \galprop{} code, Bayesian inference and the {\sc BAMBI}/{\sc SkyNet} codes. In section~\ref{method}, we discuss the propagation model used, its parameters and prior ranges (including source abundances), the iterative procedure we adopt to scan both the propagation parameters and the abundances, and the data sets adopted (including the likelihood function used). In section~\ref{results}, we present our results in terms of Bayesian posterior probability distributions and give the posterior mean and best fit parameters, along with associated errors. In section~\ref{discussion} we discuss our findings. Section~\ref{conclusions} gives our conclusions. In Appendix~\ref{validation}, we validate our neural networks/{\sc SkyNet} approach against a full (non-accelerated) scan.

\section{Theory and Algorithms}\label{theory}
%%%%%%%%%%%%%%%%%%%%%%%%%%%%%%%%%%%%%%%%%%%%%%%%%
\subsection{CR Propagation (\galprop{} model)} 
\label{sec:galprop}
A brief review of CR production and propagation relevant to the present paper is given in \citetalias{2011ApJ...729..106T} and
more information can be found in a review by \citet{SMP2007}.

The theoretical understanding of the CR propagation in the interstellar medium (ISM) became a framework that the \galprop{}
model for CR propagation is built around. \galprop{} numerically solves the system of partial 
differential equations describing the particle transport 
with a given source distribution and boundary conditions for all species of CRs.

In spite of its relative simplicity, the diffusion equation is remarkably successful at modeling transport processes in the ISM. 
The processes involved include diffusive reacceleration and, for nuclei, nuclear spallation, secondary particle production, radioactive decay, electron K-capture and stripping, in addition to energy loss from ionization and Coulomb interactions. 
%For CR electrons and positrons, important processes are the energy losses due to ionization, 
%Coulomb scattering, bremsstrahlung (with
%the neutral and ionized gas), inverse Compton (IC) scattering, and 
%synchrotron emission.

The \galprop{} source (injection) abundances are taken first as the solar system abundances, which are iterated \citep{Moskalenko2008} 
to achieve an agreement with the propagated abundances as provided by ACE at $\sim$200 MeV/nucleon \citep{Wiedenbeck2001}
assuming a propagation model. The source abundances derived for two propagation models, diffusive reacceleration and plain diffusion,
were used in many \galprop{} runs.

Galactic properties on large scales, including the diffusion coefficient, halo size,
Alfv\'en  velocity and/or convection velocity, as well as
the mechanisms and sites of CR acceleration, can be probed by measuring stable and radioactive secondary CR nuclei. The ratio of the halo size to the diffusion coefficient can be constrained by measuring the abundance of stable secondaries such as $_{5}$B. Radioactive isotopes ($^{10}_{4}$Be, $^{26}_{13}$Al, $^{36}_{17}$Cl, $^{54}_{25}$Mn) then allow the resulting degeneracy to be lifted \citep[e.g.,][]{Ptuskin1998,SM1998,Webber1998,MMS2001}.
However, the interpretation of the peaks observed in the 
secondary-to-primary  ratios (e.g., $_5$B/$_6$C, [$_{21}$Sc+$_{22}$Ti+$_{23}$V]/$_{26}$Fe) 
around energies of a few GeV/nucleon, remains model-dependent.

Closely connected with the CR propagation, but not related to the present paper, is the production of the
Galactic diffuse \gray{s} and synchrotron emission \citep{2013MNRAS.436.2127O}.
Proper modeling of the diffuse \gray{} emission, including the 
disentanglement of the different components, requires well developed 
models for the interstellar radiation field (ISRF) and gas densities, together with the 
CR propagation \citep[see, e.g.,][]{SMP2007,DiffusePaperII}. 
Global CR-related properties of the Milky Way galaxy are 
calculated in \citet{Strong2010}.

CR propagation in the heliosphere is described by the \citet{Parker1965} equation.
The modulated fluxes significantly differ from the interstellar spectra below energies of $\sim$20-50 GeV/nucleon, but correspond to the ones actually measured by balloon-borne and spacecraft instruments.

Spatial diffusion, convection with the solar wind, drifts, 
and adiabatic cooling are the main mechanisms that determine transport of CRs to the inner heliosphere. These effects have been incorporated into realistic (time-dependent, three-dimensional) models \citep[e.g.,][]{Florinski2003,Langner2006,Potgieter2004}.
The ``force-field'' approximation that is ordinarily used \citep{GleesonAxford1968}, instead characterizes the modulation effect as it varies over the solar cycle using a single 
parameter -- the ``modulation potential''.  Despite having no predictive power, the force-field approximation is a useful low-energy parameterization
of the modulated spectrum for a given interstellar spectrum. 
A new stochastic 2D Monte Carlo (HelMod) code \citep{2012ApJ...745..132B} is being developed that would allow 
an accurate calculation of the heliospheric modulation for an arbitrary epoch and is fully compatible with \galprop{}.

The \galprop{} project now has nearly 20 years of development behind it. The key idea behind  \galprop{} is that all CR-related data, including direct measurements, \gray{s}, sychrotron radiation, etc., are subject to the same Galactic physics and must therefore be modeled simultaneously.
The original FORTRAN90 code has been public since 1998, and a rewritten C++ version was produced in 2001. The latest major public 
release is v54 \citep{GalpropWebrun}. An 
advanced subversion is available through a WebRun at the dedicated website\footnote{http://galprop.stanford.edu/webrun}.  The website also
contains links to all galprop publications and has detailed information on CR
propagation and the \galprop{} code.

We refer to \citetalias{2011ApJ...729..106T} and the dedicated website for a detailed description of the code. 
In this work we use a development version of the \galprop{} code which is
described in \citet{Moskalenko2015}, and references therein.
The development version has the possibility to vary the
injection spectrum independently for each isotope. It
also includes the dependency tree pre-built from the nuclear reaction network and
made for each species to ensure that its dependencies are propagated before the
source term is generated.  This way, special cases of $\beta^-$-decay (e.g.,
$^{10}$Be$\to^{10}$B) are treated properly in one pass of the reaction
network, instead of the two passes required before.
This reduces the runtime of the \galprop{} code by up to a half.

\subsection{Statistical Framework}
%#########################################################################

Here we summarise briefly the underlying statistical framework, referring the reader to \citetalias{2011ApJ...729..106T} for full details \citep[see e.g.][for an overview of Bayesian methods]{Trotta2008}. 
Bayesian inference works by evaluating the posterior probability 
distribution function (pdf) for the parameters of interest, which is the normalised product of the prior pdf (summarising our state of knowledge before we see the data) and the likelihood function (which contains the information supplied by the data). 
Denoting by $\params$ the vector of parameters and by $\data$ the 
data, Bayes Theorem reads

\be \label{eq:bayes}
P(\params|\data) = \frac{P(\data | \params)P(\params)}{P(\data)},
\ee 

\noindent
where $P(\params|\data)$ is the posterior, $P(\data | \params) = \like(\params)$ is 
the likelihood function (when 
considered as a function of $\params$ for the observed data $\data$) 
and $P(\params)$ is the prior. The quantity in the denominator of eq.~\eqref{eq:bayes} is the 
Bayesian 
evidence (or model likelihood), a normalizing constant that does not 
depend on $\params$ and can be 
neglected when interested in parameter inference. 

Together with the model for the data \citep[entering the likelihood, possibly specified hierarchically, see e.g.,][]{Shariff:2015yoa} the priors for the parameters which enter Bayes' theorem, eq.~\eqref{eq:bayes}, must be specified. 
Priors should summarize our
state of knowledge and/or our theoretical prejudice about the parameters 
before we 
consider the new data, possibly informed by the posterior from a previous 
measurement.

The problem is then fully specified once we give the likelihood 
function (see section~\ref{sec:like} below). 
The posterior distribution $P(\params|\data)$ is determined numerically 
by drawing samples from it using an appropriate sampling scheme (see section~\ref{sec:bambi}).

\subsection{The {\sc BAMBI} algorithm} 
\label{sec:bambi}

In order to explore efficiently the propagation model parameter space for a higher spatial and energy resolution than adopted in \citetalias{2011ApJ...729..106T} (hence with a higher computational cost per likelihood evaluation, see section~\ref{sec:model}), in this work we upgrade our sampling techniques.  We use the {\sc BAMBI} algorithm~\citep{Graff2012}, which implements the nested sampling algorithm
{\sc MultiNest}~\citep{Feroz2008,Feroz2009,Feroz2013}, as described by \cite{Skilling2004,Skilling2006}, and the neural network training algorithm {\sc SkyNet}~\citep{Graff2014}
to learn the likelihood function on-line during the sampling and thus accelerate the sampling procedure. We briefly describe each algorithm below.

\subsubsection{{\sc MultiNest}}

{\sc MultiNest} is a highly efficient implementation of the nested sampling technique.
This technique is aimed at computing the Bayesian evidence, but is able to produce
samples from the posterior in the process of doing so~\citep[for details, see][]{Feroz2008}.
In nested sampling, a set of `live' points is initially sampled from the prior distribution.
The point with lowest (log-)likelihood, $\like_{\textrm{min}}$, is then replaced by a new point that is sampled from the prior under the constraint that
$\like_{\textrm{new}}>\like_{\textrm{min}}$. To facilitate this sampling from
the constrained prior, {\sc MultiNest} encloses the set of live points within a set of
(possibly overlapping) ellipsoids from which new samples can be taken analytically.
The ellipsoidal decomposition
is chosen in order to minimize the sum of the volumes and is well-suited to sampling from posterior
distributions that exibit curving degeneracies and/or multi-modality. If subsets of the
ellipsoid set do not overlap in parameter space, these can be identified as separate modes
and evolved independently. The sampling converges when the Bayesian evidence is
computed to within a user-specified tolerance.

{\sc MultiNest} takes advantage of parallel computing architectures by
allowing each CPU to compute a proposal replacement point simultaneously. As the run progresses, the
actual sampling efficiency (fraction of accepted samples over number of  proposal replacements) drops as the ellipsoidal approximation is less accurate and the
likelihood constraint on the prior is harder to meet. By computing $N$ proposal samples concurrently, we can obtain speed increases of up to a factor of $\simeq N$. This linear speed-up however flattens once $N \approx 1/\epsilon$, where $\epsilon$ is the efficiency of the algorithm (i.e., the number of accepted samples over the number of likelihood evaluations). Past this point, a further increase in the number of CPUs does not result in any appreciable speed advantage. The actual values used in our scans are given in section~\ref{sec:BAMBI}.

In
addition to providing the log-likelihood and prior, the user only needs to tune a few parameters for any specific implementation. These are the
number of live points (higher for higher-dimensional parameter spaces, and/or multi-modal posteriors), the target efficiency (controlling the degree of shrinkage of the ellipsoids), and the tolerance (controlling the precision to be achieved on the evidence). The number of
live points needs to be sufficient that all posterior modes are sampled (ideally
with at least one live point in the initial set) and we use 2000
for our analyses (which does not suffer from multi-modality). The target efficiency affects how conservatively the
ellipsoidal decomposition is made and a value of 0.5 was found to be
sufficient; smaller values will produce more accurate evidence values (irrelevant to the present study) but require more
samples. Lastly, we chose a tolerance of 0.5 in the evidence calculation, as recommended in~\cite{Feroz2009}.
\subsubsection{{\sc SkyNet} and  {\sc BAMBI}}
\label{sec:BAMBI}

{\sc SkyNet}~\citep{Graff2014} is an algorithm for training of artificial neural networks -- computational models that are used to approximate one or several target functions and that depend on a number of free input parameters. In our application, the input parameters are the free parameters in the model, $\Theta$, and  the target function is the log-likelihood, $\like$. {\sc SkyNet} implements a feed-forward neural network, where scalar values are passed from one layer to the next over
weighted connections with non-linear activation functions. {\sc BAMBI} is a framework that joins up {\sc MultiNest} with {\sc SkyNet}: accepted samples
from the {\sc MultiNest} run are fed as training samples into {\sc SkyNet}, which uses them to train the neural network on-line (i.e., as the posterior sampling progresses).

Training is performed using a fast, approximate second-order algorithm to find the neural network weights
that best approximate a value of $\like$ for a given input $\Theta$. This method efficiently finds
an optimal set of weights and is designed to minimize overfitting to the training data. $\ell$-$2$ norm regularization
aids the algorithm in finding the global optimum. A test data set, distinct from the training data, is used to stop
training when the algorithm begins to overfit to the training data. The algorithm is described in more detail in~\cite{Graff2014}.

The user must specify the size of the network, both in the number of hidden layers and the number of nodes in each.
We use a network with a single hidden layer of 200 nodes. This was verified to give a sufficiently accurate approximation, as shown in
Appendix~\ref{validation}. The sigmoid activation function, $f(x)=1/(1+\exp(-x))$, is used for the hidden layer and a
linear activation function, $f(x)=x$,  is used for the output layer.

Once {\sc SkyNet}'s training has reached sufficient accuracy on likelihood values provided by {\sc MultiNest},
within {\sc BAMBI} the network is tested for the
accuracy of its predictions. If the root-mean-square error is below a user-defined threshold, the network will
be used for calculating future likelihood calls by {\sc MultiNest}. Since the trained network is effectively an analytic interpolating function, calls to the neural-network approximated likelihood are almost instantaneous, thus greatly reducing the computational cost. If the predictions are insufficiently accurate, then more samples will be generated using the full likelihood
function and training will resume once enough new samples have been collected. This setup is explained further
and examples are provided in~\cite{Graff2012}. Setting the accuracy threshold too low will require more samples from
the original likelihood and longer network training, while setting it too high can produce unreliable likelihood approximations
that affect the accuracy of the posterior sampling. We use a tolerance of 0.8, which led to convergence in an acceptable amount of time, although it also led to some spurious maxima in the likelihood. These were removed by post-processing the posterior samples with a full evaluation of GALPROP (which can be done in an exact parallel way at post-processing stage, and thus can benefit from massive parallel processing). 
Our two main {\sc BAMBI} scans (see Sec. \ref{method}) were doubly parallelized. For the light elements (Be--Si) we used 96 CPUs, split over 12 MPI nodes, with each \galprop{} evaluation using 8 openMP thread. For the second scan, over $p, \bar p$ and He only, we were able to use 144 CPUs, with 18 MPI processes using 8 openMP cores each. In this configuration, full convergence of these scans required approximately 2 million GALPROP calls each, totalling 35 CPU years in the light element case, and 5.5 CPU years for protons and helium. Over 99\% of the computing power was used for GALPROP likelihood evaluations, with the remaining $\sim 1\%$ spent on {\sc BAMBI} training.  In both cases the neural networks performed approximately 20\% of the likelihood evaluations, saving around 10 CPU years, or 4.5 months of real computing time. More details are presented in Appendix~\ref{validation}. 

\section{Method} \label{method}

\subsection{Propagation Model and Parameters}\label{sec:model}

%#########################################################################

The aim of this study is to simultaneously constrain the propagation parameters,
as in \citetalias{2011ApJ...729..106T}, as well as the CR source abundances, since the latter are 
model-dependent.

Our benchmark model for this study is the diffusive reacceleration (hereafter DR) model, 
which is by far the most commonly used propagation model used with GALPROP
\citep[e.g.,][and references
therein]{Moskalenko2002,SMR2004,Abdo2009midlatitudes,Vladimirov:2012,DiffusePaperII,2015ApJ...799...86A,2016ApJ...819...44A}. 
The distribution of Galactic CR sources is based on pulsars \citep{Lorimer2004}.
For this study, we use $f_{CR}(R) = (R/R_0)^\alpha e^{-\beta(R-R_0)}$, i.e., 
normalized to 1 at $R=R_0=8.5$~kpc, where $\alpha=0.475$, and $\beta= 2.166$.
The profile is constant for $R>10$~kpc and cuts off at $R=15$~kpc.  The
flattening in the outer Galaxy is 
suggested from Fermi studies \citep{2010ApJ...710..133A,2011ApJ...726...81A}.

In this model the spatial diffusion coefficient is given by
\begin{equation}
D_{xx} = \beta D_{0} \left( \frac{\rho}{\rho_0}\right)^\delta~,
\label{Dxx}
\end{equation}
\noindent
where $D_{0}$ is a free normalization at the fixed 
rigidity $\rho_0 = 4\times 10^3$ MV. 
For Kolmogorov diffusion the power law index is
$\delta=1/3$; however, we allow $\delta$ to freely vary. 
Re-acceleration is modeled as a momentum-space diffusion
where the coefficient $D_{pp}$ is related to the spatial coefficient $\Dxx$ 
\citep{Berezinskii1990,Seo1994} with
\begin{equation}%=====================================================+
\label{eq.2}
\Dpp\Dxx = {4 p^2 \valf^2\over 3\delta(4-\delta^2)(4-\delta) w}\ ,
\end{equation}%=======================================================+
where $w$ characterizes the level of turbulence (we take $w = 1$ since only the quantity
$\valf^2 /w$ is relevant);  the Alfv\'en velocity $\valf$ is allowed to vary freely.

The CR injection spectrum is modelled as a broken power-law, with 
index below ($-\nu_0$) and above ($-\nu_1$) the break as free parameters. This
is known to be necessary in DR models in order to compensate for the large
bump at low rigidities, a consequence of the large Alfv\`en velocities needed
to fit the B/C ratios below 1 GeV. While the location of this break is
typically fixed for a given study (around $\rho_{br} = 10$ GV), we allowed it
to vary in our scan. Other models are able to reproduce the B/C ratio without
a low-energy break in the injection spectra, but at the cost of an {\it ad hoc} 
break in the diffusion coefficient; these will be examined in detail
in an upcoming study.  Because we are using high-energy ($>$TeV) data, we
must also include a second break in the injection spectrum at 220 GV, and thus
a third freely-varying index $-\nu_2$  \citep[see a discussion of the possible origin of this break in][]{Vladimirov:2012}.  We also allow
for a different injection spectrum for protons and heavier elements by setting
the power-law indices of the proton injection spectrum to $\nu'_i = \nu_i +
\delta_\nu$ for $i \in \{0,1,2\}$.

The other free model parameters are the halo height $z_h$
and the normalization $N_p$ of the propagated CR proton 
spectrum at 100~GeV. This yields a total of 10 free propagation parameters, summarized in Table~\ref{tab:params}; we label these 
\be
\paramsP = \{N_p, D_{0}, \delta, \valf, z_h, \nu_0, \nu_1, \nu_2, \rho_{br}, \delta_\nu \}. 
\ee
For each parameter in $\paramsP$ we use a uniform prior distribution, whose range is informed by the results of \citetalias{2011ApJ...729..106T}. Although we let more
parameters vary in this analysis\footnote{Specifically, $\rho_{br}$, $\nu_0$, $\delta_{\nu}$ -- see Table~\ref{tab:params}, as well as the 10 abundance parameters.}, we do not expect the
posterior distributions to stray very far from the determination of \citetalias{2011ApJ...729..106T}. 
The prior ranges (informed by the results we obtained in \citetalias{2011ApJ...729..106T}) are given in Table~\ref{tab:params}, and are discussed in greater
detail in Sec. \ref{sec:priors}.

Thanks to the speed up from {\sc BAMBI} and the \galprop{} code improvements mentioned in Sec. \ref{sec:galprop}, we are able to use a finer grid in this work than in \citetalias{2011ApJ...729..106T}, giving better accuracy.  We found that a spatial resolution of $\Delta
r=1$~kpc and $\Delta z=0.2$~kpc and an energy grid with $E_{i+1}/E_i =
1.2$ was a reasonable compromise between accuracy and speed. The full set of numerical parameters that we adjusted is shown in Table \ref{restable}. 

Due to their smaller inelastic
cross sections, secondary antiprotons probe different length scales than the
light elements; the diffusion parameters that characterize their propagation
can therefore be different and indeed, we found in our test scans that the
same parameter set would not allow a good simultaneous fit to the high and low
mass data. We therefore split the propagation scan into two: one,
propagating only protons, antiprotons and helium; and one ``light-element''
scan, propagating elements from beryllium up to silicon. This has the further
advantage that the \ppbarHe scans do not require computation of the full
nuclear network for each likelihood evaluation, allowing them to run quickly and in parallel with the light
element scans. Thus, at the chosen resolution, our light element scan took approximately 9.8 CPU minutes per evaluation (or 1.22 minutes when parallelized over 8 CPUs), while the \ppbarHe case was sped up to 1.25 CPU minutes (or 9.4 seconds in real time).

The nuclear chain that we use for the light element scans begins at $^{30}$Si and proceeds down to
protons.  The source abundances of nuclei $6\le Z\le 14$ have an important 
influence on the B/C and $^{10}$Be/$^9$Be ratios used in this study. We
therefore let the abundances of the ten most important elements vary freely,
with prior ranges determined by the measured CR abundances from ACE
data at a few 100 MeV/nucleon \citep{George2009}. The isotopes that are
allowed to vary are $^{1}$H, $^{4}$He, $^{12}$C, $^{14}$N, $^{16}$O, $^{20}$Ne, $^{22}$Ne, Na, $^{24}$Mg, $^{25}$Mg,$^{26}$Mg, $^{27}$Al, $^{28}$Si, $^{29}$Si and $^{30}$Si; their prior ranges
are presented in Table~\ref{tab:params}. The abundances $X_i$ are scaled to the proton injection abundance $X_{\mathrm{H}}$, whose absolute
normalization, $N_p$, is fixed by its \textit{final} flux at Earth, at the
reference energy $E_{\mathrm{ref} }= 10^2$ GeV. We label the 10-dimensional abundance
parameters set $\paramsA$, defined with respect to $X_{\mathrm H} \equiv 1.06\times 10^6$.

For each of the experiment that provides data below a few GeV/nucleon, we must introduce an additional nuisance parameter $\phi_j$ ($j=1,\dots, 5$) to account for solar modulation. Furthermore, we introduce a set of parameters $\tau_j$ ($j=1,\dots, 5$) designed to mitigate the possibility that the fit is dominated by unknown 
systematic errors in the data, as explained in detail in Section \ref{sec:like}, and following the procedure introduced in \citetalias{2011ApJ...729..106T}. We denote the joint set of nuisance parameters by $\nuis$.

\begin{figure}
\includegraphics[width=.5\textwidth]{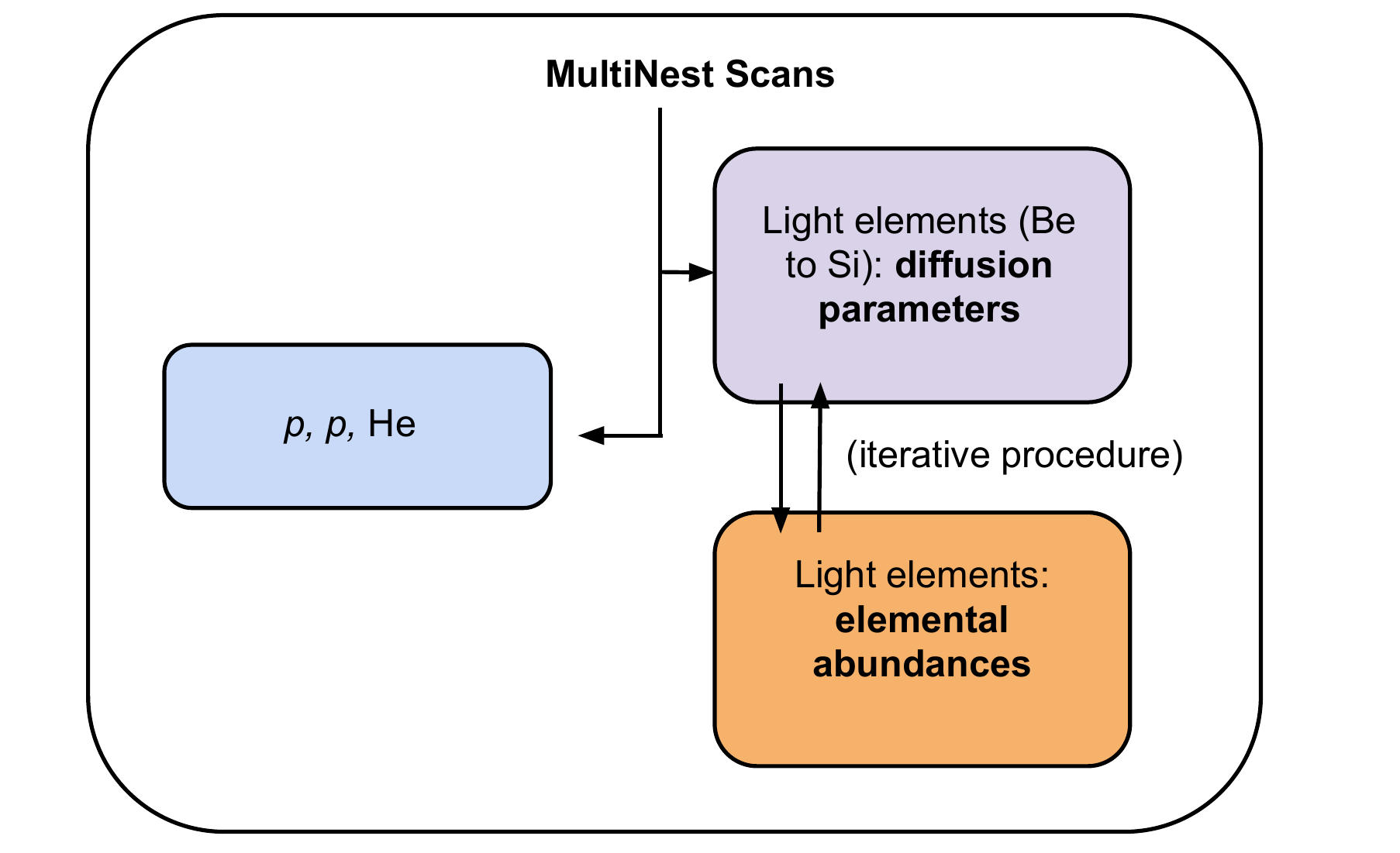}
\caption{The sets of neural network-assisted nested sampling scans that we perform in this work. We separate \ppbarHe (left) from the light elements (right) into two separate runs. For the light elements, we also vary the elemental abundances in separate, faster runs, which are performed iteratively with the propagation parameter scans. Since the \galprop{} output is linear in the injection abundances, this allows extremely rapid convergence of the abundances. We will keep the same color code throughout the text: blue for \ppbarHe results, magenta for light elements, and orange for the abundances.}
\label{fig:galbayes_procedure}
\end{figure}

Adding the abundance parameters constitutes a significant enlargement of the parameter space to be sampled: our full parameter space has 30 dimensions, and it would be computationally very costly to sample it simultaneously, even with {\sc MultiNest} and {\sc BAMBI}. Instead, we take advantage of the fact that for a given set of propagation parameters the final CR composition depends {\em linearly} on the injection abundance of each
isotope. Thus the likelihood as a function of $\paramsA$ for fixed $\paramsP$ is obtained quickly by linear rescaling of the CR spectra with $\paramsA$. This requires 
to run GALPROP only once per nuclear species ($O(10)$ runs) and therefore the posterior for $\paramsA$ conditional on $\paramsP$ can be explored very quickly. Then we fix the abundances to their posterior mean, and sample from the posterior of $\paramsP$ conditional on $\paramsA$. In all cases, we leave free the applicable nuisance parameters (solar modulation potentials $m_j$ and error rescaling parameters $\tau_j$). This procedure is then iterated with new abundances determined using propagation parameters fixed to the posterior mean of the scan over
propagation parameters (Figure~\ref{fig:galbayes_procedure}). This effectively amounts to implementing a Gibbs sampling scheme as follows: 
\begin{align}
\paramsA' & \sim P(\paramsA | \data, \paramsP) \\ 
\paramsP' & \sim P(\paramsP | \data, \paramsA'),
\end{align}
where a prime denotes the updated value of the parameter set. 

We start our procedure with a
scan over the abundance parameters $\paramsA$, fixing the propagation
parameters to the posterior means of a low-resolution test scan over the
propagation parameters $\paramsP$ using the same isotopic abundances as in
\citetalias{2011ApJ...729..106T}. This was followed by a propagation parameter scan at full numerical precision using the results of the first abundance scan,  after which we performed a final abundance scan, which yielded no significant variation with respect to the first scan -- and thus no need for a third iteration.
The structure of our three scans is illustrated in Figure~\ref{fig:galbayes_procedure}.

\subsection{CR Data and Modulation}\label{CR_data}
%#########################################################################

The data selection is based on similar principles as in \citetalias{2011ApJ...729..106T}.  We use the
most accurate CR data sets available, preferably taken near the solar minimum to
reduce the effect of solar modulation. Table~\ref{tab:PropData} lists the data
we use in the analysis \citep[obtained from a database by][]{Maurin2013}.

To reduce the number of nuisance parameters we limit our data to instruments which cover many different CR species.  As in the first paper we use data from ACE-CRIS
\citep{George2009} for the lowest energies.  Those data
agree well with data from other instruments while providing better
statistics and elemental coverage.  At intermediate energies, the HEAO3-C2 data
\citep{Engelmann1990} provides good statistics while also agreeing with
observations of other instruments.  The recent elemental data observed by PAMELA
\citep{Adriani2014} has better statistics, but was not available at the start
of this analysis; nor were the recent determination of the $p$ flux by AMS-02 \citep{2015PhRvL.114q1103A}. Our
results should not be affected by the former because the data from the HEAO and PAMELA
instruments are compatible. We return to the recently-released proton data from AMS-02 in Section~\ref{results}.

  For higher energies we decided to use only
CREAM data \citep{Ahn2008,Ahn2009,Yoon2011}, since its small energy binning was compatible with our method of evaluating the likelihood from a single energy point per bin, in contrast with the wide binning of, e.g., TRACER \citep{Ave2008,Obermeier2011,Obermeier2012}. 
For additional constraints on the propagation injection spectrum we also use
H and He data from PAMELA at intermediate energies \citep{Adriani2011} and
CREAM at higher energies \citep{Yoon2011}.  The PAMELA data was used because
of its superior statistics and the high energy CREAM data can be used without
an additional modulation nuisance parameter.

For $^{10}$Be/$^9$Be, we include ACE \citep{Yanasak2001}, which yields the most accurate measurements at low energies.
These  data are in agreement 
with Voyagers 1 and 2  \citep{Lukasiak1999}, as well Ulysses \citep{Connell1998} 
data. 
 ISOMAX \citep{Hams2004} has given two data points at higher energies (per nucleon), with very large error  
bars, which we nonetheless include. 

Like in \citetalias{2011ApJ...729..106T} we fit to the CR data in the whole energy range from few tens
MeV/n to a few TeV/n and account for heliospheric modulation below
$\sim$20 GeV/n.  We employ here the same method as in \citetalias{2011ApJ...729..106T} and use a simple force-field
approximation \citep{GleesonAxford1968}, which is characterized with the value
of the modulation potential.  To avoid the uncertainty associated with the
specific choice of the modulation potential we allow some flexibility to the
fits and include it as a free nuisance parameter (one free parameter per experiment).
Gaussian priors with mean and standard deviation motivated by ballpark
estimates of the modulation potential are used to avoid unphysical or
implausible values. Because
CREAM data start above 20 GeV/n we do not include a modulation parameter for
that experiment as it is irrelevant.

\subsection{The Likelihood Function} \label{sec:like}
%#########################################################################

We denote by $\params = \{\paramsP, \paramsA \}$ the joint set of CR propagation parameters and abundances, and by $\nuis = \{\phi, \tau \}$ the joint set of nuisance parameters. For a given value of $\{\params, \nuis \}$ we use \galprop{} to compute the CR spectrum as a function of 
energy, $\Phi_Y(E , \params, \nuis)$ for species $Y$. To mitigate against undetected systematics, we follow the procedure 
described in e.g., \citet{Barnes2003}. 
For each data set we introduce in the likelihood a 
parameter $\tau_j$ ($j=1,\dots,5)$, whose function is to rescale 
the variance of the data points in order to account for possible 
systematic uncertainties (see \citetalias{2011ApJ...729..106T} for a more detailed description). The role of the set of parameters $\tau = \{ \tau_1, \dots, \tau_5 \}$, 
which we call ``error bar rescaling parameters'', is to allow for the 
possibility that the error bars reported by each of the experiments 
underestimate the true noise. 
Furthermore, $\tau$ also takes care of 
all aspects of the model that are not captured by the reported 
experimental error: this includes also theoretical 
errors (i.e., the model not being completely correct), errors in the 
cross section normalizations, etc. 

Assuming Gaussian noise on the observations, we take the following likelihood 
function for each observation of species $Y$ at energy $E_i$

\begin{eqnarray} 
\label{eq:Gauss_like_rescaled}
& P(\hat{\Phi}_Y^{ij} |& \params, \nuis) = \\
&& \frac{\sqrt{\tau_j}}{\sqrt{2\pi}\sigma_{ij}} \exp\left( -\frac{1}{2}\frac{ \left(\Phi_Y(E_i, \params, \phi) - \hat{\Phi}_Y^{ij} \right)^2}{\sigma_{ij}^2/\tau_j} \right),
\nonumber
\end{eqnarray}
\noindent
where $\Phi_Y(E_i, \params, \nuis)$ is the prediction from the CR 
propagation model for species $Y$ at energy $E_i$, $ \hat{\Phi}_Y^{ij}$ is 
the measured spectrum, and $\sigma_{ij}$ is the reported standard deviation. 
The index $i$ runs through the data points within each 
data set $j$. 
We assume bins are independent, such that the full likelihood function 
is given by the product of terms of the form given above:  
\be \label{eq:like1}
P(\data | \params,\nuis) = \prod_{j=1}^{5}\prod_{i=1}^{N_j} P(\hat{\Phi}_Y^{ij} | \params, \phi).
\ee

\subsection{Choice of Priors} \label{sec:priors}

The full posterior distribution for the CR propagation 
model parameters $\params$, the variance rescaling 
parameters $\tau$ and the modulation parameters $\phi$ is  written
 
\be \label{eq:fullposterior}
P(\params, \vartheta | \data) \propto P(\data | \params, \phi, \tau) P(\params)P(\tau)P(\phi).
 \ee

\noindent
The likelihood $P(\data | \params, \tau, \phi)$ is 
given by Eqs.~\eqref{eq:Gauss_like_rescaled} and \eqref{eq:like1}. 
  
The priors $P(\params)$, $P(\phi)$ and $P(\tau)$ in 
Eq.~\eqref{eq:fullposterior} determined in the following way.
Priors on the model parameters $P(\params)$ are taken as uniform on $\params$, with ranges given in Table~\ref{tab:params}. 
As shown below, the posterior is close to Gaussian and well-constrained for $\params$; the results should thus be fairly independent of the choice of priors.

We take a Gaussian prior on each of the modulation parameters. This is informed by the values provided by 
each experiment (see Table~\ref{tab:params}), 
in order to avoid physically unreasonable values. 
A description of the CR data sets is given in Section \ref{CR_data}.
 
The $\tau_j$ are scaling parameters in the likelihood; 
the applicable prior is therefore given by the Jeffreys' prior, which is uniform 
on $\log\tau_j$ (see \citealt{Barnes2003} or \citealt{Jaynes2003}). 
We thus adopt the proper prior 

\be
P(\log\tau_j) = \left\{ 
\begin{array}{c l}
  2/3 \text{ for}  & \mbox{ for } -3/2 \leq \log\tau_j \leq 0  \\
  0 &\mbox{ otherwise}
\end{array}
\right.
\ee 

\noindent
corresponding to a prior on $\tau_j$ of the form

\be
P(\tau_j) \propto \tau_j^{-1}.
\ee

Including the nuisance parameters $\phi$ and $\tau$ (which are then marginalized over) in our analysis yields a more robust fit (as $\tau$ can absorb the effects of potential systematic effects in the data and $\phi$ incorporates solar modulation), while simultaneously giving more conservative constraints on the CR parameter space, since we fully account for degeneracies with all values of the nuisance parameters that are compatible with the data.

%%%%%%%%%%%%%%%%%%%%%%%%%%%%%%%%%%%%%%%%%
\section{Results} \label{results}
\begin{figure*}[h]
%\begin{minipage}{2.0\textwidth}
\begin{tabular}{c c c }
\includegraphics[width=.35\textwidth]{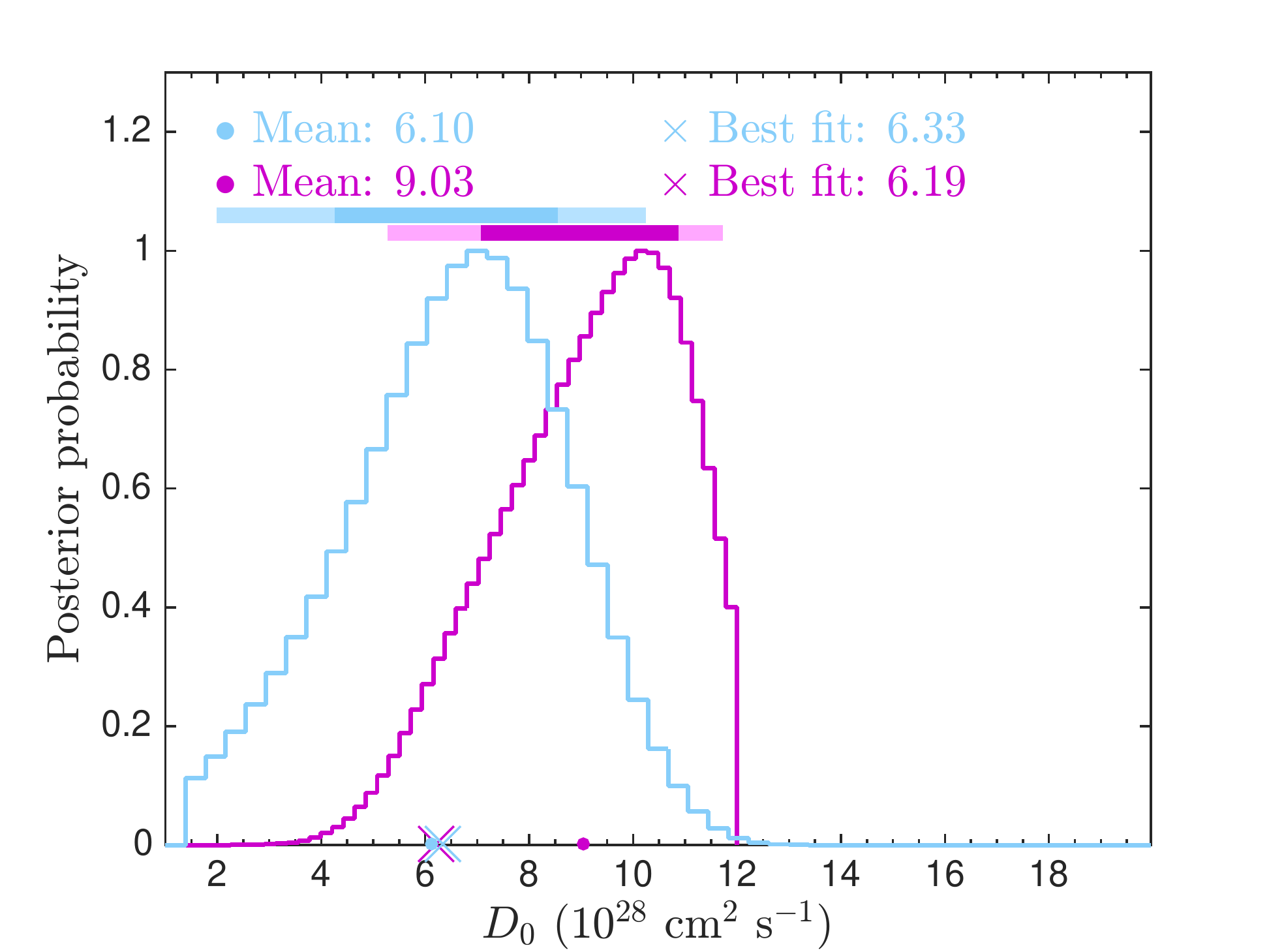} & \includegraphics[width=.35\textwidth]{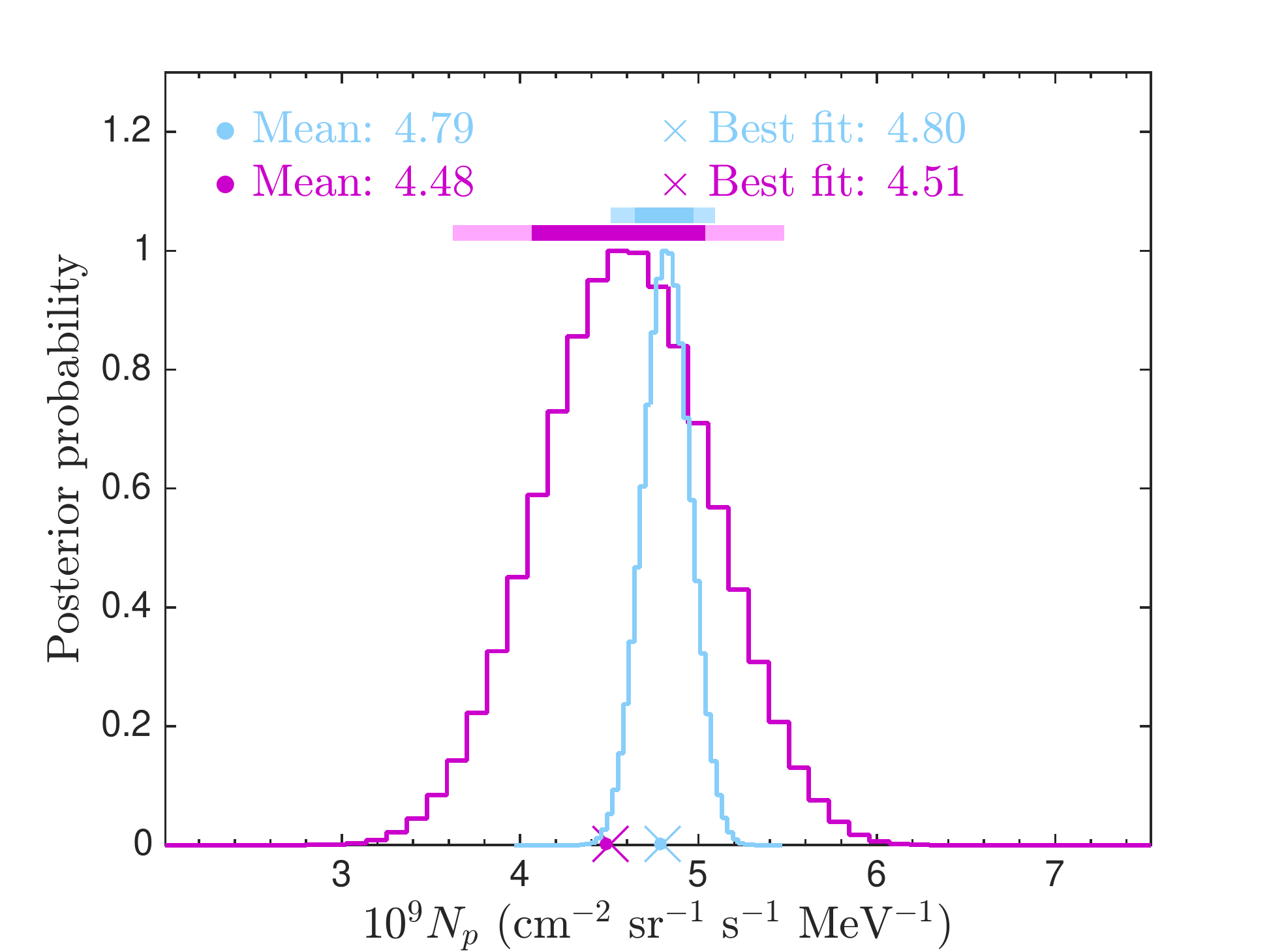} &\includegraphics[width=.35\textwidth]{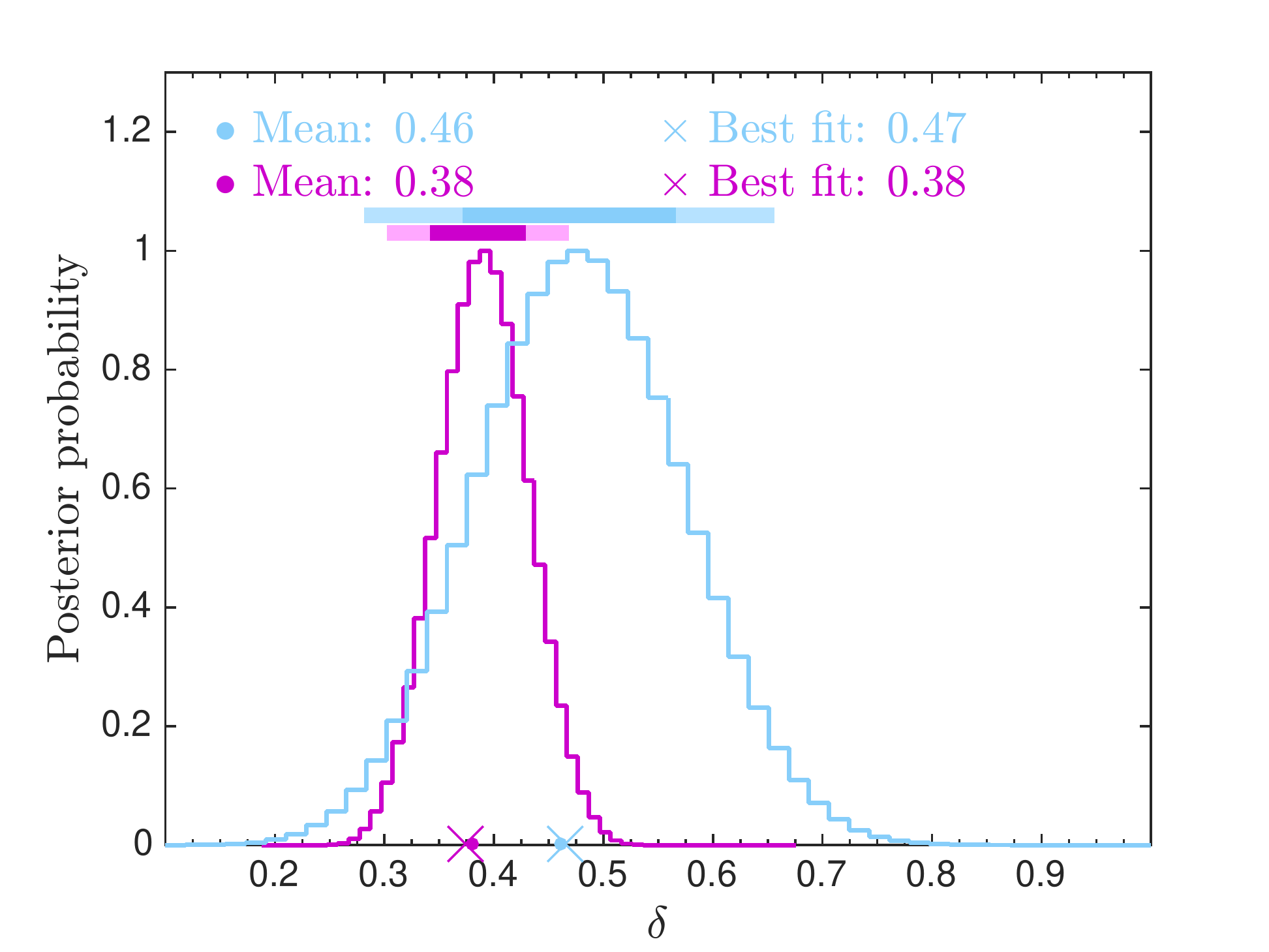} \\
\includegraphics[width=.35\textwidth]{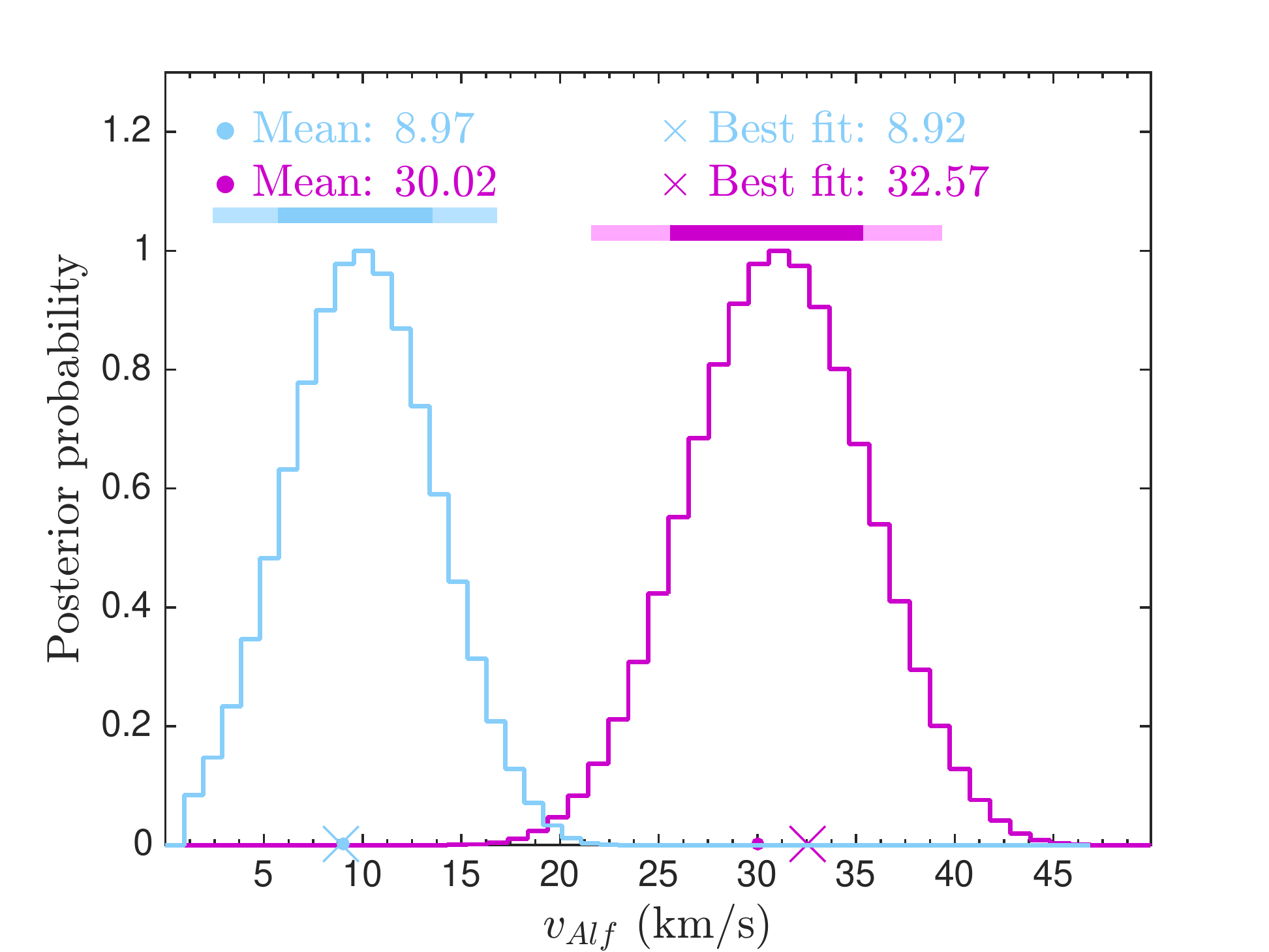} & \includegraphics[width=.35\textwidth]{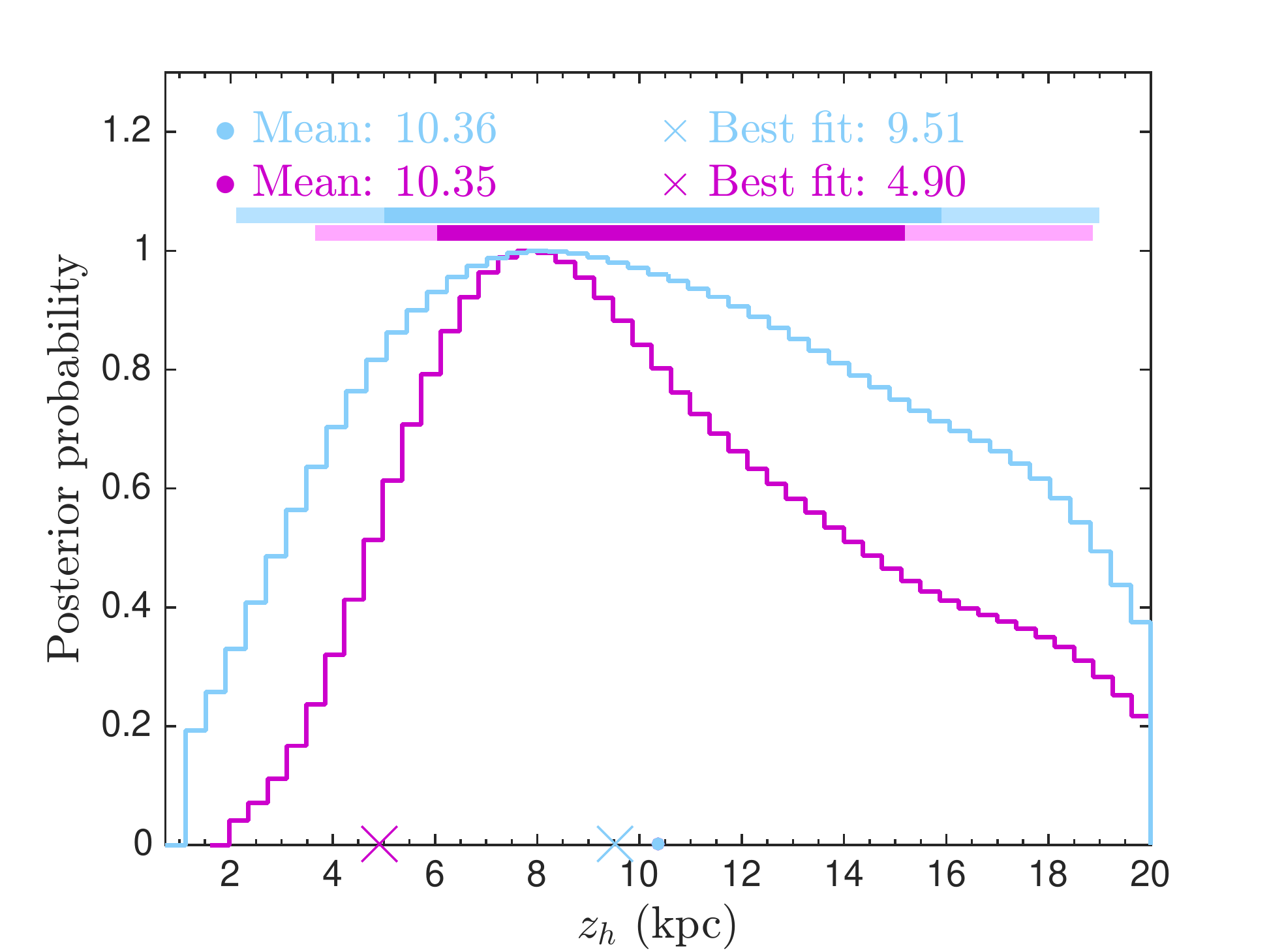} &\includegraphics[width=.35\textwidth]{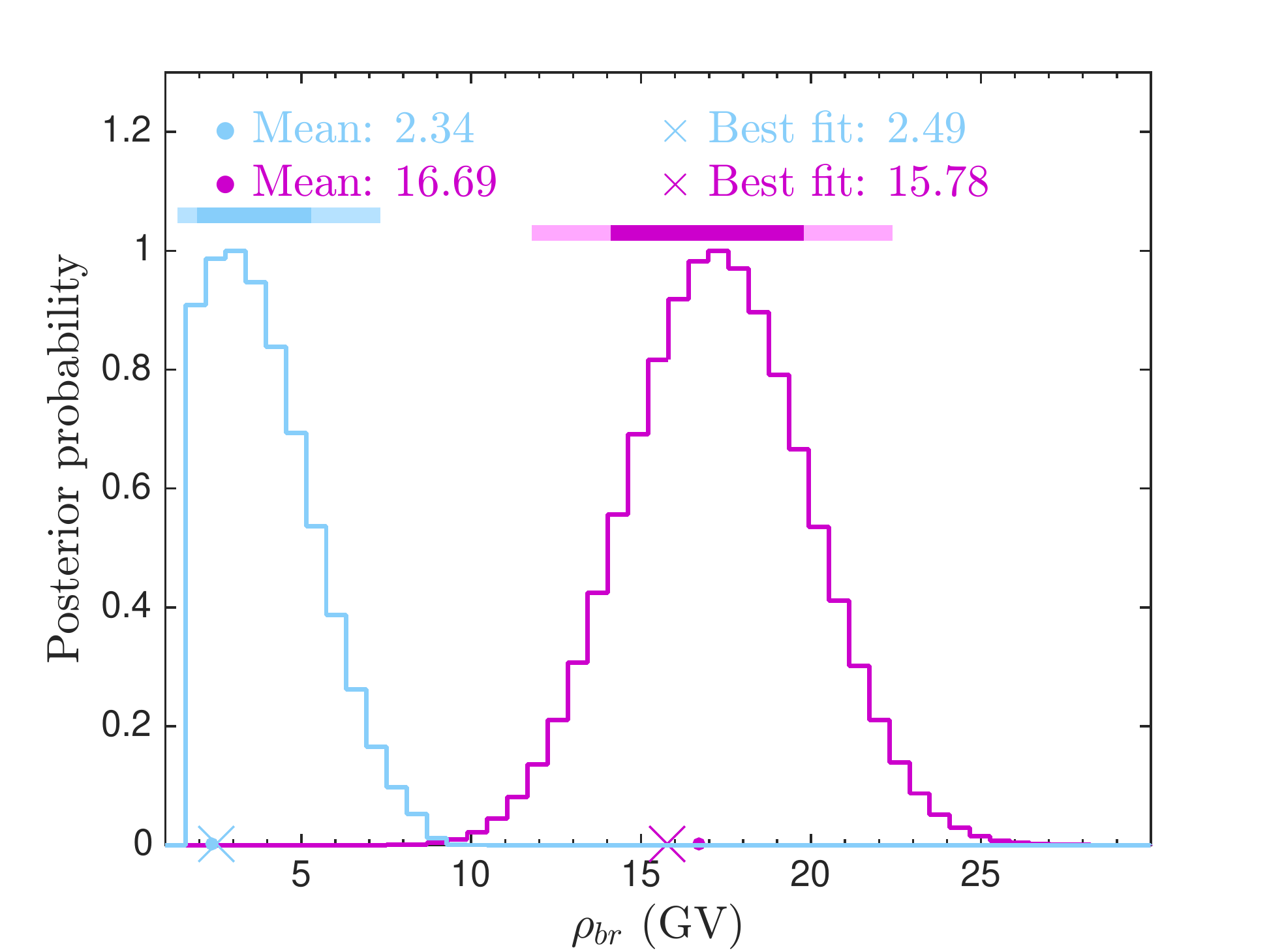} \\
\includegraphics[width=.35\textwidth]{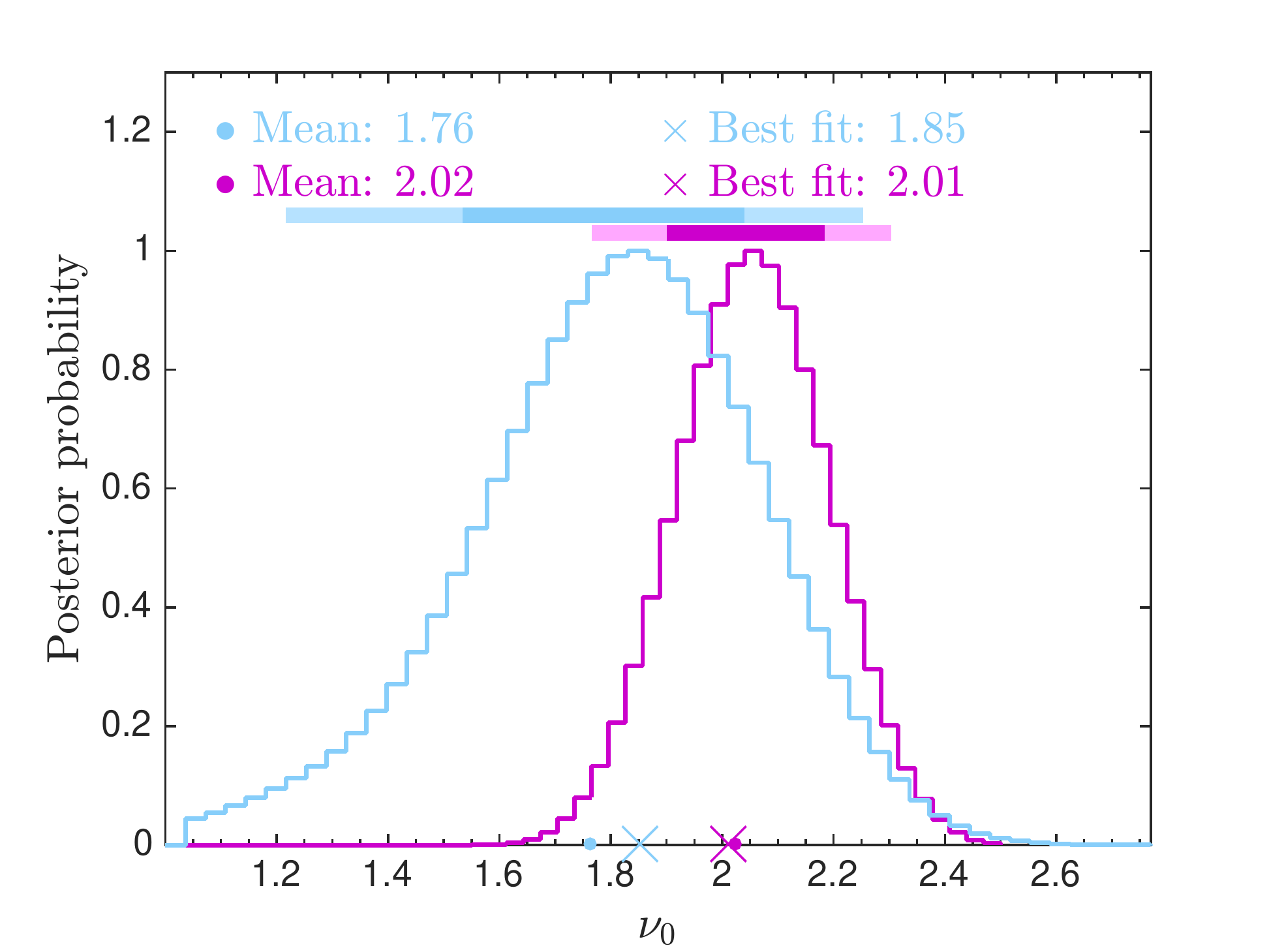} & \includegraphics[width=.35\textwidth]{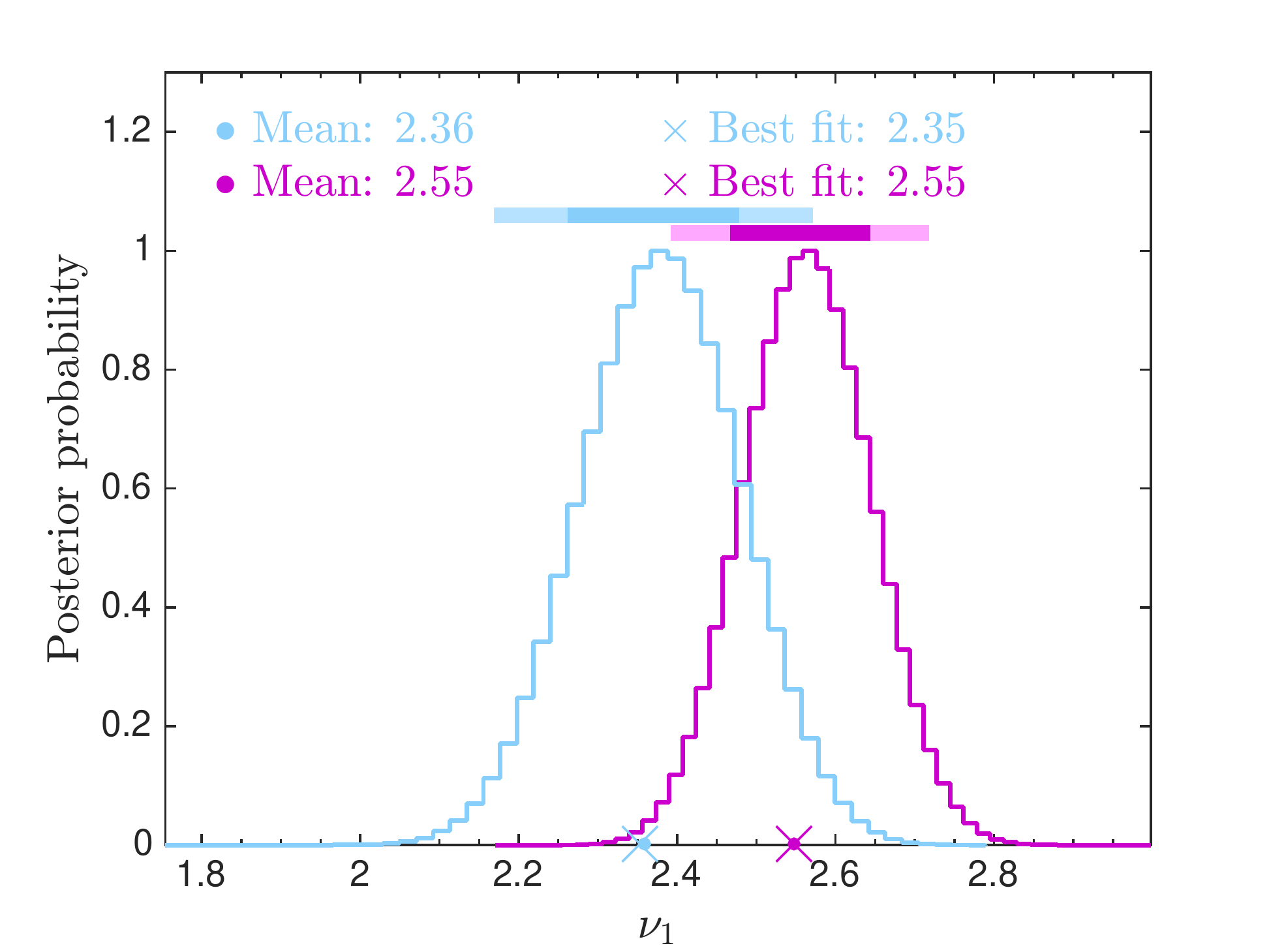} &\includegraphics[width=.35\textwidth]{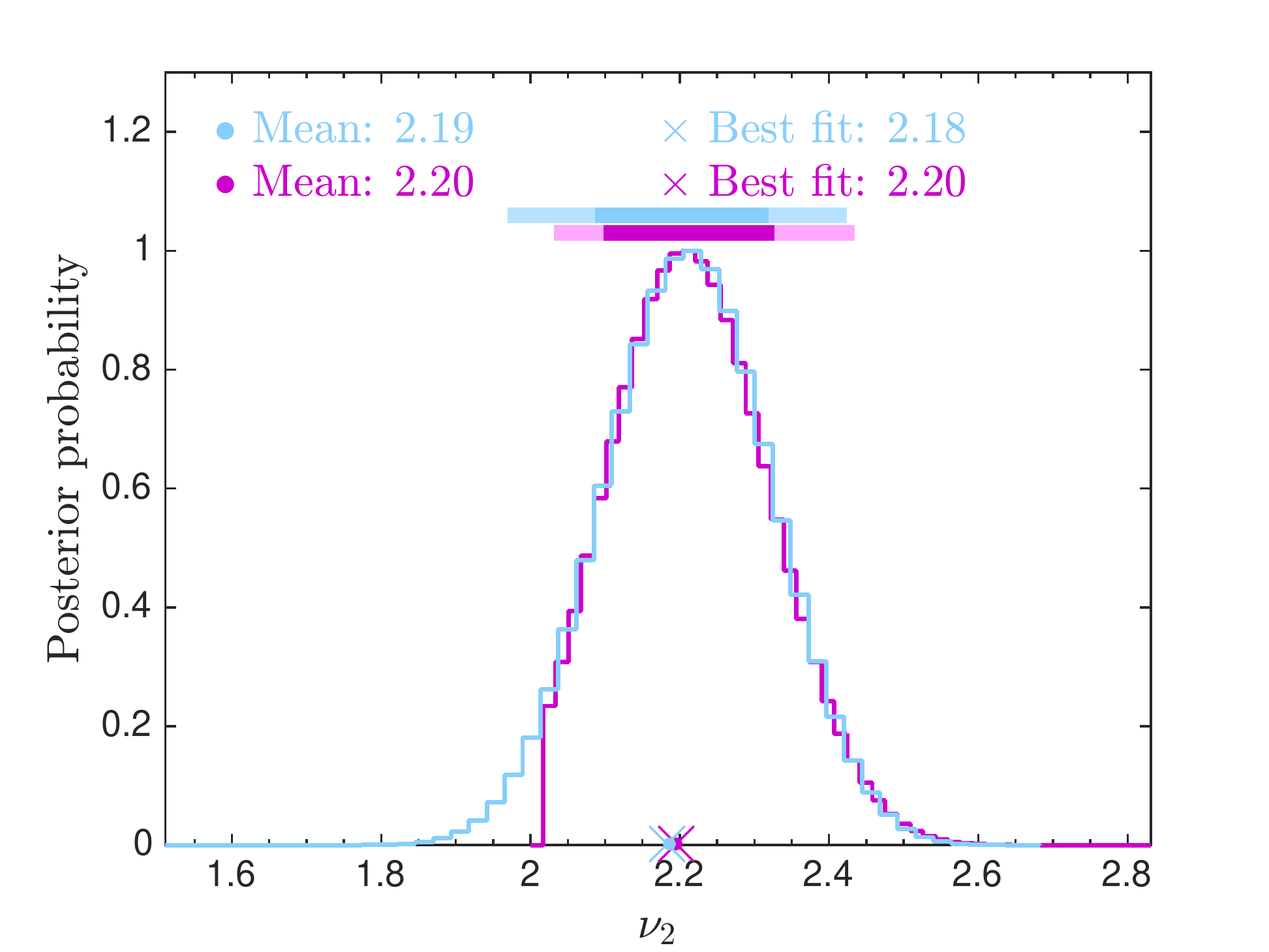} \\
 \includegraphics[width=.35\textwidth]{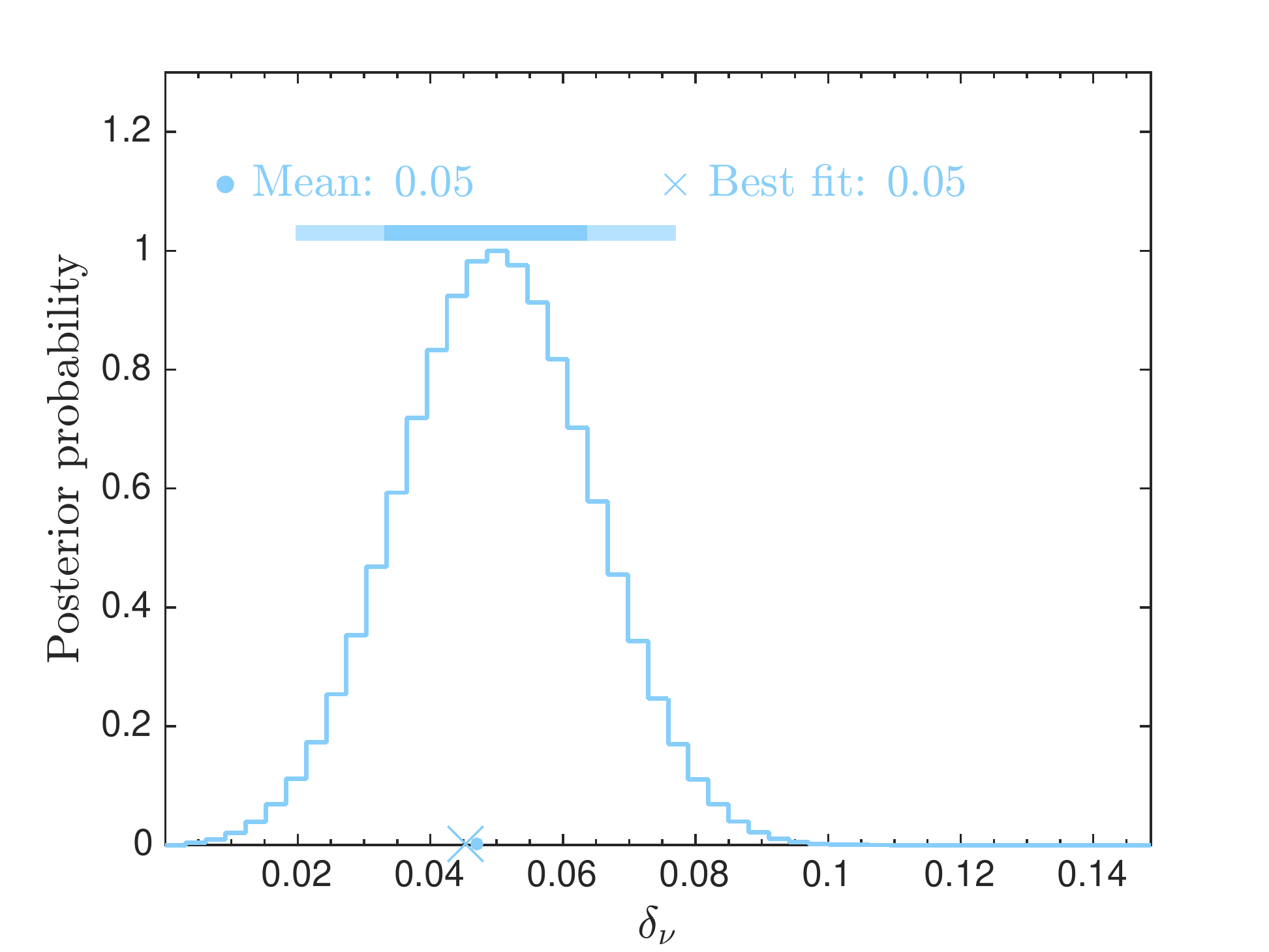} &\includegraphics[width=.35\textwidth]{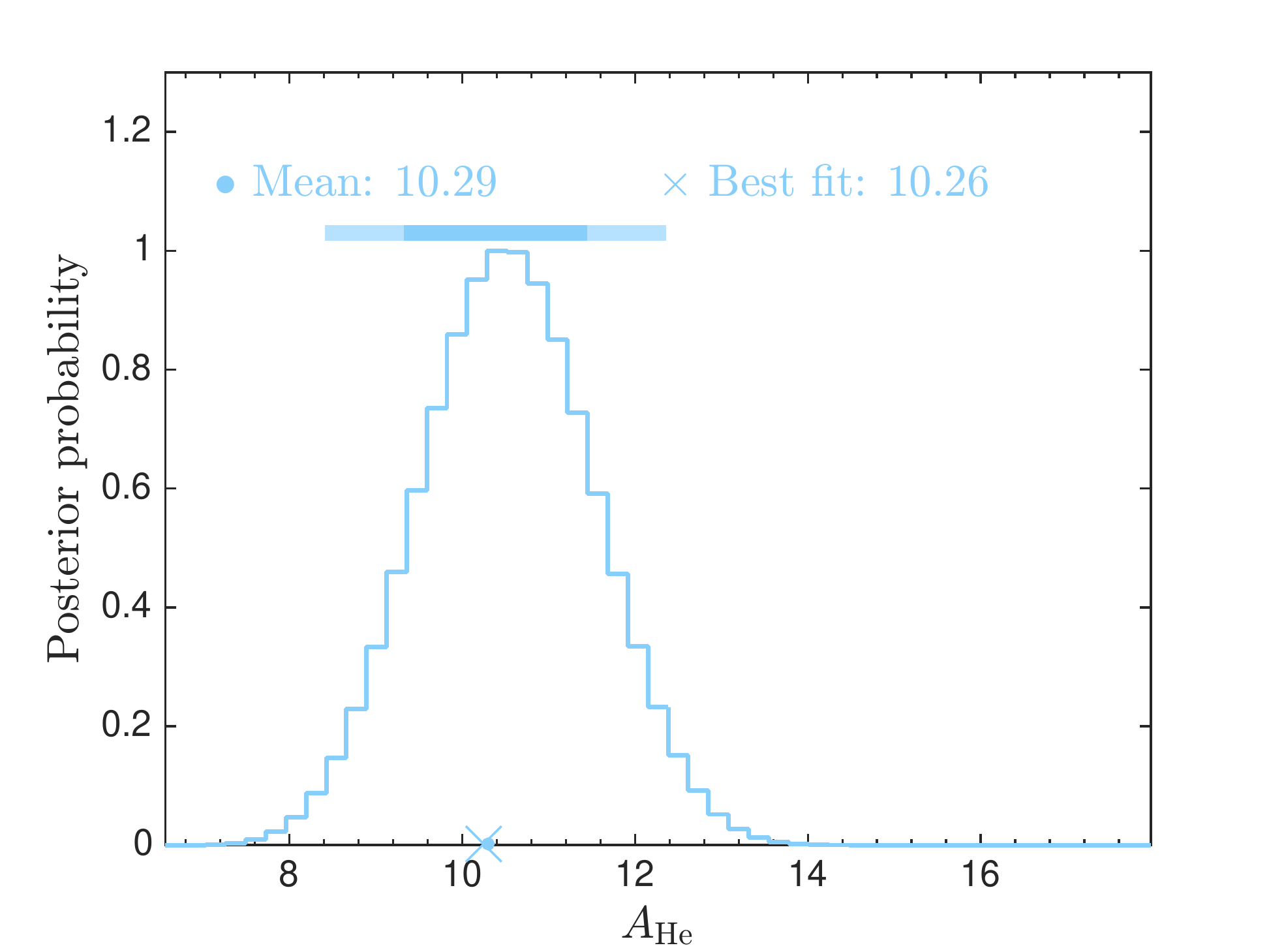} &\\
\end{tabular}
\caption{One-dimensional marginalized posterior distributions, showing 1 and 2-sigma
credible intervals, for the propagation parameters that were varied in the propagation scan. Light blue: the constraints from \ppbarHe scan, using PAMELA and CREAM data only; Purple: light-element
scan, fitting Be, B, C, N and O data. (Given in Table~\ref{tab:PropData}).
While most of the propagation parameters overlap between runs, there is a
clear ($> 2\sigma$) separation seen in the Alfv\'en speed and in the
low-energy injection break rigidity $\rho_{br}$. Differences in the $D_0
-z_h$ plane can be clearly seen in Figure~\ref{fig:2dpost}.  The injection
index for $p$ and He is also consistently lower below the 220 GV
break, suggesting a harder source injection spectrum.}\label{fig:1dpost}
%\end{minipage}
\end{figure*}

\begin{figure*}[h]
\begin{tabular}{c c c c}
\includegraphics[width=.25\textwidth]{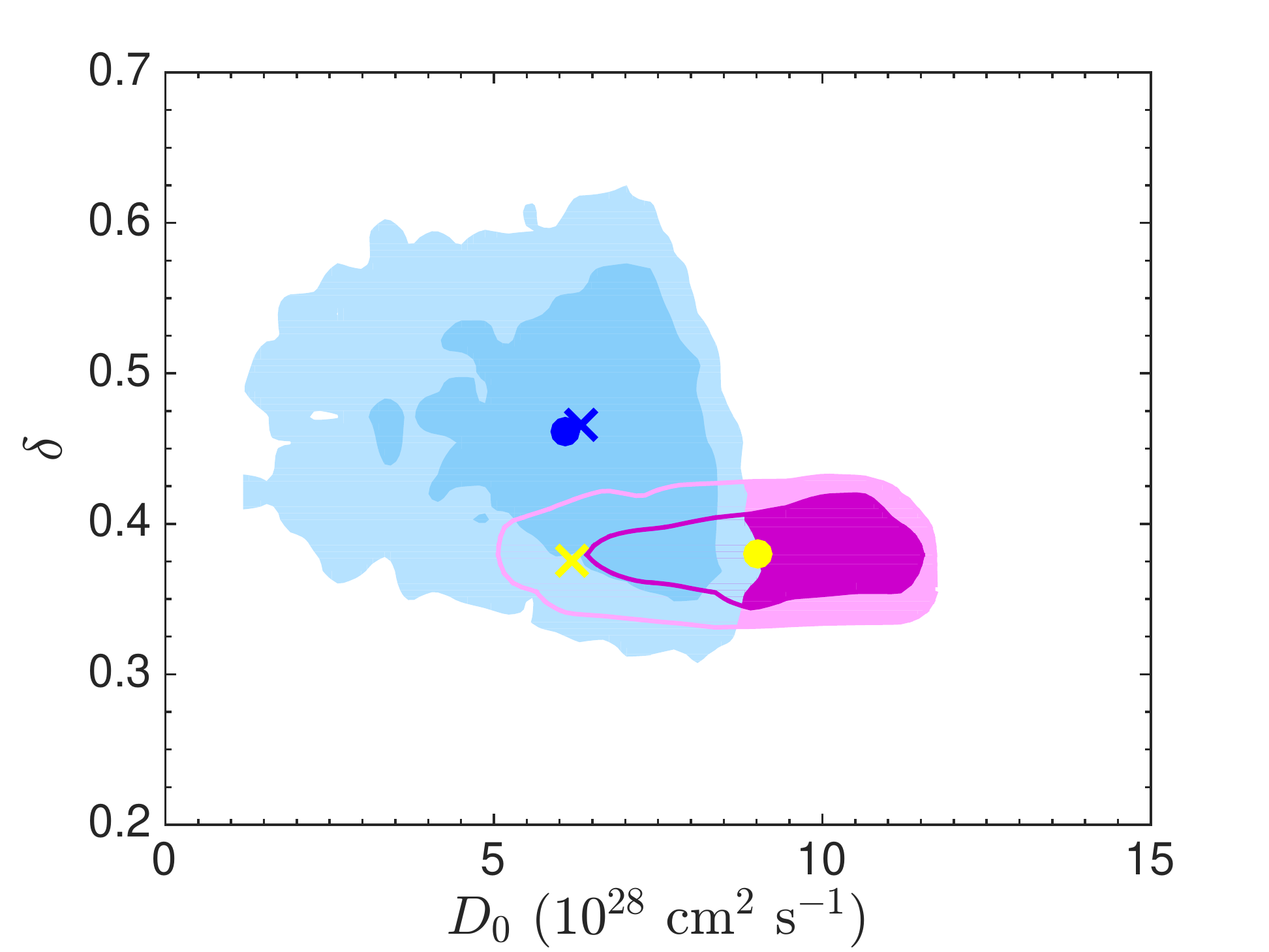} & & & \\
\includegraphics[width=.25\textwidth]{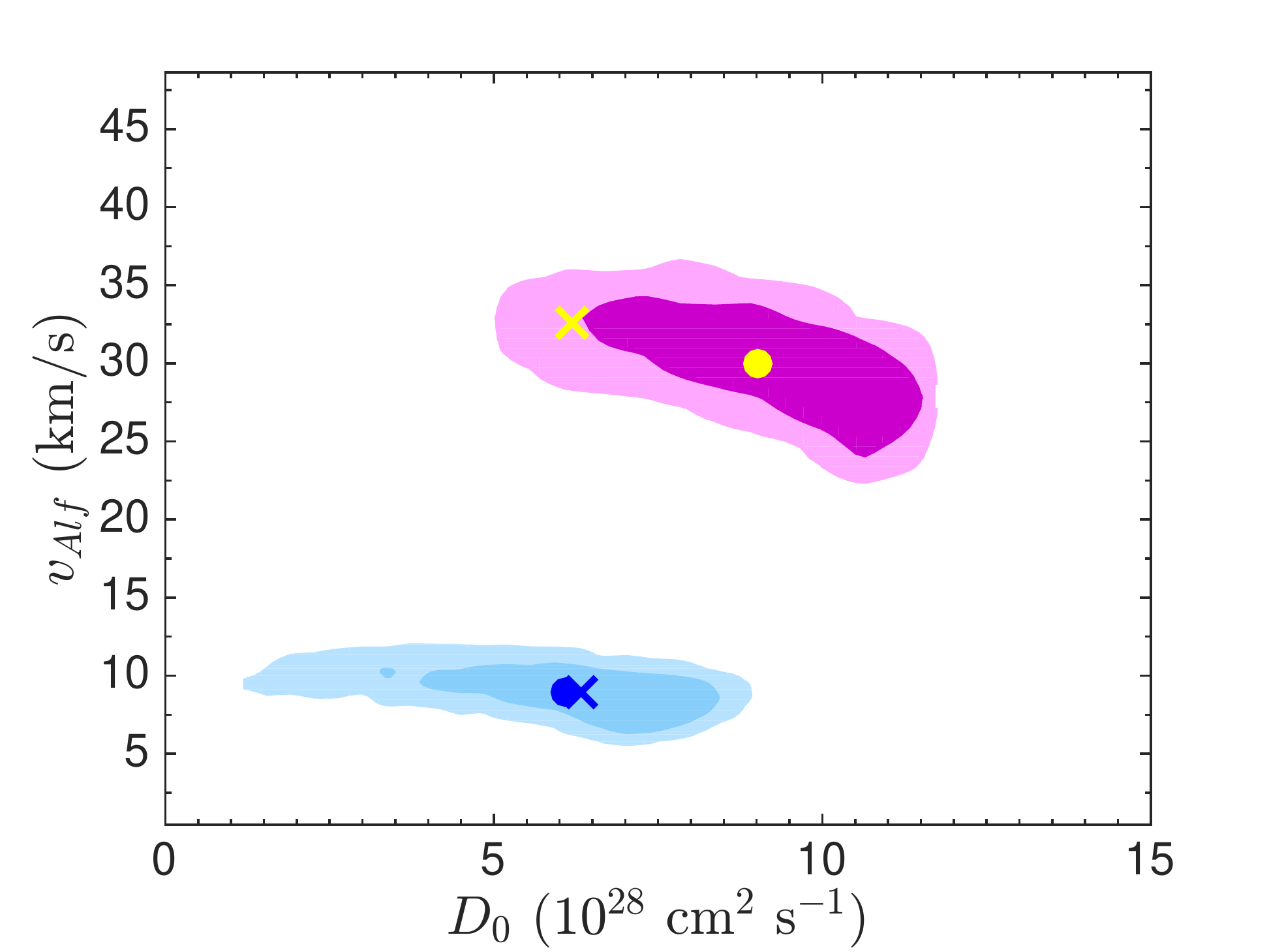} &\includegraphics[width=.25\textwidth]{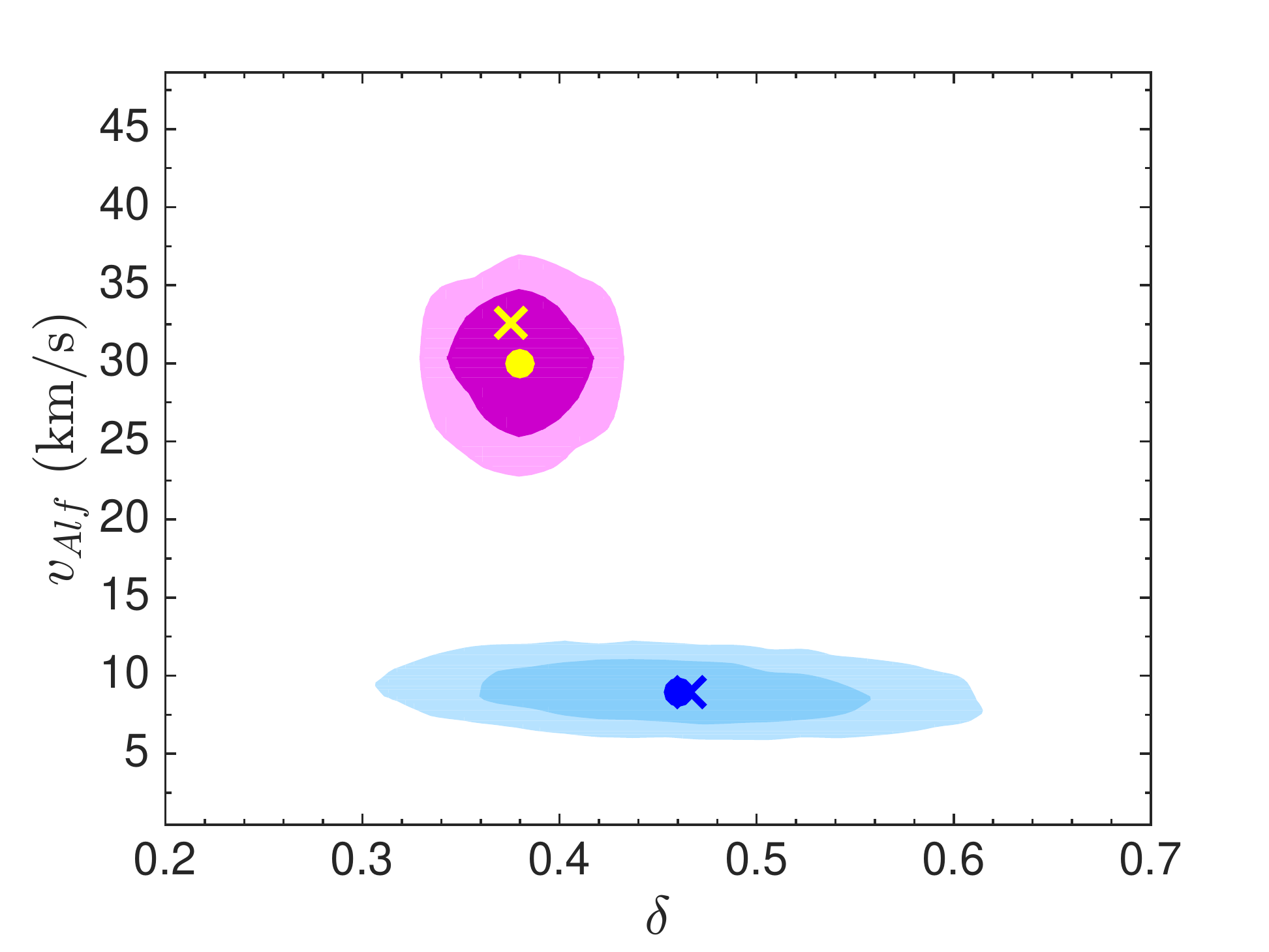}  && \\
\includegraphics[width=.25\textwidth]{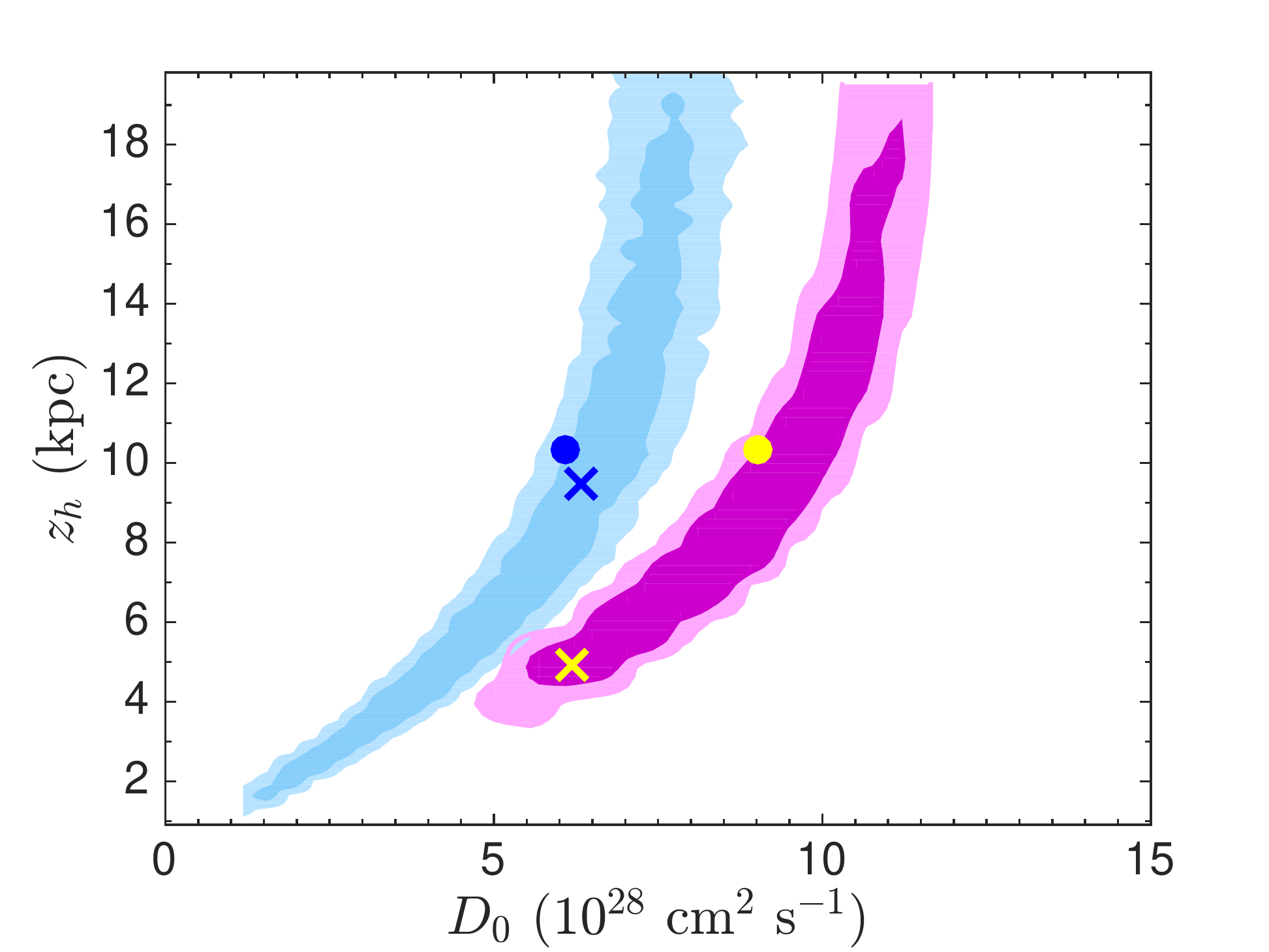} &\includegraphics[width=.25\textwidth]{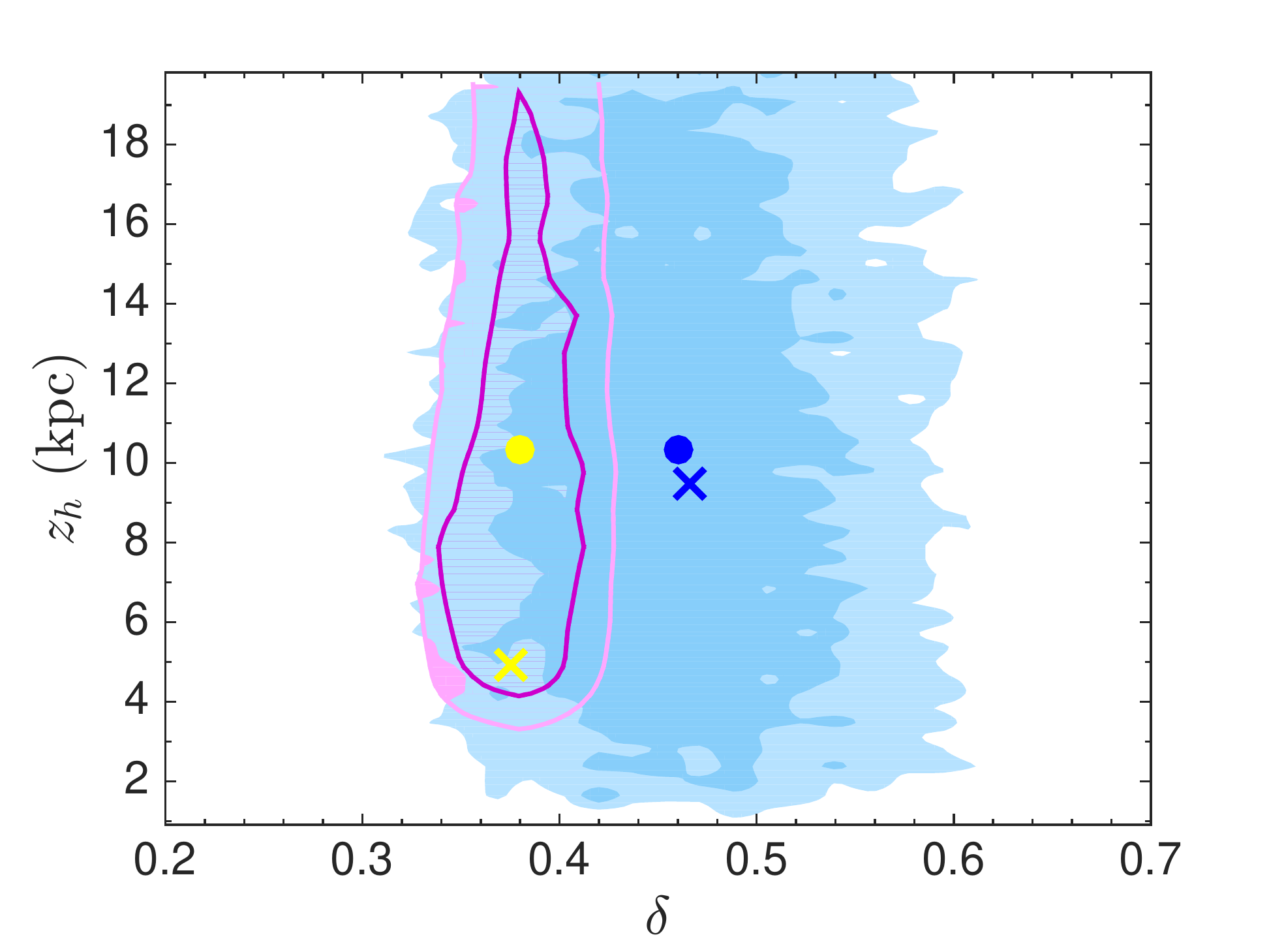}  & \includegraphics[width=.25\textwidth]{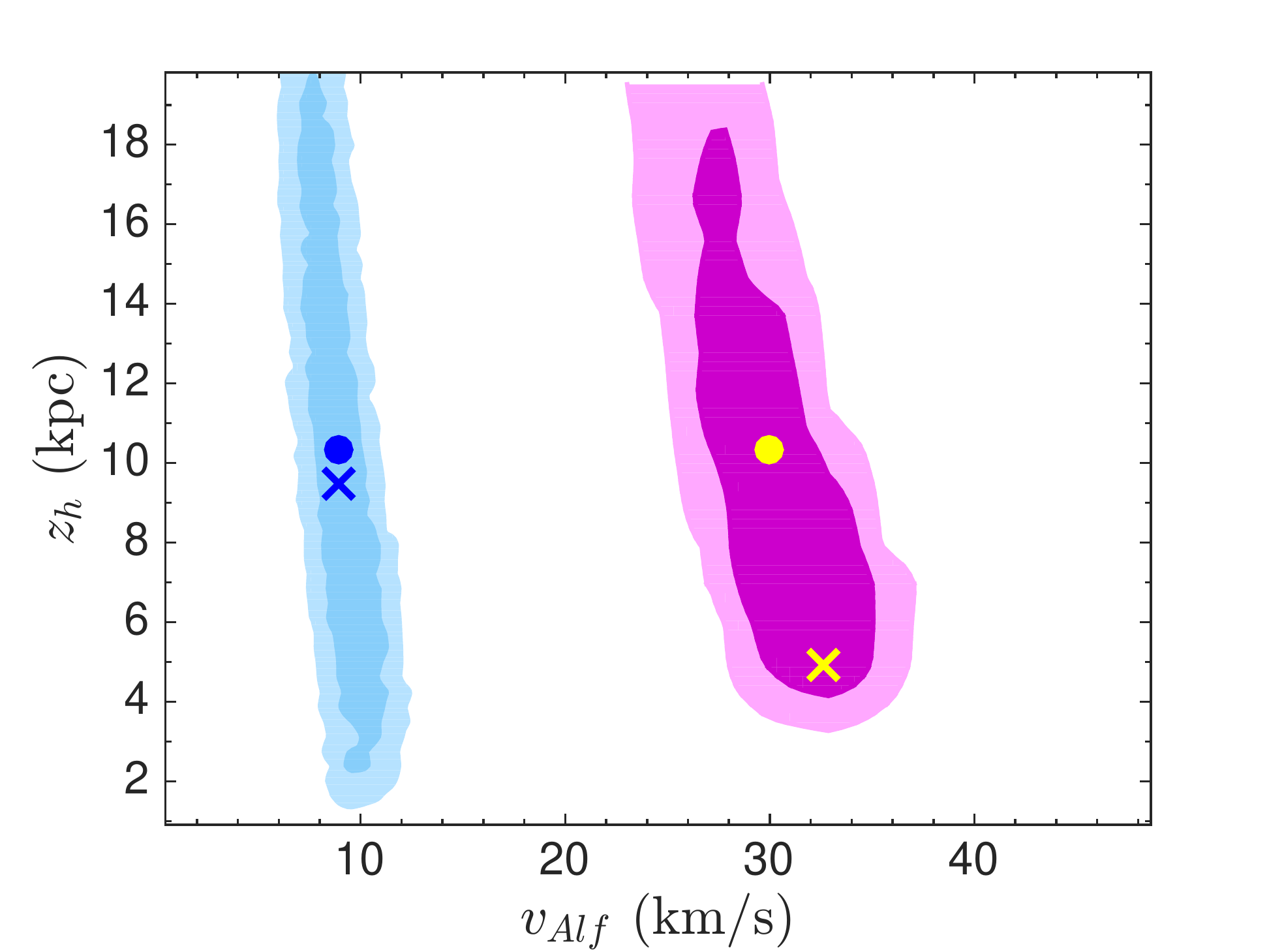} & \\
\includegraphics[width=.25\textwidth]{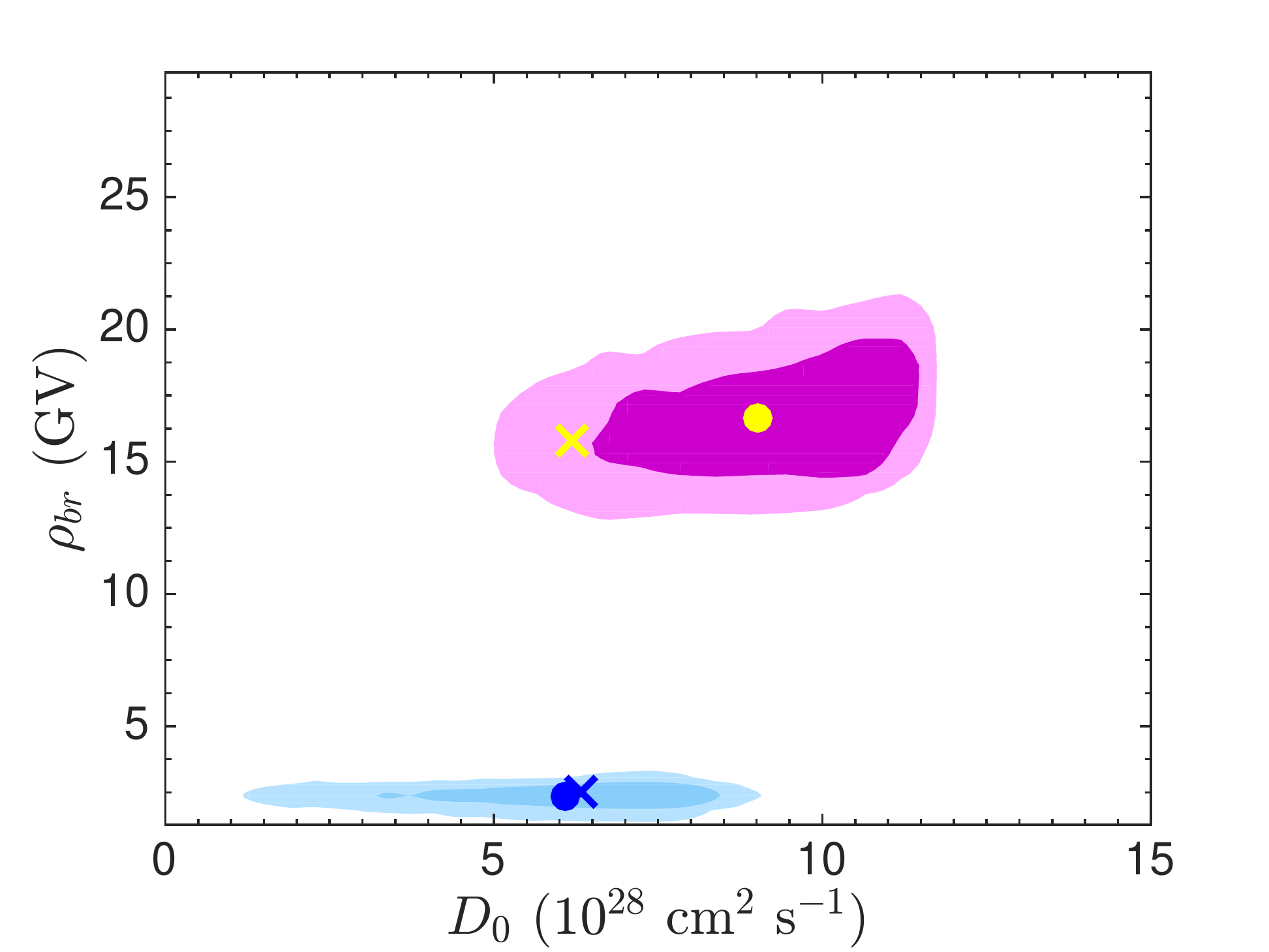} &\includegraphics[width=.25\textwidth]{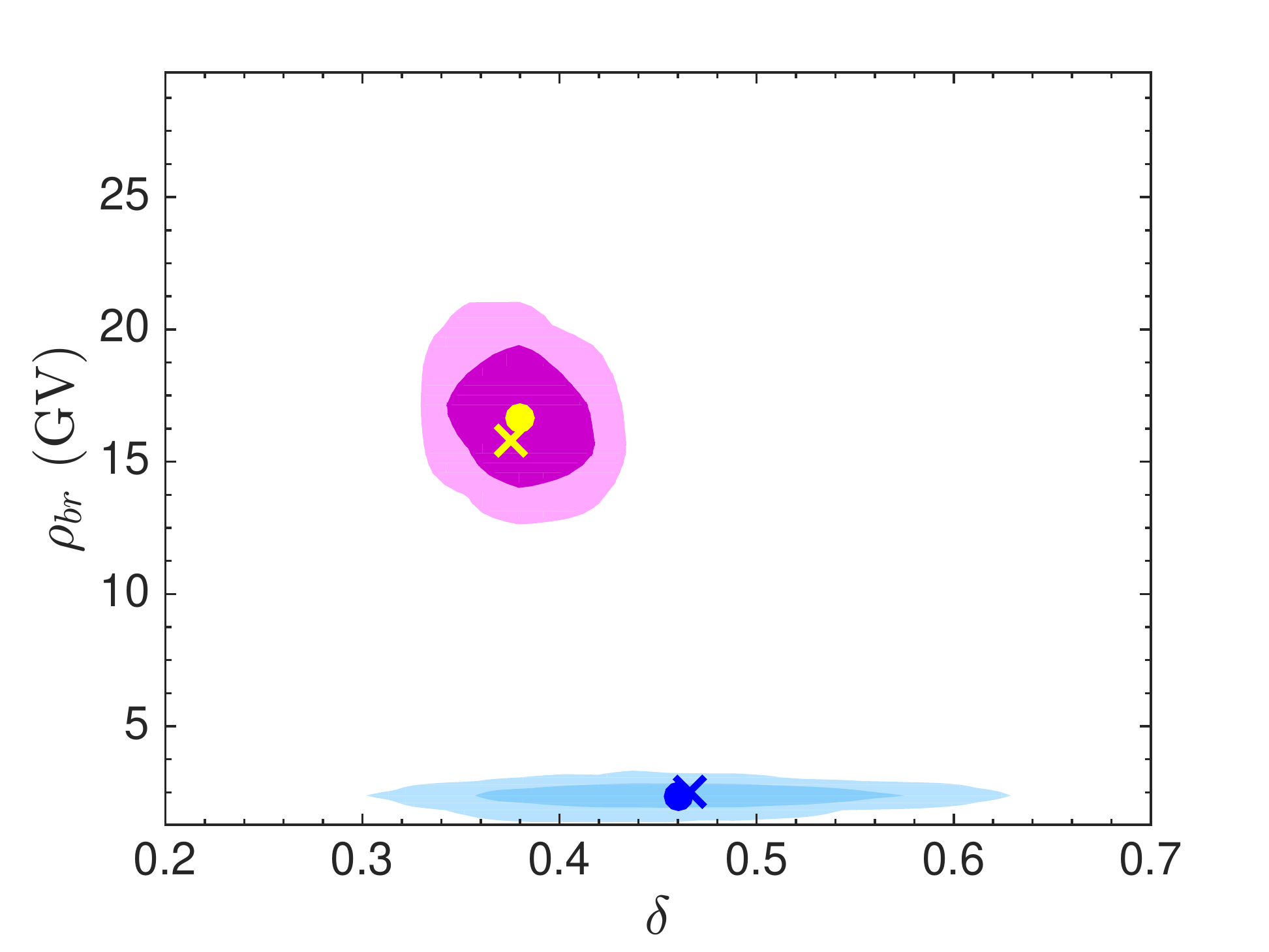}  & \includegraphics[width=.25\textwidth]{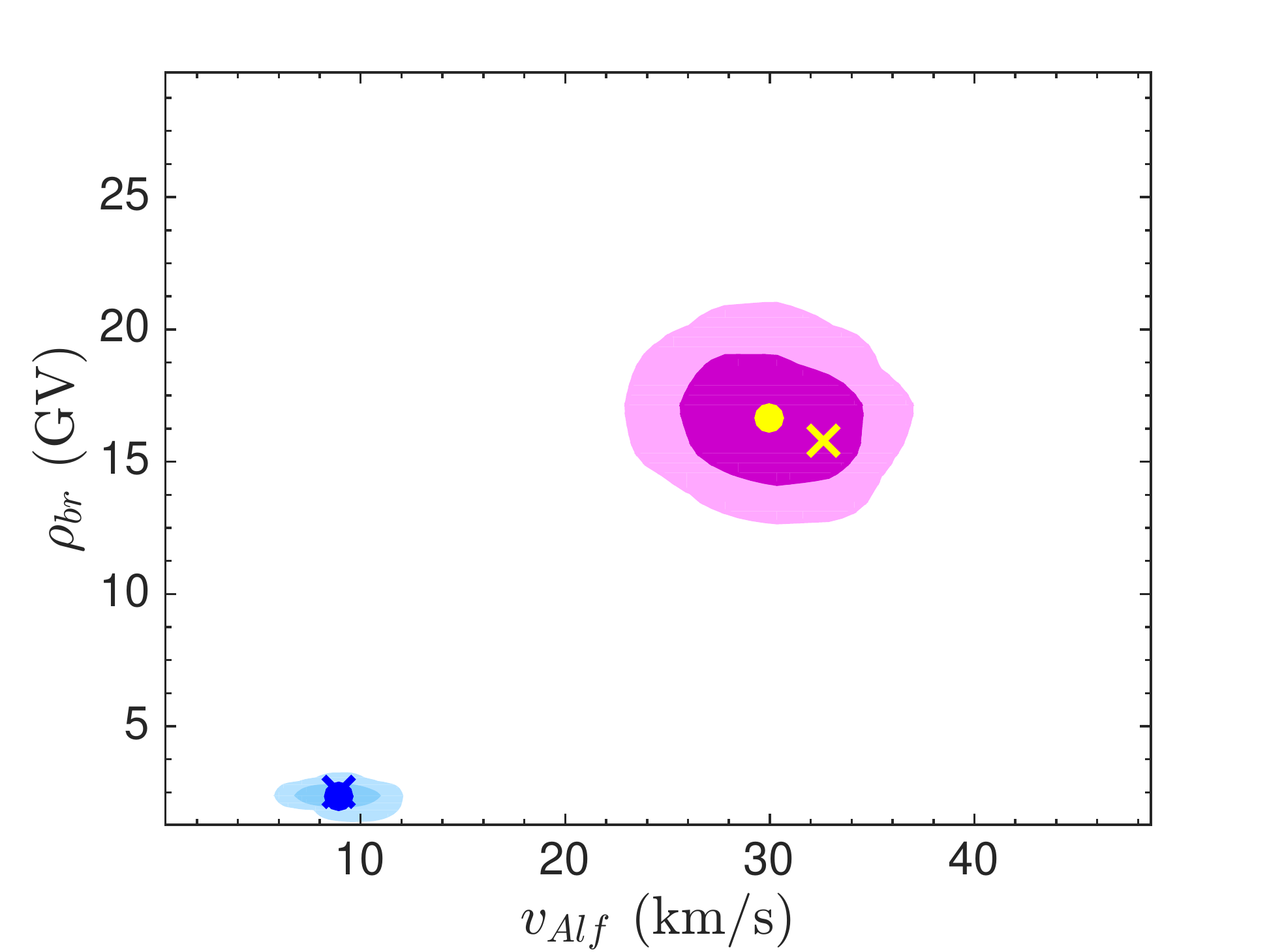} & \includegraphics[width=.25\textwidth]{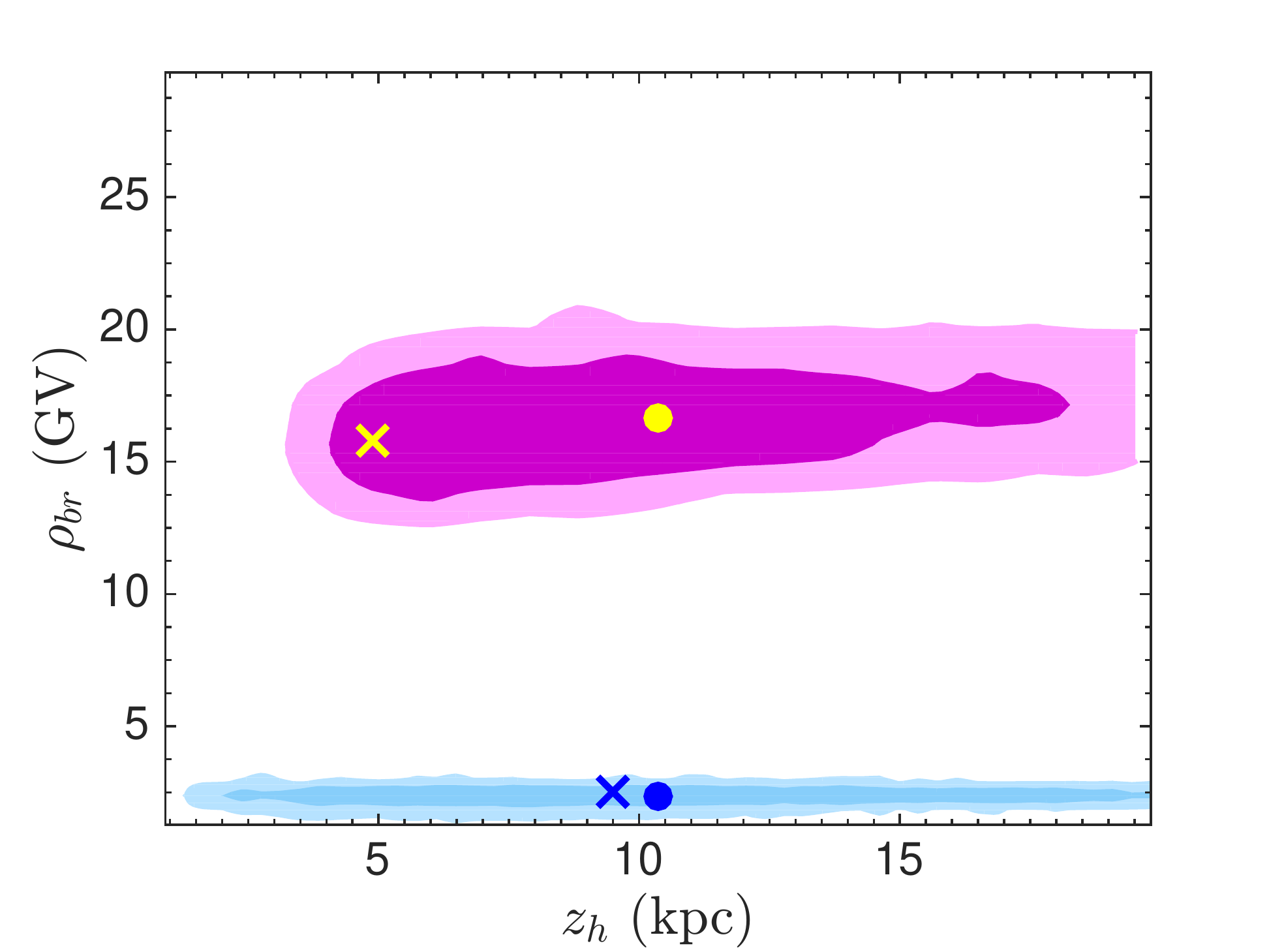}  \\
\end{tabular}
\caption{Two-dimensional posterior distributions, showing 1 and 2-sigma credible intervals for the $p$, $\bar p$ and He scan (blue), and for the light element (Be--Si, magenta). The posterior mean in each case is shown as a dot and the best fit as a cross.}\label{fig:2dpost}
\end{figure*}

%%%%%%%%%%%%%%%%%%%%%%%%%%%%%%%%%%%%%%%%%%%
We report the best fit and posterior mean locations, along with confidence intervals, in Table \ref{tab:resultstable}.
In Figure \ref{fig:1dpost} we show the one-dimensional posterior distributions
obtained for the propagation parameters in our full Multinest/BAMBI scan. We
present two-dimensional 68\% and 95\% (highest posterior density) credible regions for the most relevant
propagation parameters in Figure \ref{fig:2dpost}. One can see that
parameters are generally well-constrained. However, it is apparent that the
measurements of the $^{10}$Be/$^9$Be ratio used here are not sufficient to break the well-known
degeneracy between $D_{0xx}$ and $z_h$. Indeed, values of the halo height $z_h$
can range between about 4 and 20 kpc, while the diffusion parameter normalization
can be in the range [5,  11] $\times 10^{28}$ cm$^2$ s$^{-1}$ in the light element
scan.  Comparing to the \ppbarHe scan we can see that the inclusion of the
radioactive-to-stable secondary ratio only marginally improves the constraint on
the halo-height, mostly from below.

While the one-dimensional posterior distribution for $\Dxx$ and $z_h$ from the
two propagation runs contain a significant overlap, the two-dimensional
distributions show a clear separation between the two
scans.  There is therefore a significant tension between using $\bar{p}$ and B
for the determination of propagation parameters.  This is also evident in the
$\valf$ posterior distributions that are clearly separated for the two scans.
These results thus strongly suggest that the propagation parameters are not
constant over the entire Galaxy and using only the B/C ratio to determine the
propagation parameters can significantly bias the results. 

The reason behind this separation can partially be gleaned from the $z_h-D_0$ posterior distributions shown in Fig. \ref{fig:2dpost}. For a fixed diffusion parameter, \ppbarHe probe a halo height that is approximately twice as large as the light elements. Since the inferred propagation parameters represent a volume-averaged quantity, these results indicate that \ppbarHe are probing a significantly larger volume than the light elements, and that the ISM properties vary quickly enough on these large distances to yield a significantly different Alfv\`en speed, diffusion coefficient and its index.

The posterior distributions for the injection spectrum for the two scans are
very similar at high energies, but the spectrum of He is systematically harder than H
below the break at 220~GV.  The low energy break of the proton and He spectrum
is also lower than that of the heavier elements.  Given that the power-law
indices of the proton injection spectrum are $\delta_{\nu}$ larger than that of
the He injection spectrum, the results indicate that the injection
spectrum of heavier elements are closer to that of the protons rather than
that of He.  

\begin{figure}[h]
\includegraphics[width=0.5\textwidth]{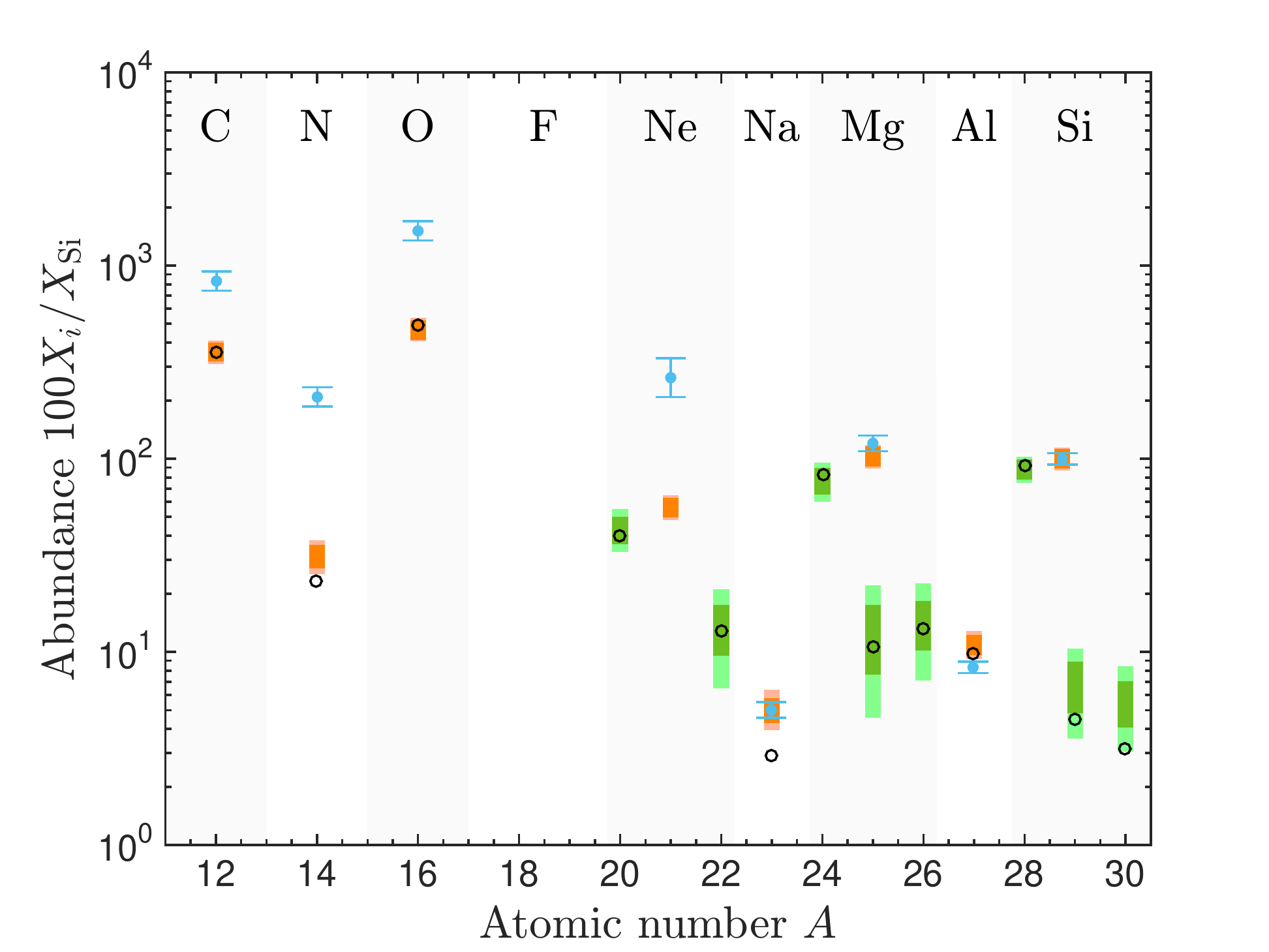}
\caption{95\% (light bars) and 68\% (dark bars) posterior intervals from our final abundance study. Total elemental abundances are in orange, while individual isotopes are in green.  We show the latest determination of the solar photospheric (blue dots) elemental abundances and errors from \citet{2009ARA&A..47..481A}, with updated heavier (A $\ge 23$) elemental abundances from \citet{2015A&A...573A..25S}. We also show previously-used values from GALPROP \citep{Moskalenko2008} with open black circles.}
\label{fig:abd}
\end{figure}

\begin{figure*}
\includegraphics[width=0.33\textwidth]{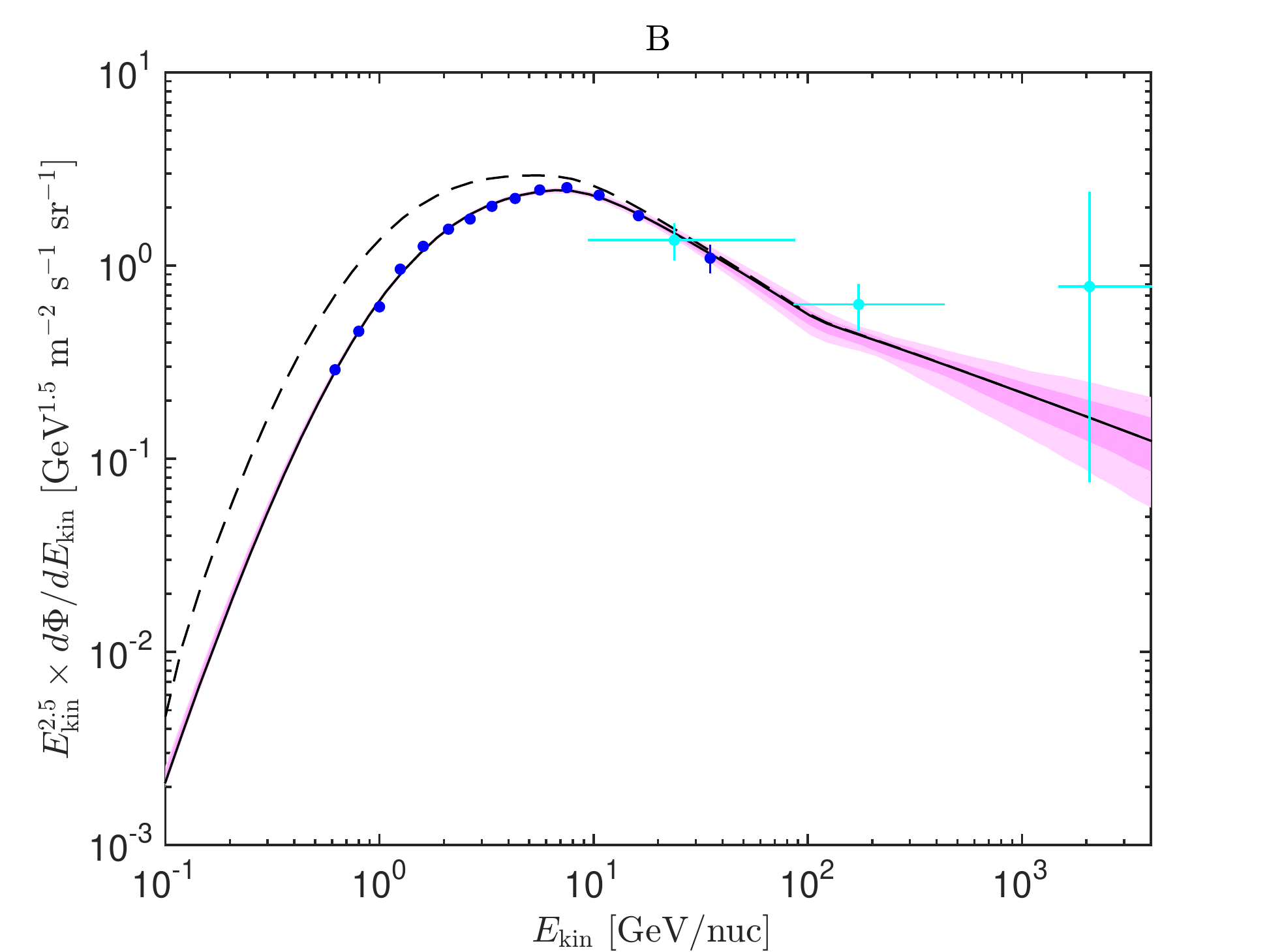}\includegraphics[width=0.33\textwidth]{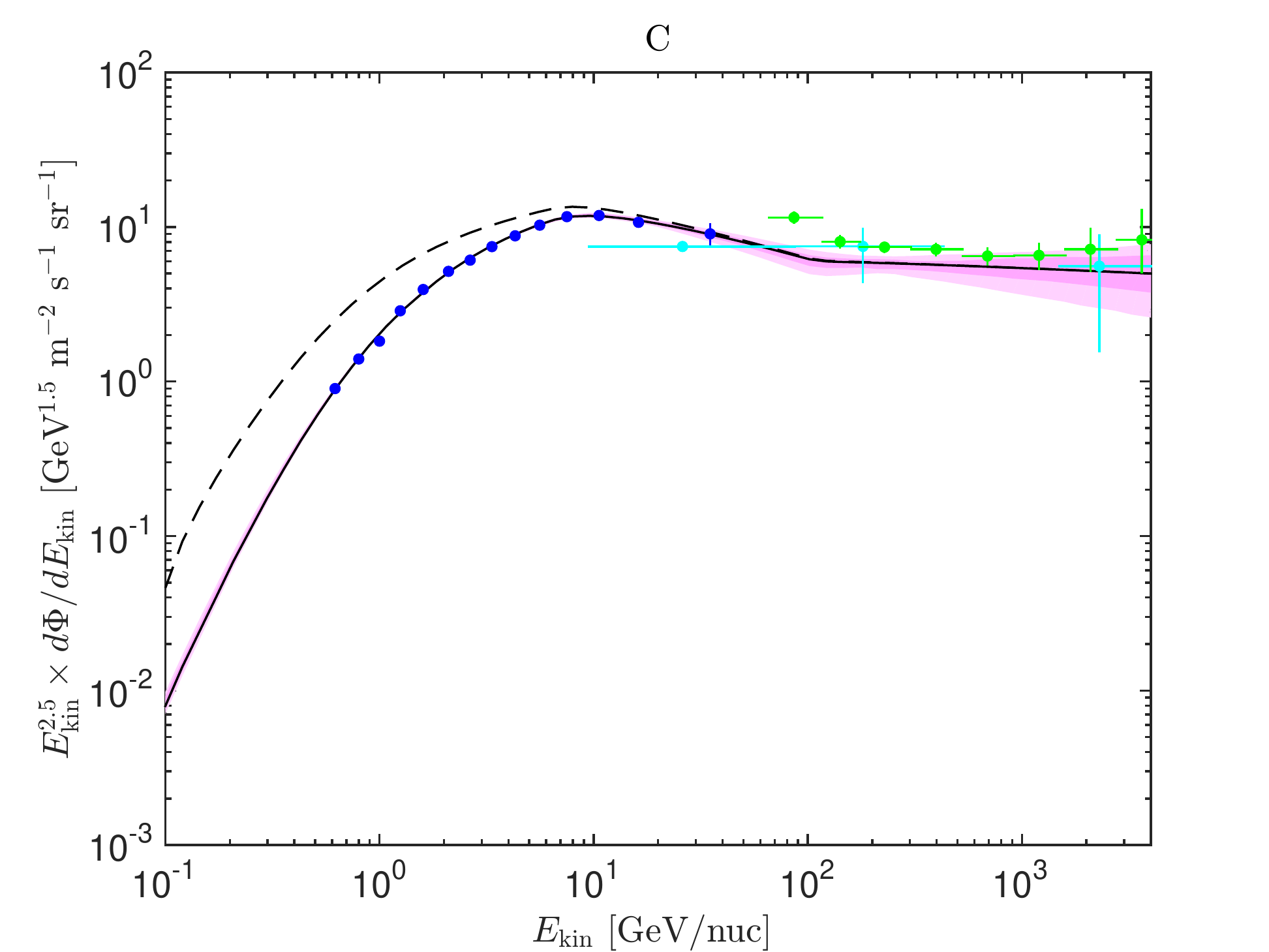}\includegraphics[width=0.33\textwidth]{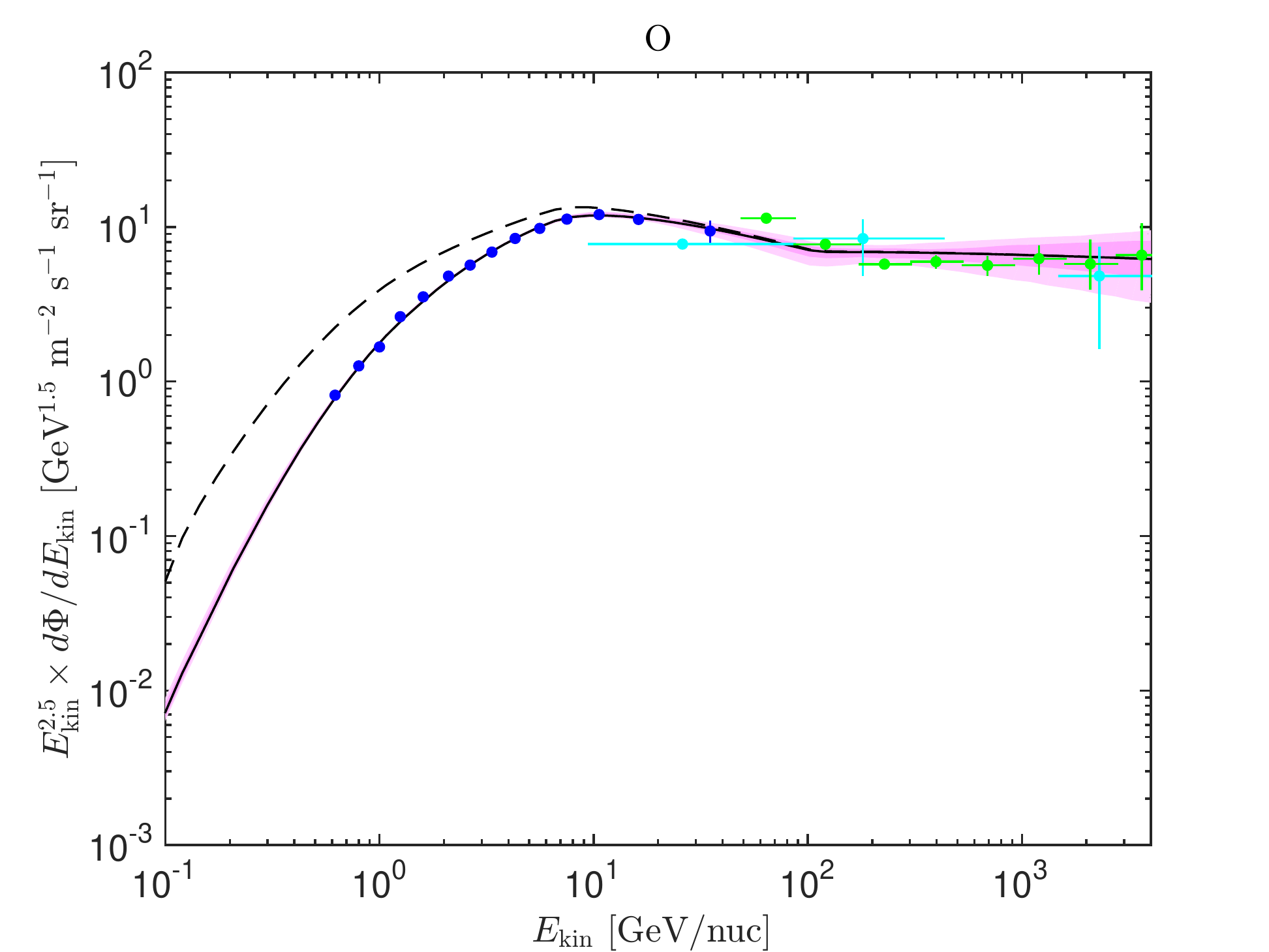}
\caption{Spectral fluxes with 68\% and 95\% posterior regions from the posteriors of our light element (Be--Si) scan, shown in magenta in Figure \ref{fig:1dpost}, and  using the HEAO modulation posteriors, . 
Data shown are HEAO (blue), CREAM (green) and TRACER (cyan). The best fit is shown as a black line, and the dashed lines correspond to the LIS (umodulated) spectra.}
\label{fig:fluxspecplot}
\end{figure*}

\begin{figure*}
\includegraphics[width=0.33\textwidth]{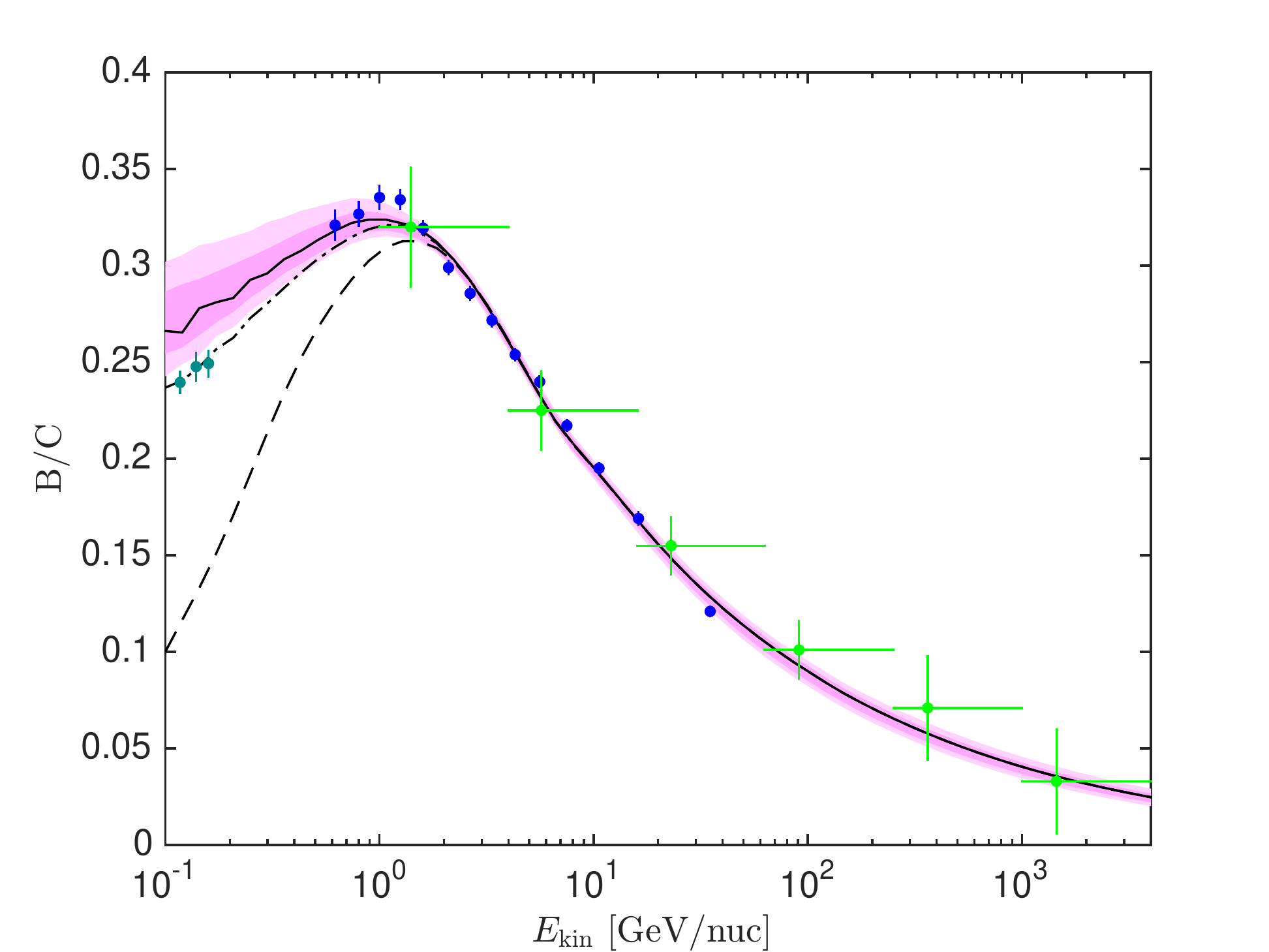}\includegraphics[width=0.33\textwidth]{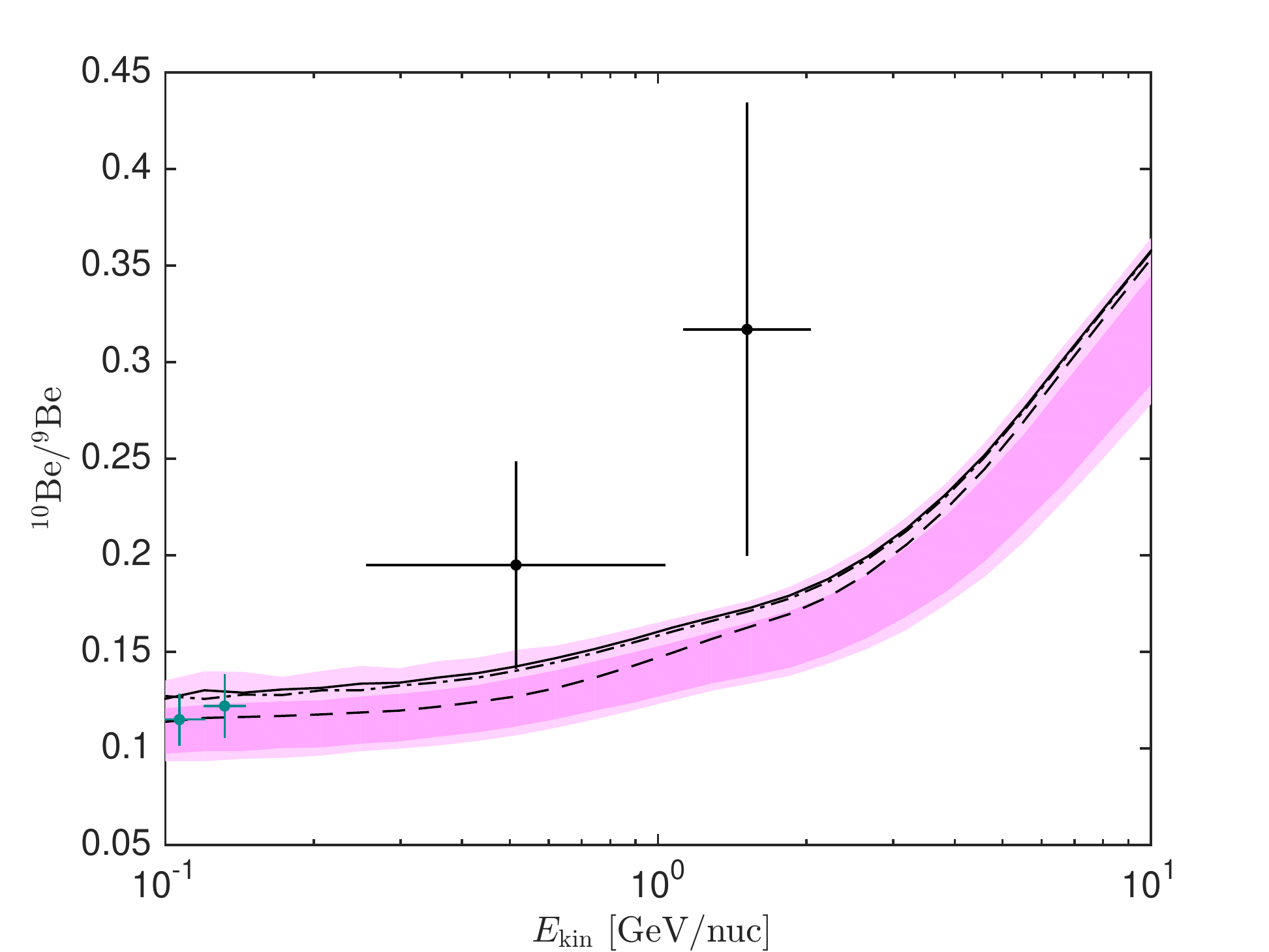}\includegraphics[width=0.33\textwidth]{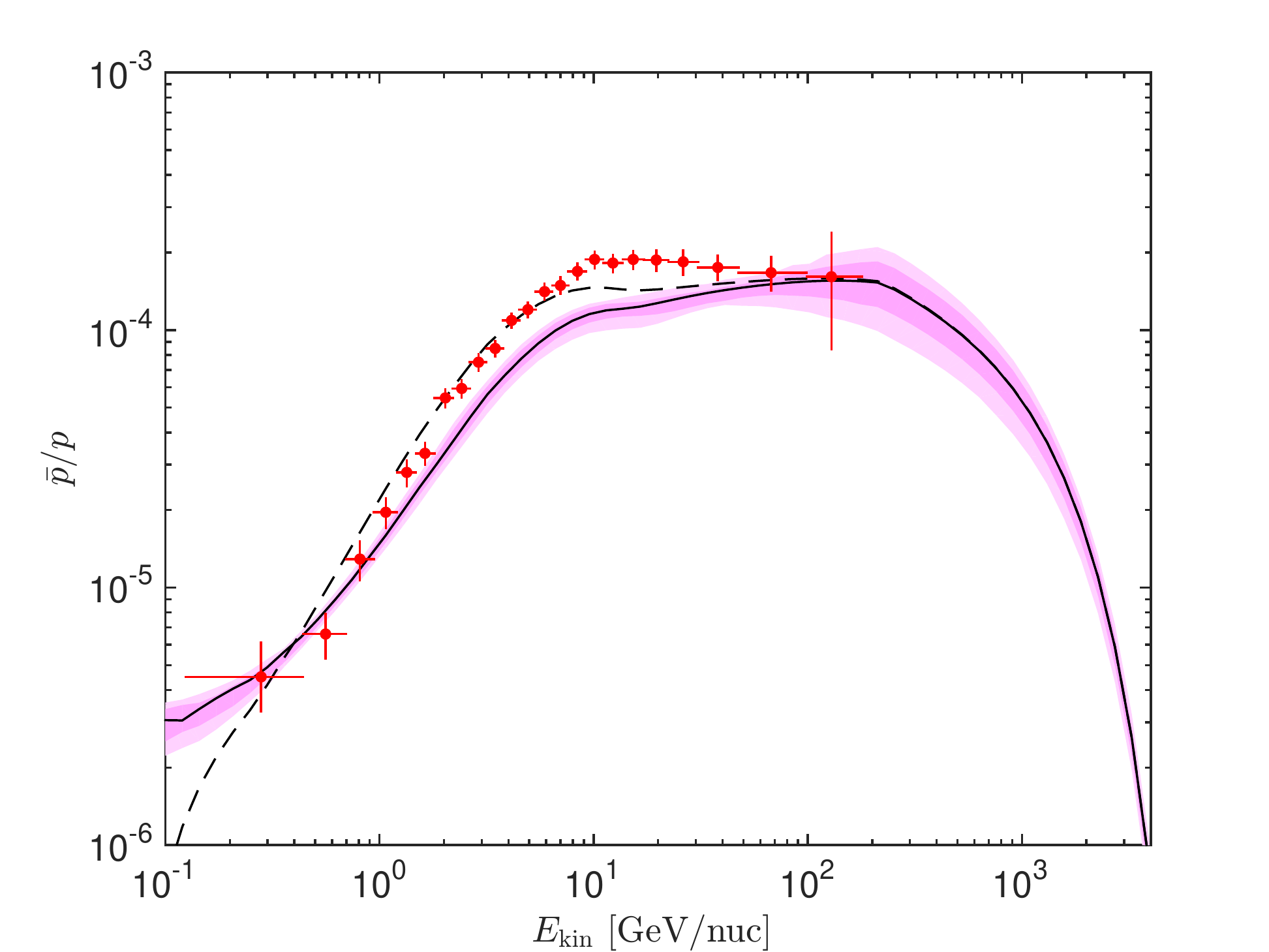}
\caption{Secondary-to-primary ratio 68\% and 95\% posterior bands from our light element (Be--Si) scan, shown in magenta in Fig. \ref{fig:1dpost}. The $\bar p/ p$ ratio is shown to indicate that using the same propagation parameters for hydrogen yields a very bad fit to the data. Data shown are HEAO (blue), CREAM (green), ACE (light blue), ISOMAX (black) and PAMELA (red). The best fit is shown as a black line, and the dashed lines correspond to the LIS (umodulated) ratios. In the left-hand panel we use the HEAO modulation posterior, and the solid line uses the HEAO best fit modulation potential. The dash-dotted line is the modulated spectrum using the best fit to the ACE-CRIS modulation potential; for clarity we do not show the posterior intervals for this case. Correspondingly, the central plot uses the ACE modulation (BF in black), and we show the best fit using the ISOMAX best fit modulation potential with a dash-dotted line.}
\label{fig:secspecplot}
\end{figure*}

\begin{figure*}
\begin{centering}
\includegraphics[width=0.4\textwidth]{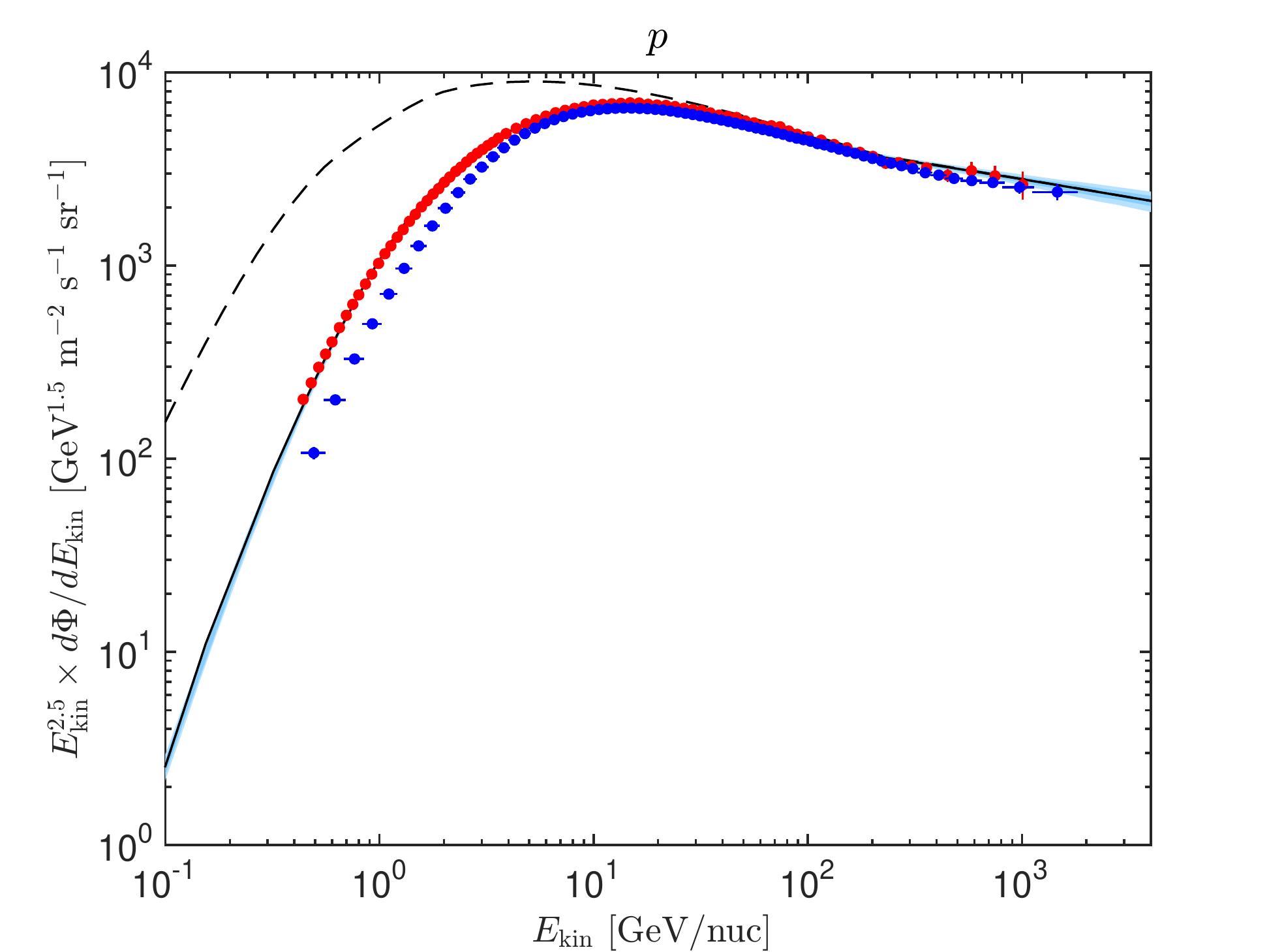} \includegraphics[width=0.4\textwidth]{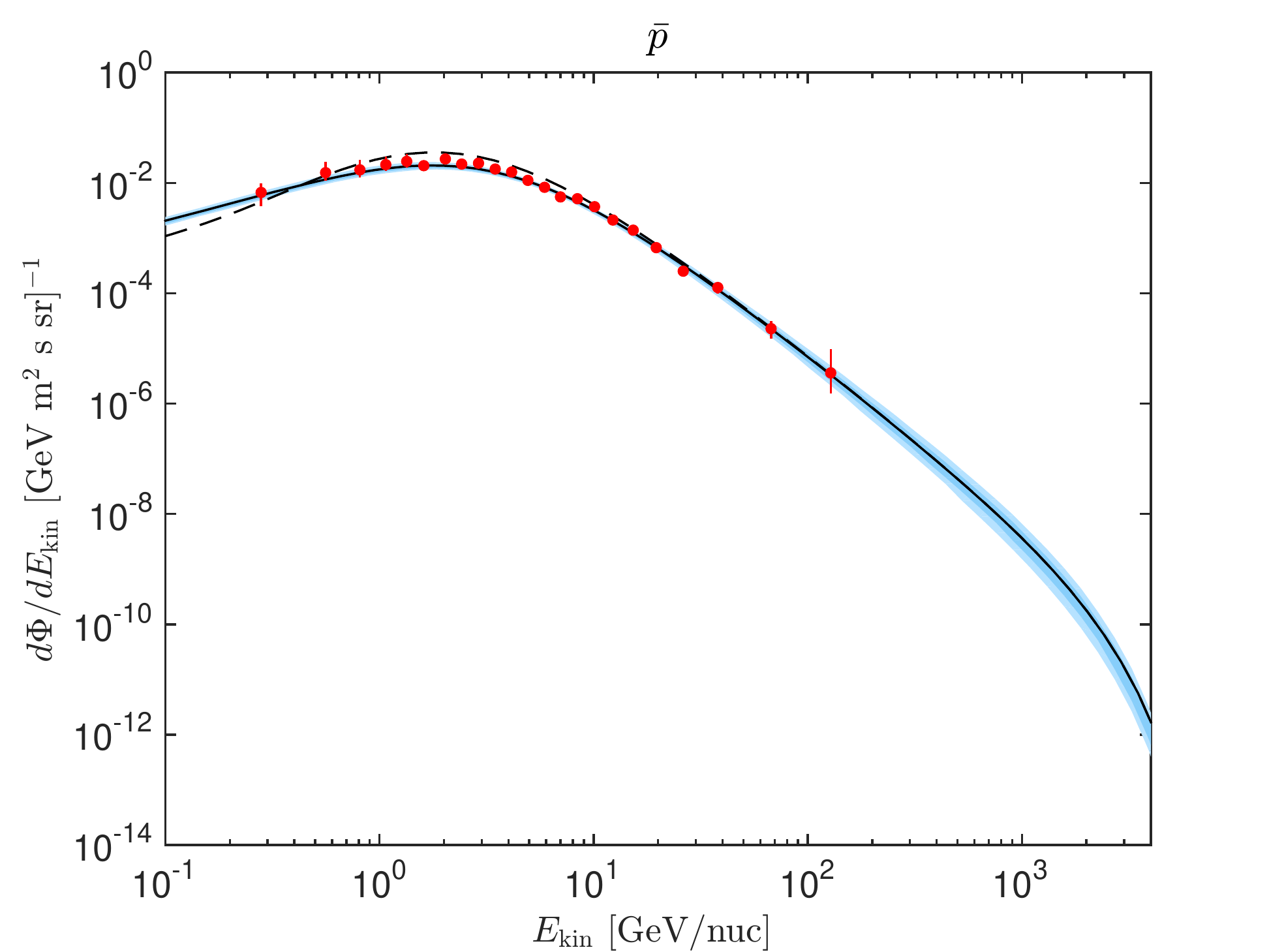} \\
\includegraphics[width=0.4\textwidth]{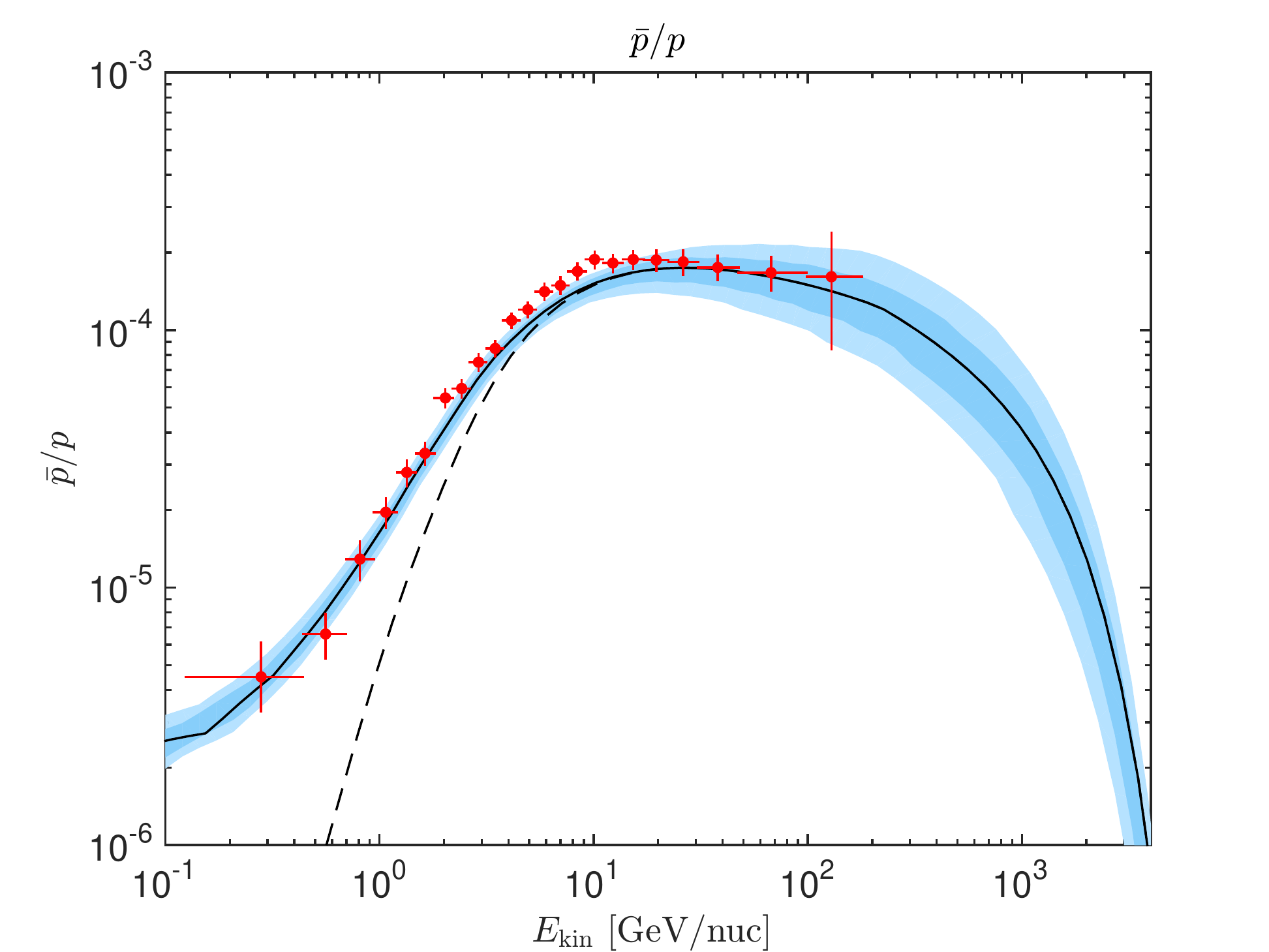} \includegraphics[width=0.4\textwidth]{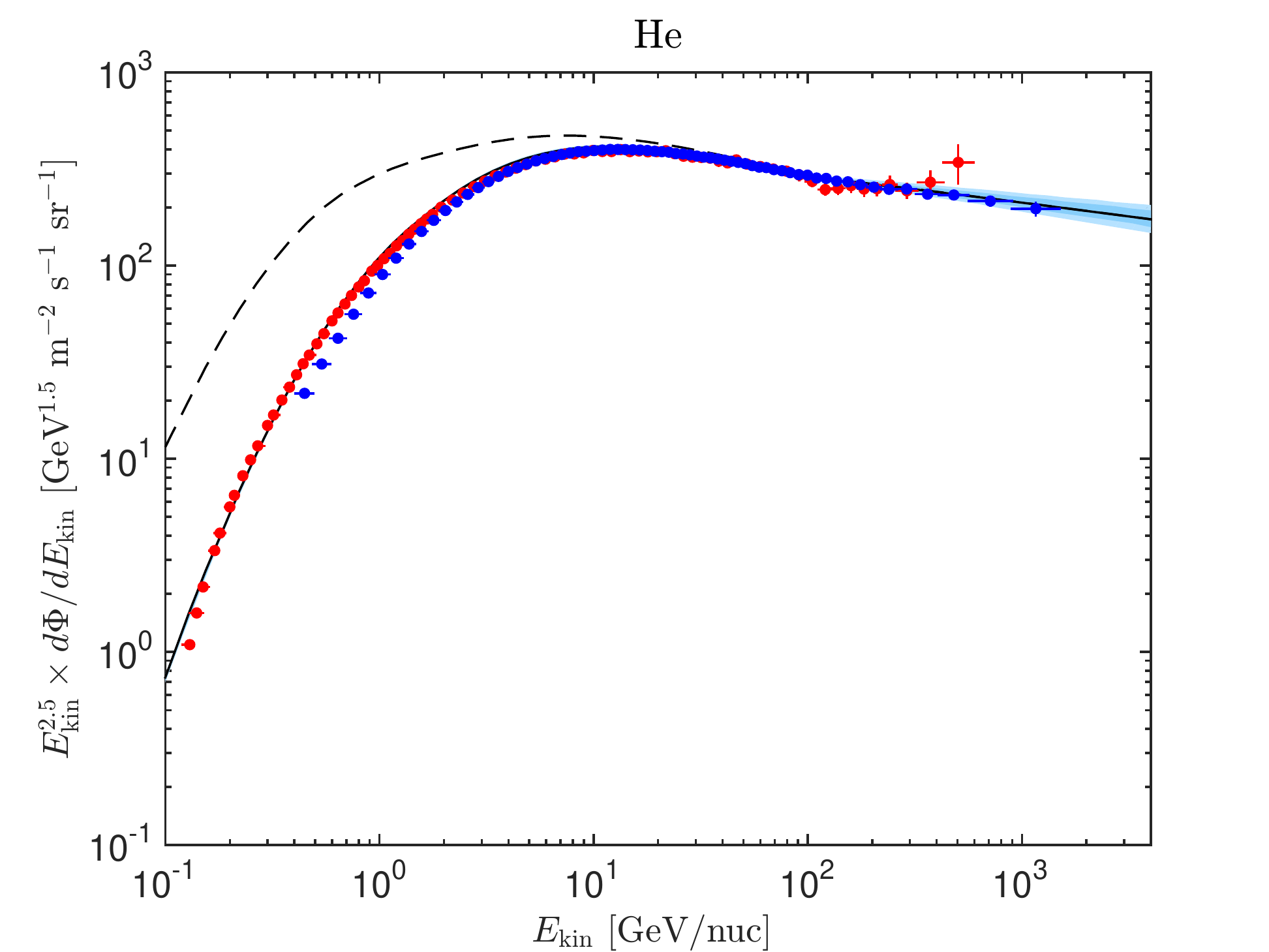}
\caption{Spectra and $\bar p /p$ ratio 68\% and 95\% posterior bands of our $\bar p,p$ He scan, shown in blue in Fig.
   \ref{fig:1dpost}. The best fit is plotted in black, and the dashed lines correspond to the LIS (umodulated) spectra. PAMELA data are shown in red. We also show recent AMS-02 \citep[blue]{2015PhRvL.114q1103A,2015PhRvL.115u1101A} for the available proton and helium flux data, which were not available at the time of our analysis (and hence are not included in the likelihood).}\label{fig:ppbarHespec}
\end{centering}
\end{figure*}

Figures \ref{fig:fluxspecplot}, \ref{fig:secspecplot}, and
\ref{fig:ppbarHespec} show 68\% and 95\% posterior intervals for our models
overlaid on some of the data used in the analysis.  The data-model agreement
is very good in all cases.  The need for the high energy break is evident in
the spectrum of protons and He in Figure~\ref{fig:ppbarHespec} but we can also see
from the spectra of heavier elements in Figure~\ref{fig:fluxspecplot} that the
high energy break improves the agreement between model and data.  The
prediction for the $\bar{p}/p$ ratio of the light element scan in
Figure~\ref{fig:secspecplot} further illustrates the tension between the two
datasets because there is a clear and significant mismatch between the data
and model prediction. Indeed, a preliminary scan which included all datasets was not able to find an acceptable fit, yielding very large error rescaling parameters with $\tau_{\rm PAMELA}$ reaching the prior box boundary at $-\log \tau_{\rm PAMELA} = 1.5$. This is an indication that the model cannot simultaneously fit the light element and \ppbarHe data.

Note that newer $\bar p$ production cross sections \citep{2015ApJ...803...54K} 
yield better description of the $\bar p$ production in proton-proton, proton-nucleus, and nucleus-nucleus interactions, but were not available at the start
of this analysis. They provide a higher $\bar p$ yield above $E_{\bar p}>100$ GeV. Meanwhile, the parameterizations used in the present paper 
\citep{Moskalenko2002,1983JPhG....9..227T,1983JPhG....9.1289T} were tuned to the $\bar p$ data at moderate energies providing a reasonable
description in that energy range.
The new cross sections are now incorporated into the \galprop{} code to be used in our future calculations.


In Table \ref{tab:abd_constraints} we provide the best fit, posterior means and confidence intervals for the abundance parameters. These are compared with solar data in Figure \ref{fig:abd}. We also show the previously-recommended values from GALPROP \citep{Moskalenko2008}. Abundances are in generally good agreement with the solar values, with partial volatiles (C, N, O, Ne) being depleted with respect to the solar abundances. This is a well-known result, as CRs are likely preferentially accelerated from refractory-rich dust grains \citep{1997ApJ...487..182M,1997ApJ...487..197E,2009ApJ...697.2083R,2010ApJ...715.1400A}. The only major change versus previous GALPROP determinations is a higher sodium abundance, which is now brought in line with solar system measurements. 

\begin{figure*}
\begin{centering}
\includegraphics[width=.33\textwidth]{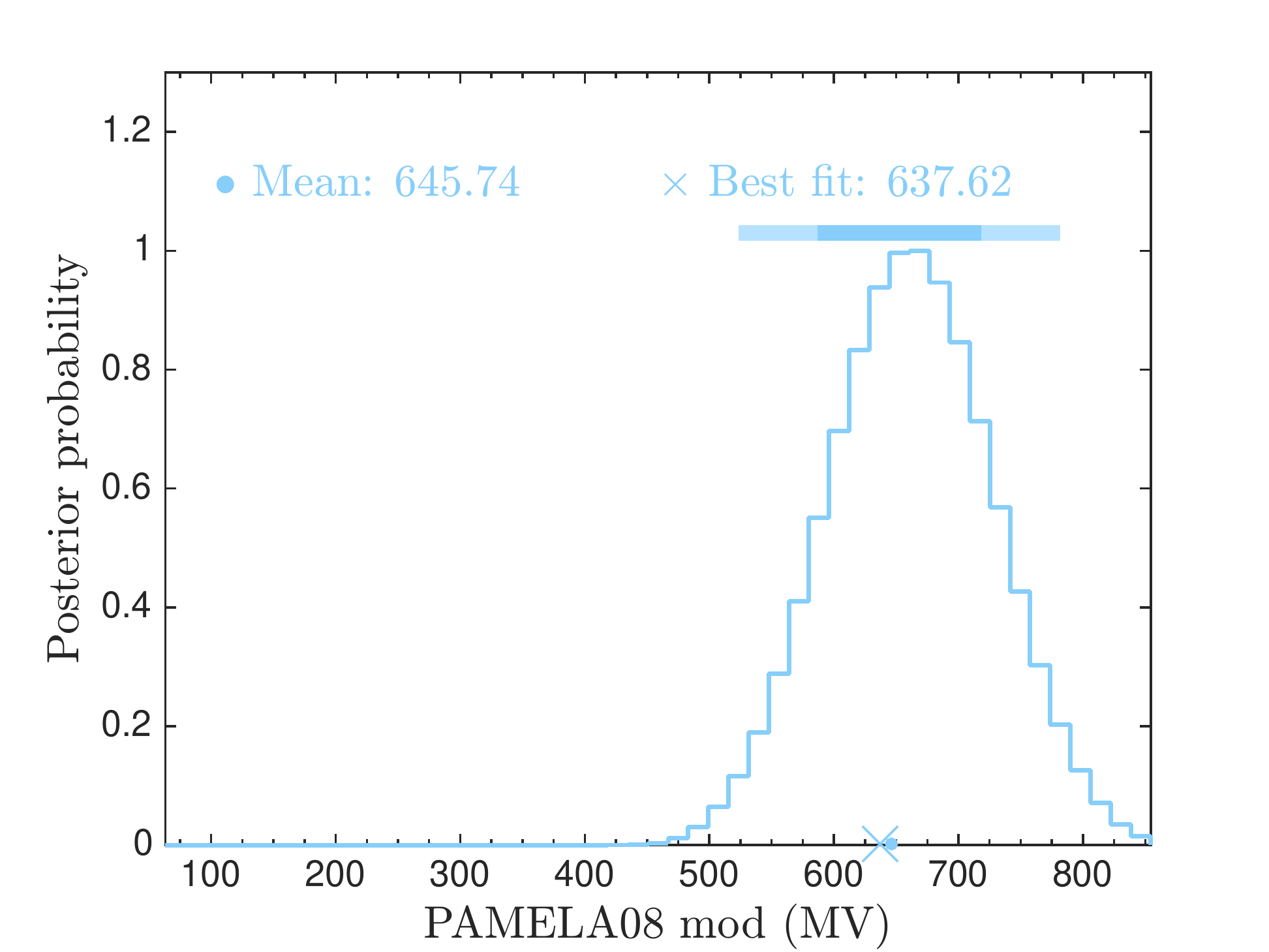}\includegraphics[width=.33\textwidth]{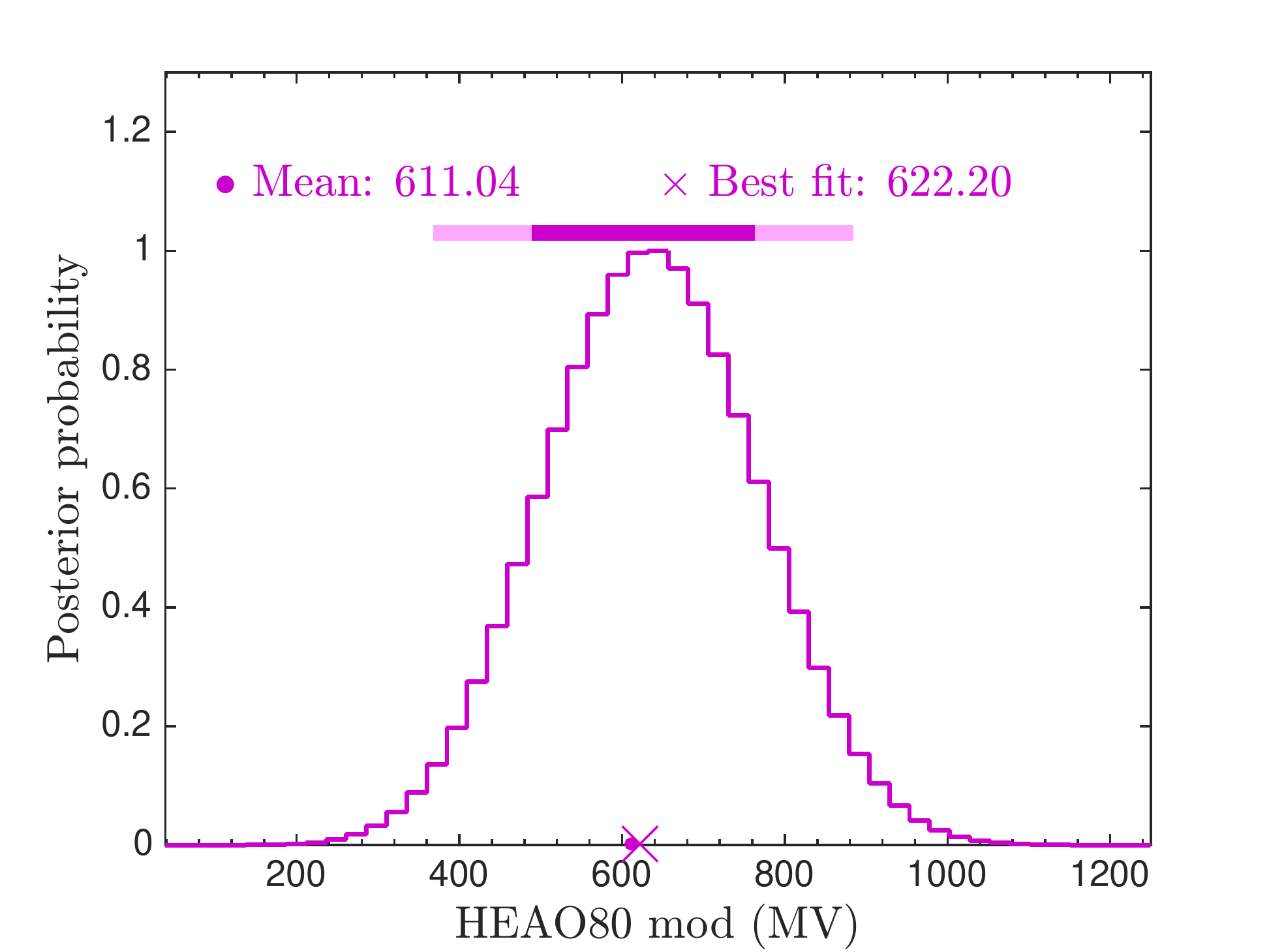} \\ \includegraphics[width=.33\textwidth]{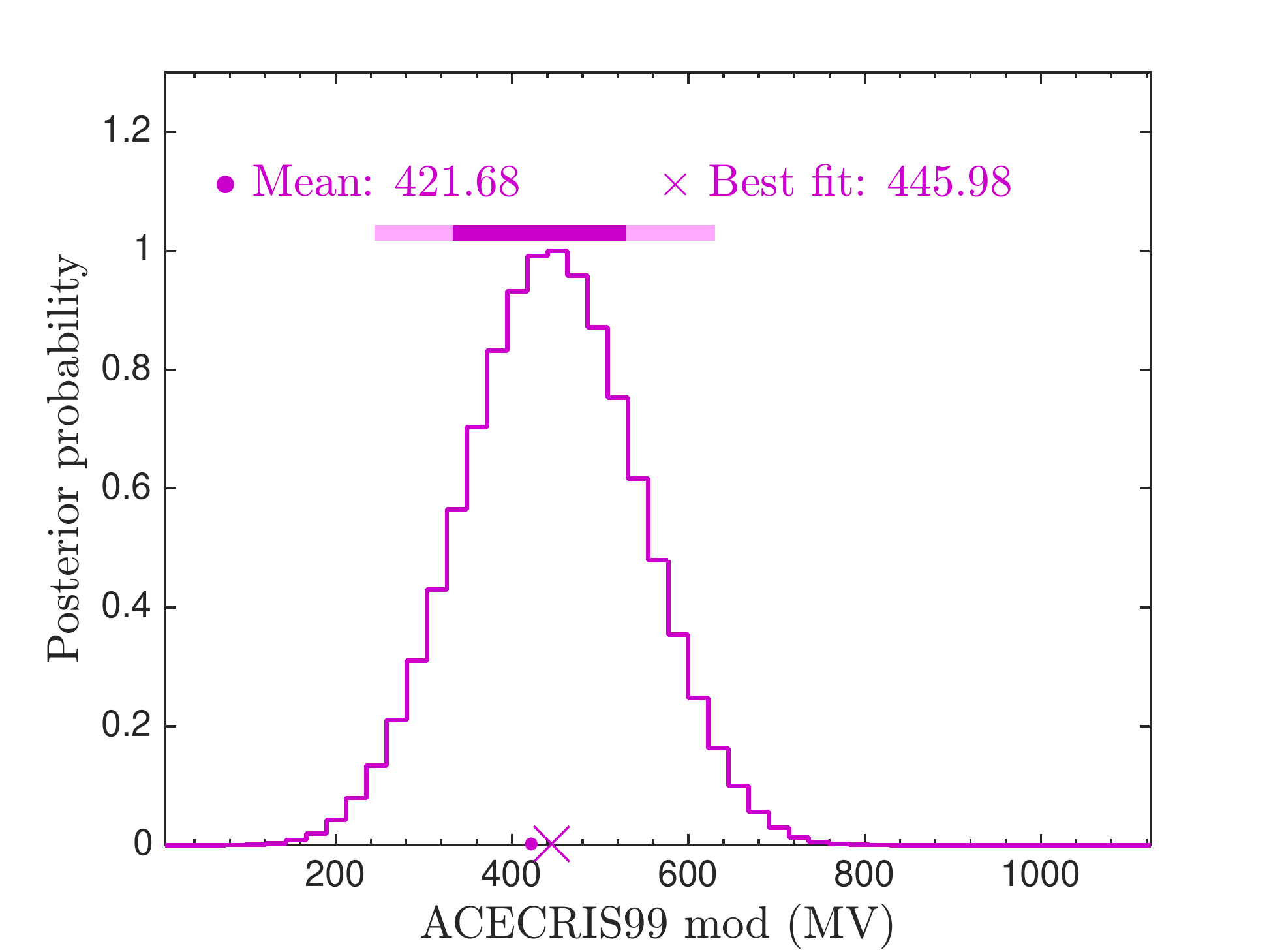} \includegraphics[width=.33\textwidth]{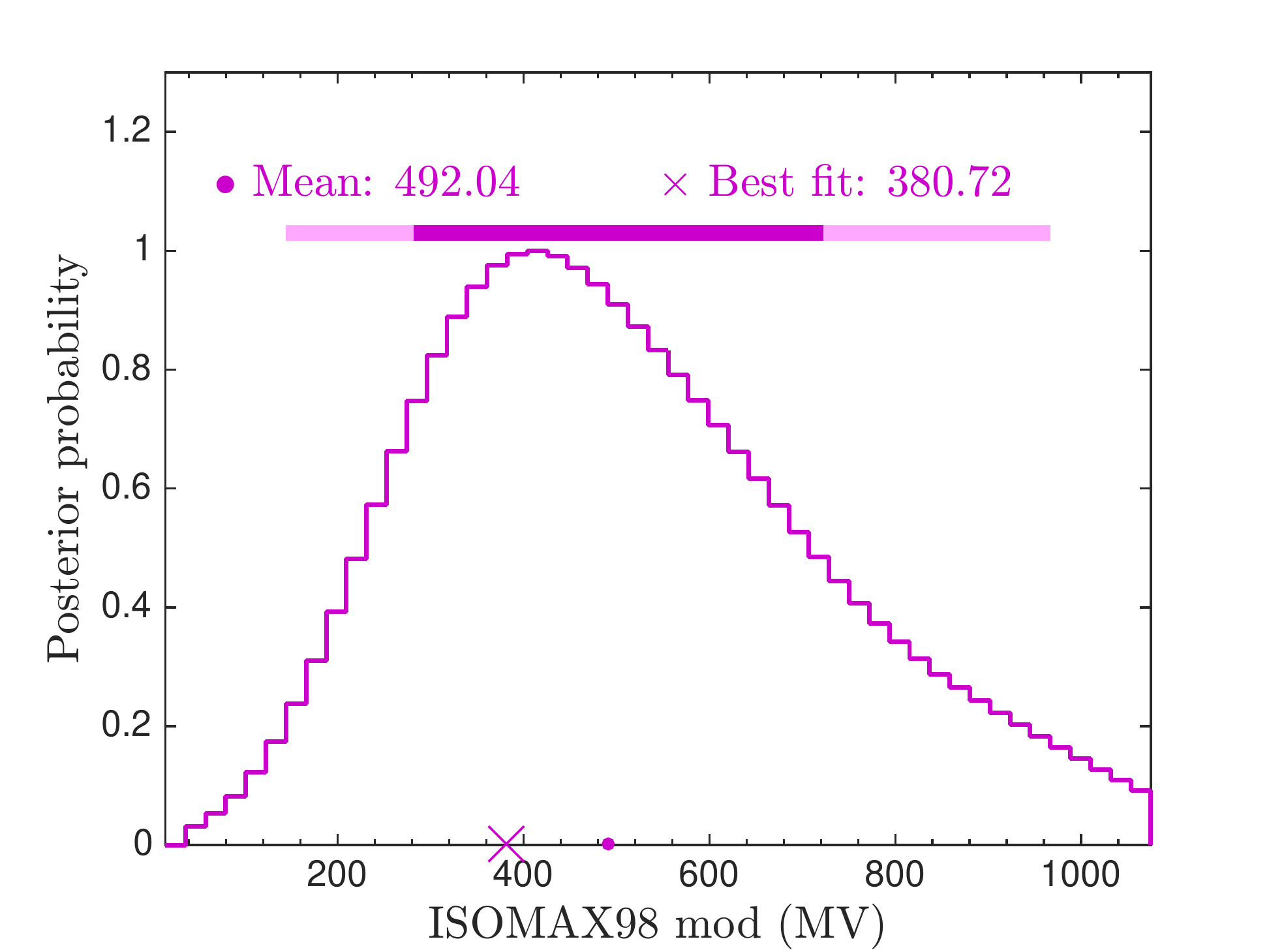}
\caption{Posterior distributions of the modulation parameters for each experiment used in the fit, with 1 and 2 sigma credible intervals.}
\label{fig:modpost}
\end{centering}
\end{figure*}

In Figure \ref{fig:modpost} we show the posterior distributions for the
modulation potentials of the three experiments that we used  whose energy
range was low enough to be affected by solar modulation.  The posterior mean values are in
good agreement with those estimated using ground-based neutron monitors
\citep{Usoskin:2011}. 

\begin{figure*}
\begin{centering}
\includegraphics[width=.33\textwidth]{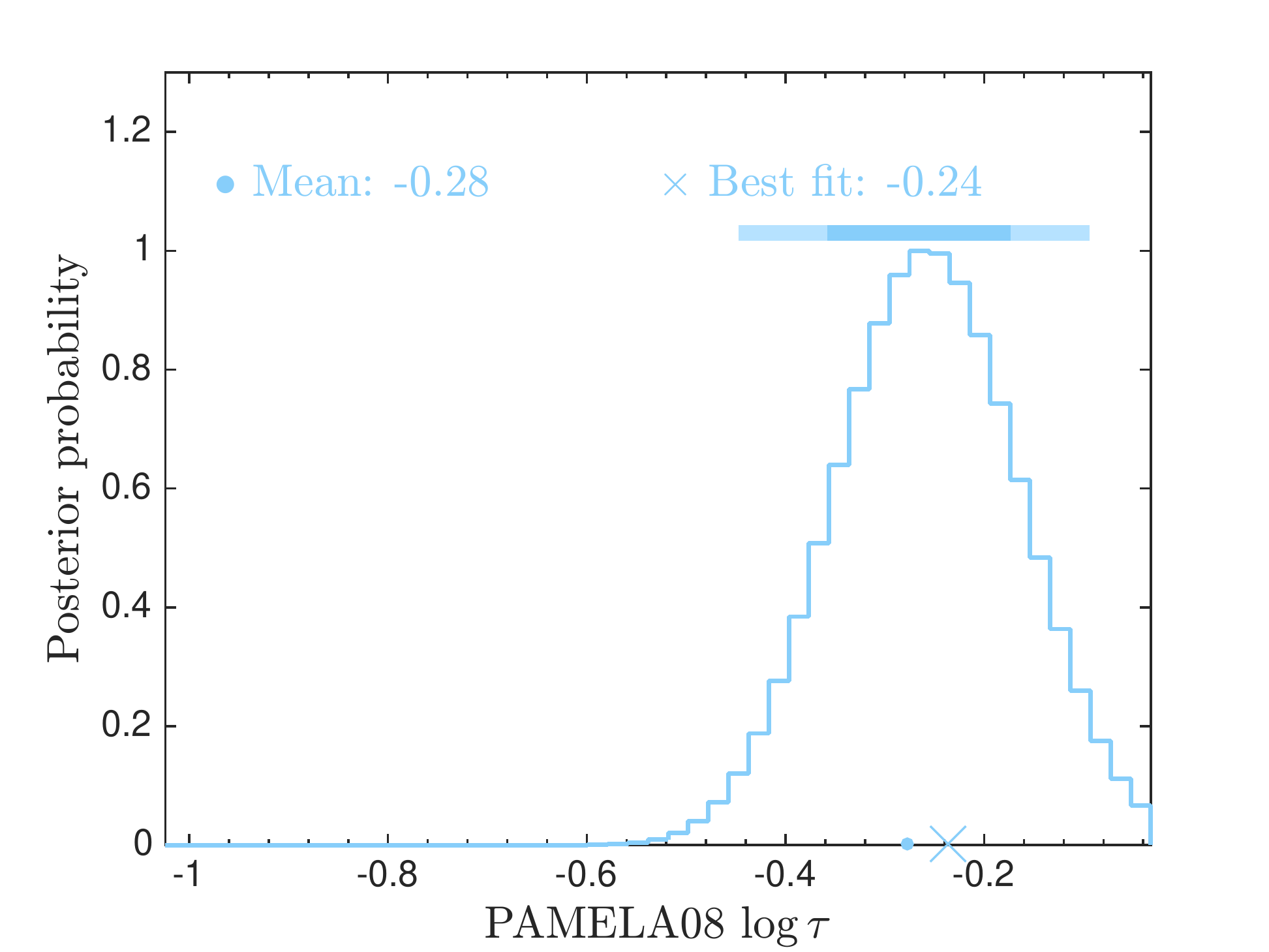} \includegraphics[width=.33\textwidth]{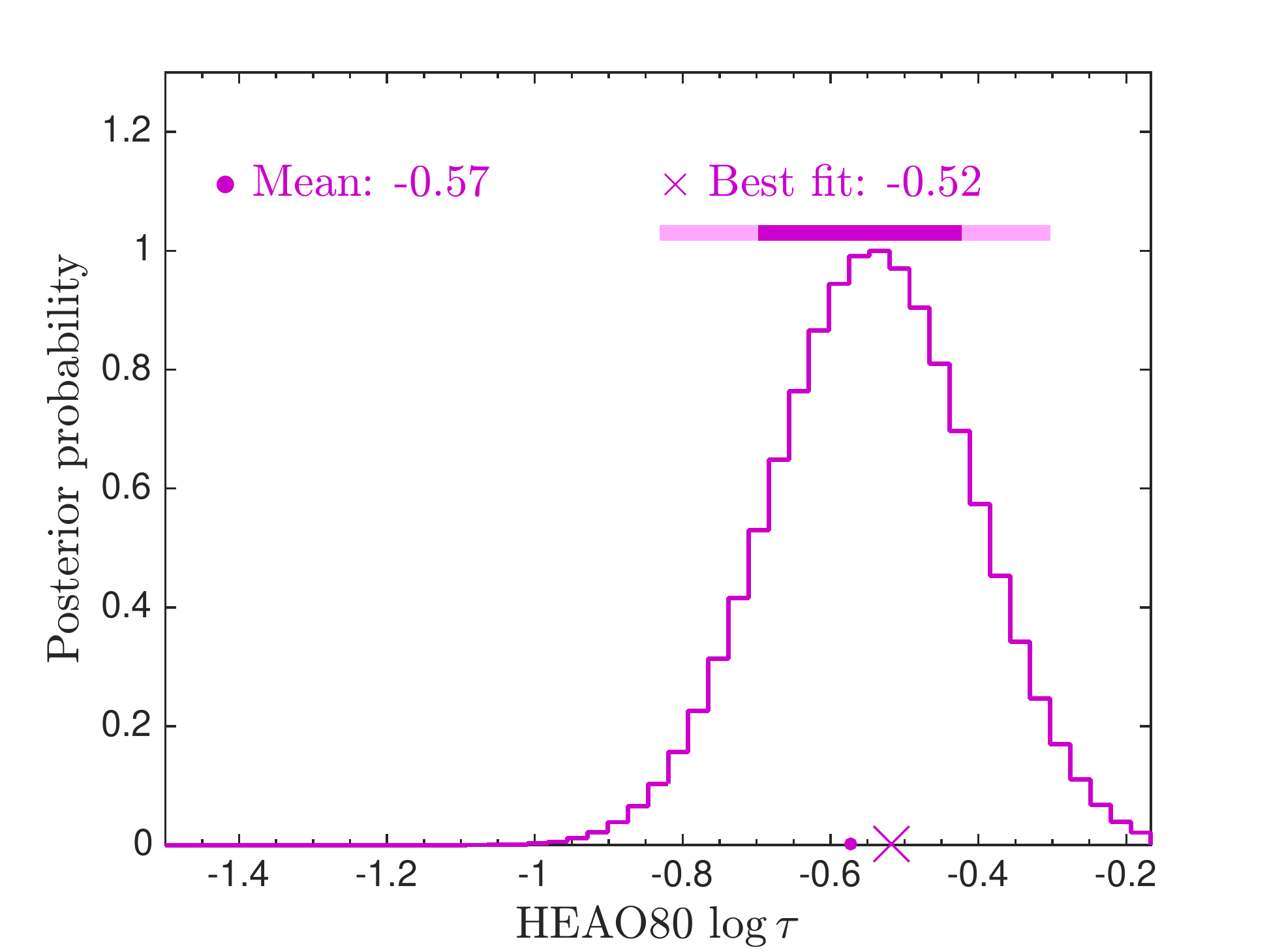}\includegraphics[width=.33\textwidth]{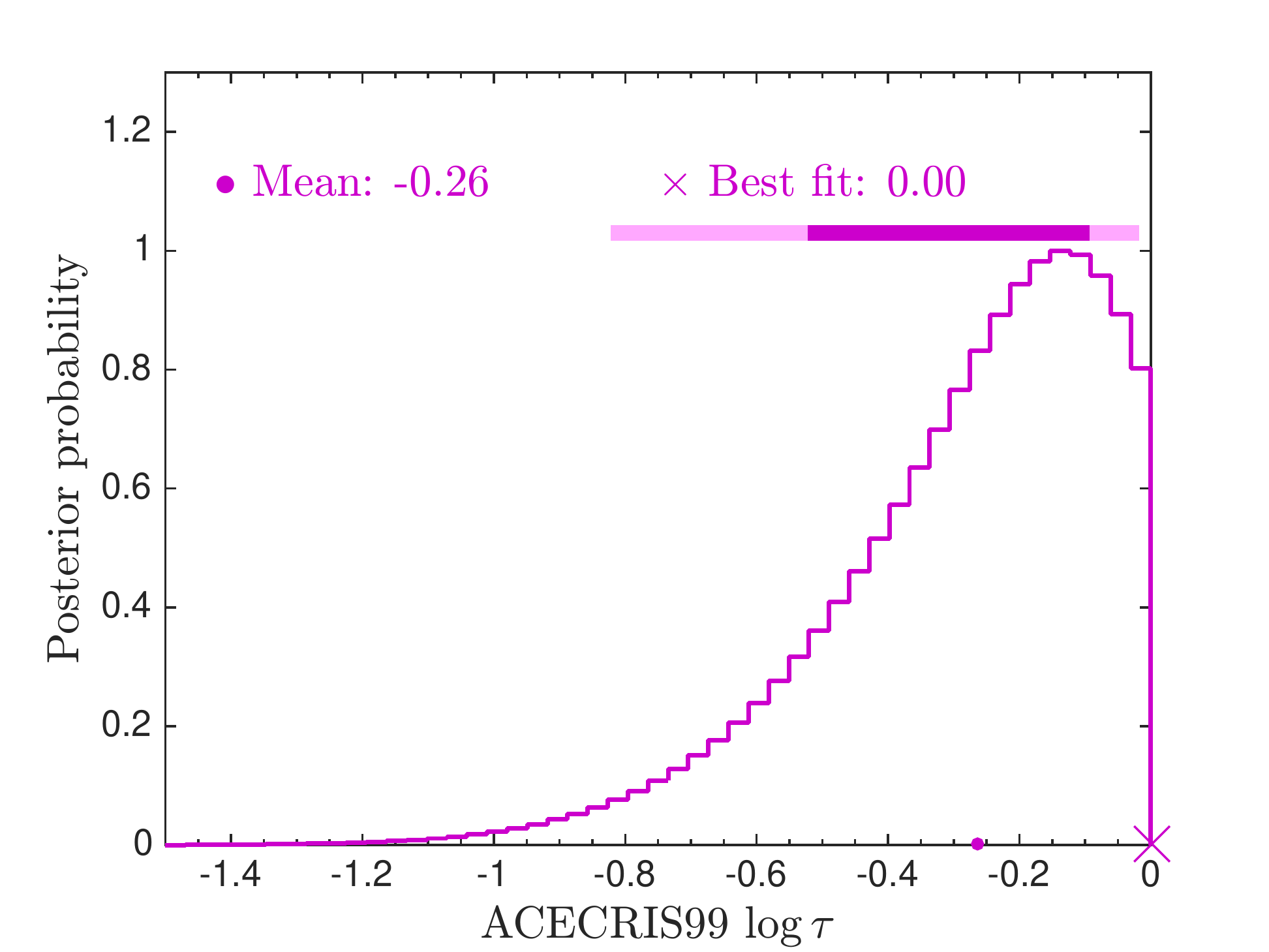} \\ \includegraphics[width=.33\textwidth]{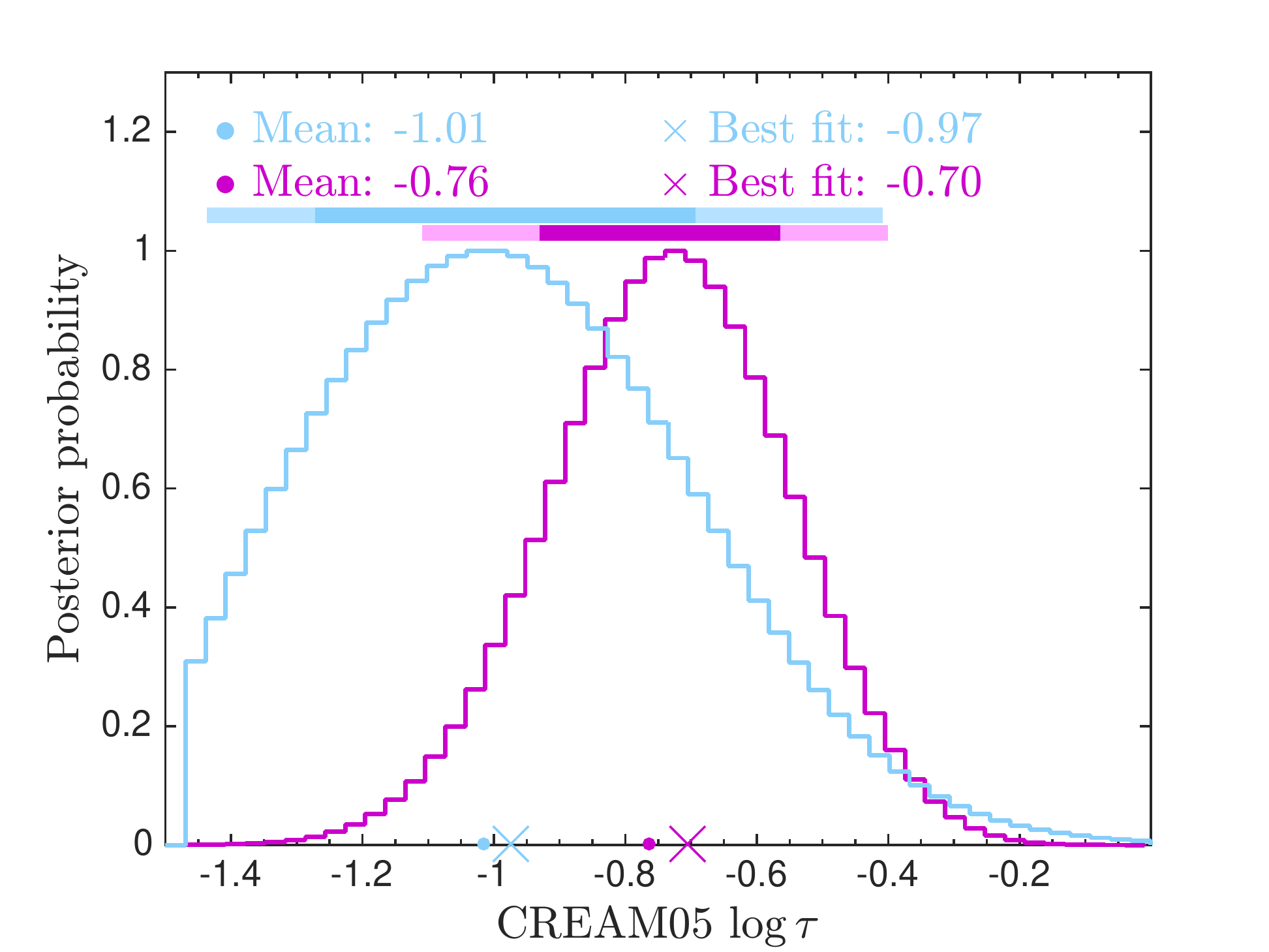}\includegraphics[width=.33\textwidth]{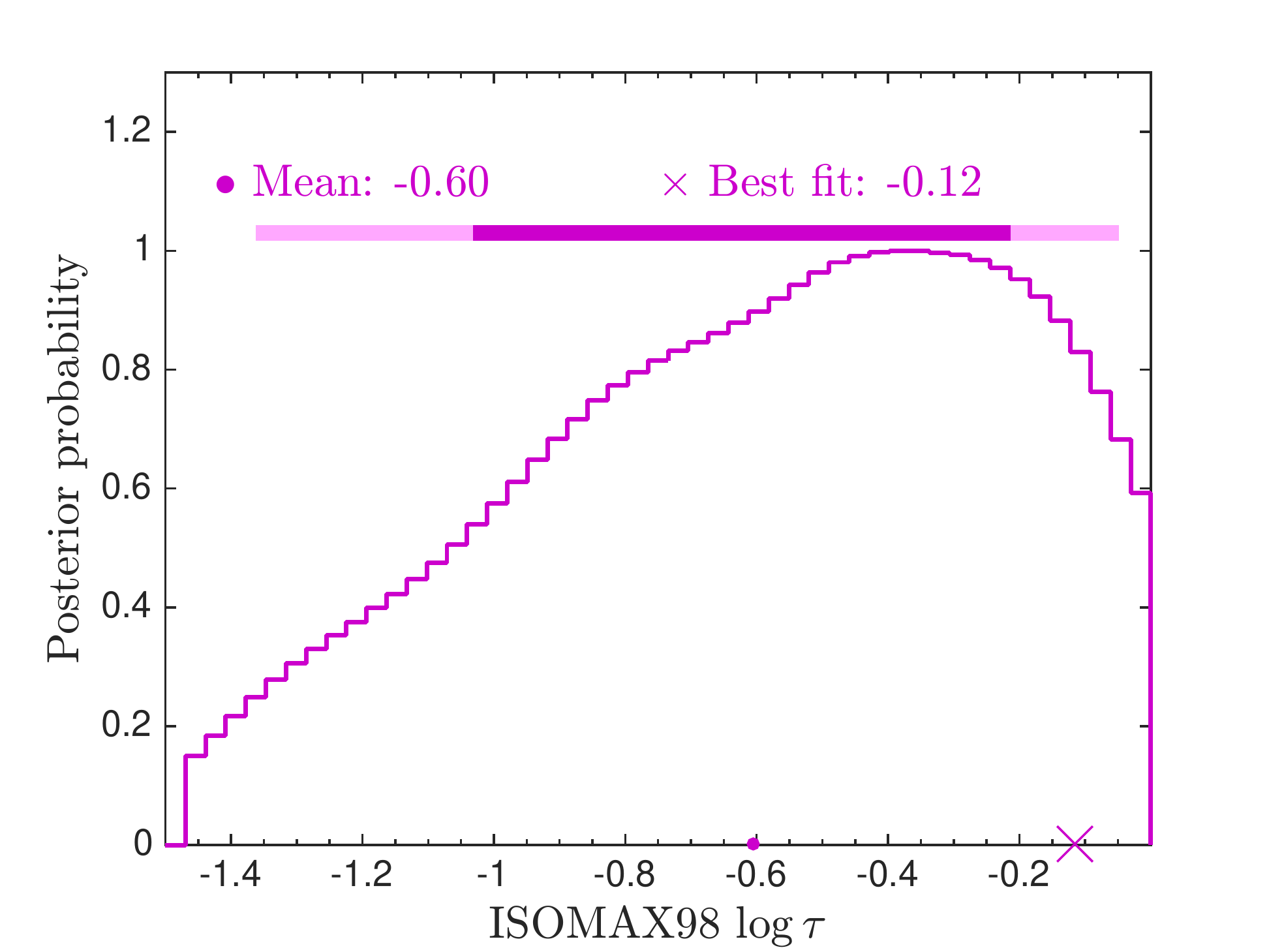}
\caption{Posterior distributions of the $\tau$ rescaling parameters, with 1
   and 2 sigma credible intervals. }\label{fig:tauparams}
\end{centering}
\end{figure*}

The $\tau$ error bar rescaling parameters for each experiment are shown in Figure \ref{fig:tauparams}. These are mainly skewed towards $\log \tau = 0$, indicating no rescaling is necessary and thus good agreement between datasets. Some tension can be seen in the CREAM data (green points in our figures), possibly owing to the wide binning. Finally, the ISOMAX rescaling parameter was effectively consistent with the entire prior range, due to the paucity of available data (2 data points). 

%\clearpage

%%%%%%%%%%%%%%%%%%%%%%%%%%%%%%%%%%%%%%%%%%%%%%%%%%%%%%%%%%%%%%%%%%%%%%%%%
\section{Discussion} \label{discussion}
%%%%%%%%%%%%%%%%%%%%%%%%%%%%%%%%%%%%%%%%%%%%%%%%%%%%%%%%%%%%%%%%%%%%%%%%%

A considerable underprediction of the $\bar p$ flux calculated in reacceleration models that are tuned
to the B/C ratio was first noticed by \citet{Moskalenko2002}. 
It has been shown that accurate antiproton measurements during the solar
minimum of 1995-1997 by the BESS instrument \citep{2000PhRvL..84.1078O} are inconsistent with existing propagation models at the $\sim$40\% level at about 2 GeV, 
while the stated measurement uncertainties in this energy range were $\sim$20\%. 
Using local CR measurements, simple energy dependence of the diffusion coefficient, 
and uniform CR source spectra throughout the Galaxy, conventional models failed to reproduce simultaneously both the secondary/primary nuclei ratio and $\bar p$ flux.
The reacceleration model designed to match secondary/primary nuclei ratios (e.g., B/C) produces too few antiprotons because matching the B/C ratio at all energies requires the diffusion coefficient to be too large. 
The models without reacceleration can reproduce the $\bar p$ flux; however, the low-energy decrease 
in the B/C nuclei ratio requires an \emph{ad hoc} break in the diffusion coefficient. 
The diffusion-convection model was constructed specifically to reproduce the $\bar p$ data, but required fine tuning. 
These results were later confirmed by \cite{2005AdSpR..35..147S}.

An attempt to find an acceptable solution for the reacceleration models was made by \citet{Moskalenko2003}.
They showed that the spectra of \emph{primary nuclei} as measured in the heliosphere may contain a 
fresh, local, ``unprocessed" component at low energies. The latter leads to an effective decrease in both the B/C ratio at low energies and the diffusion coefficient,
thus increasing the production of antiprotons. The paper associated the fresh component with the Local Bubble and
independent evidence for supernova activity in the solar vicinity in the last few Myr was taken as a support to this idea. 

\citet{Ptuskin2006} found that the diffusive reacceleration model with Iroshnikov-Kraichnan spectrum of interstellar turbulence $\delta=0.5$
\citep{1964SvA.....7..566I,1965PhFl....8.1385K} and wave damping helps to alleviate the problem, though does not solve it completely.
The main idea of that paper is that the dissipation of waves due to the resonant interaction with CR particles may terminate the slow Kraichnan-type cascade 
below wavelengths $10^{13}$ cm thus leading to the increase in the diffusion coefficient at low rigidities.
No significant effect of CR damping was found in the case of the fast Kolmogorov cascade. 

These early papers \citep{Moskalenko2002,Moskalenko2003,Ptuskin2006} compared the predicted $\bar p$ flux to 
the data collected during the two balloon flights of the BESS instrument \citep{2000PhRvL..84.1078O}.
The total number of collected antiprotons was between 51 and 64 per energy bin in four bins ranging from 1.52 GeV to 3.00 GeV. 
Some of these antiprotons could be mismodeled secondaries produced in the atmosphere above the instrument. The discrepancy with
the predictions of the reacceleration model could also imply possible unaccounted systematic errors of the data analysis. However, 
direct measurements in space by PAMELA experiment \citep{Adriani2010} made during the next solar minimum, confirmed the earlier 
BESS measurements with doubled statistics in the same energy range. 
Simultaneously, the PAMELA measurements of the B/C ratio \citep{Adriani2014} yield a value of $\delta=0.397\pm0.007$ 
for the index of the diffusion coefficient that is close to the classical value of $\delta=1/3$, hinting at the Kolmogorov spectrum of 
interstellar turbulence. Furthermore, the preliminary AMS-02 results for the B/C ratio reported by the PI Professor 
S.~Ting\footnote{https://indico.cern.ch/event/381134/timetable/\#20150415} agree with PAMELA data and indicate a somewhat flatter index.

Agreement between BESS and PAMELA on the $\bar p$ measurement and a hint that the index of the diffusion coefficient
is close to the Kolmogorov value support the idea that the discrepancy with the predicted $\bar p$ flux is inherent
and not due to experimental uncertainty.
Our first scan of the parameter space \citepalias[see e.g.][]{2011ApJ...729..106T} quantitatively confirms this finding.
Our new results (Section \ref{results}) show significant tension between a set of propagation parameters derived from a standard secondary to 
primary ratio B/C, and those derived from $p, \bar p$, He data, as can be explicitly seen in Fig.~\ref{fig:secspecplot}.
This tension may, in fact, reflect the properties of significantly different Galactic volumes probed by different species.

To illustrate this point, let us calculate the effective propagation distance for different CR species.
For the interaction time scale we have
\be
\tau \sim [\sigma_r n c]^{-1},
\ee 
where $\sigma_{r}$ is the total reaction cross section, $n\sim1$ cm$^{-3}$ is the average gas number density in the Galactic disk, 
and $c$ is the speed of light. The effective propagation distance can be estimated as
\be
\left<x\right> \sim \sqrt{6D\tau} \sim \left(\frac{6D_0}{\sigma_r n c}\right)^{1/2} \left(\frac{\rho}{\rho_0}\right)^{\delta/2}.
\label{eq:x}
\ee 
In the case of nuclear species, the total reaction cross section is approximately
\be
\sigma_{r}(A)\approx 250\ {\rm mb}\ (A/12)^{2/3},
\ee 
which is made to roughly reproduce the cross sections published by \citet{1996PhRvC..54.1329W}, and corrected by Wellisch (private comm.),
and we took $\sigma_{r}(^{12}$C$)\approx 250$ mb.
In the case of $p$ and $\bar p$, $\sigma_{r}^p\approx\sigma_{r}^{\bar p}\approx 40$ mb. 
The exact numbers are not very important as we are seeking for a rough estimate of the diffusion volume for different species at the rigidity of a few GV.

Table~\ref{tab:params_constraints} gives the results of the propagation parameters scan.
For $p, \bar p$, He scan, we have $D_0^p\approx 6\times10^{28}$ cm$^2$ s$^{-1}$ at $\rho_0=4$ GV, and $\delta\approx 0.46$.
For the light elements (Be--Si), we have $D_0^A\approx 9\times10^{28}$ cm$^2$ s$^{-1}$ at $\rho_0$, and $\delta\approx 0.38$.
The superscripts $p$ and $A$ are added to distinguish between the values derived from $\bar p$ and B propagation parameters scans.
The spectral indices are somewhat different, but we can use a single index of $\delta\approx 0.4$ in our estimates.

Substitution of these values into Eq.~(\ref{eq:x}) gives:
\begin{eqnarray}
\left<x\right>_A &\sim& 2.7\ {\rm kpc}\ \left(\frac{A}{12}\right)^{-1/3} \left(\frac{\rho}{\rho_0}\right)^{\delta/2},\label{xA}\\
\left<x\right>_p &\sim& 5.6\ {\rm kpc}\ \left(\frac{\rho}{\rho_0}\right)^{\delta/2}.\label{xp}
\end{eqnarray}
Even though the value of the diffusion coefficient derived from $p, \bar p$, He is a factor of 1.5 smaller than that for the light nuclei, 
the former probes an area ($\propto \left<x\right>^2$) of the Galaxy that is four times larger. This ratio does not depend on $\delta$. The volume 
probed by the lighter species includes a considerable area in the inner Galaxy, where the SNR rate and the OB star distribution reach their 
maxima (at a distance of about 5 kpc from the Galactic center).
It is thus natural to expect that more turbulent interstellar medium has a smaller diffusion coefficient. 

This is only an estimate, but it gives some idea of the typical distances. Even though CRs can in principle 
 come from larger distances, their number density would be negligible compared to locally-produced CRs of the same species. 
This estimate is consistent with the typical lifetime of CRs in the Galaxy assuming a uniform diffusion coefficient in the disk and halo.
The best fit halo size derived from the $^{10}$Be/$^9$Be ratio is $z_h\approx5$ kpc in the case of the light elements, 
and $z_h\approx10$ kpc from $p, \bar p$, He scan (Table~\ref{tab:params_constraints}), i.e., larger than the effective distances given by Eqs.~(\ref{xA})-(\ref{xp}). 
Their posterior means are even larger, $z_h\approx10.35$ kpc with $1\sigma$ error bars of 4.2 kpc and 4.9 kpc correspondingly.

Our results are, therefore, the first to definitively show that by separating the two data sets, one can fit them with two different reacceleration parameter sets. 
The significantly lower Alfv\`en speed $v_{Alf}\propto B/\sqrt{\rho_{ISM}}$, $8.9 \pm 1.2$ km s$^{-1}$ ($p$, $\bar p$, He) vs.\ $30.0 \pm 2.5$ km s$^{-1}$ (Be--Si), may hint at a smaller $B/\sqrt{\rho_{ISM}}$, possibly owing to a denser ISM plasma as one approaches the inner Galaxy.

Variations of the propagation parameters throughout the Galaxy is not the only possible reason of the discussed differences. 
Source (SNe) stochasticity \citep{2001ICRC....5.1964S} may contribute to the local fluctuations in fluxes of individual CR species.
Freshly accelerated CR particles from relatively recent SN explosions may or may not lead to the increased local production of secondary species.
As was already mentioned, the presence of local sources of low-energy primary nuclei could lead to effects that mimic
the propagation parameters variations \citep{Moskalenko2003}. 
In particular, the value of the effective diffusion coefficient $D_0^A$ could be reduced, i.e., made consistent with $D_0^p$, by invoking an
additional component of the locally produced \emph{primary} CR nuclei. 
Eqs.~(\ref{xA}), (\ref{xp}) indicate that such sources should be located within 1--2 kpc. 
Besides the Local Bubble, other obvious candidates are the local (Orion) arm and the Perseus arm, where the SN rate is 
higher than in the interarm region \citep{1999MNRAS.302..693D}.

If instead, the value of $D_0^p$ is required to be made consistent with $D_0^A$,
then CR sources should produce additional antiprotons. Production of secondary nuclei in the SNR shocks was proposed
by \citet{2003A&A...410..189B}. Antiprotons are also secondary and thus can be produced in the same process \citep{2009PhRvL.103h1103B},
see also \citep{2013PhRvD..87d7301K,2014PhRvD..90f1301M,2014PhRvD..89d3013C}. However, this argument is circular
unless we assume that there is a distinct type of CR sources that is nuclei ($Z>2$) deficient and that this type of sources has enough
material nearby to produce additional antiprotons in significant amounts.
The first hypothesis of the local sources producing mostly \emph{primary} CR nuclei, therefore, appears more reasonable. 

Even though the structure of our Galaxy has been the subject of research since the invention of the telescope in the beginning of the 17th century,
only now we are starting to learn about its effects on CR fluxes.

The Galaxy is clearly not axially symmetric, yet the quality of the 
CR and diffuse emission data available until recently did not warrant propagation calculations beyond simple 
cylindrically symmetric geometry \citep{SMP2007}. The full 3D setup was available in GALPROP since the very
beginning \citep{SM1998,2001ICRC....5.1964S}, but it was mostly used to test the cylindrically symmetric 2D solution. In fact, 
the entirely uniform, so-called the Leaky-Box model, was completely dominating CR modeling in the 20th century.

Not surprisingly, the discussions on the influence of the Galactic structures on the intensity of CRs began about a decade ago.

The effects of the solar system's passage through the spiral arms on the global climate (ice ages)  
were discussed by \citet{2003NewA....8...39S}. These ideas were further developed in \citet{2009PhRvL.103k1302S} in connection 
with the so-called positron excess reported by PAMELA \citep{2009Natur.458..607A} and earlier by the HEAT experiment 
\citep{1997ApJ...482L.191B}. 
Clear evidence of the increased CR density in the spiral arms is provided by the {\it Fermi}-LAT residual maps
\citep{DiffusePaperII}, produced by subtracting the GALPROP diffuse $\gamma$-ray predictions from the {\it Fermi}-LAT skymaps.
The most significant excesses coincide with the tangential directions to the spiral arms which presumably contain freshly accelerated CRs.
There is currently no fully consistent model that would incorporate the details of the Galactic structure. This is mostly connected with
the difficulty of recovering the 3D structure of our Galaxy, such as the distributions of gas, magnetic field, SNRs, and starforming regions 
using astrophysical observations. Dependence on the temperature gradient in the ISM is discussed in \citet{2016AdSpR..57..519E}, and possible effects of
the details of the Galactic structure on CR propagation are actively discussed in the literature using a simplified description
\citep[e.g.,][]{Johanneson2015,Orlando2015,Porter2015,2015APh....70...39K,2016arXiv160103072B,2015arXiv151007801B}. A nearby source \citep[see e.g.][]{2015JPhG...42k5202E} would also lead to similar problems.

The most complete ever scan of the parameter space for CR injection and propagation is another landmark of the present paper.
Calculations of the CR source abundances were done in the past \citep[e.g.,][]{Engelmann1990,1996ApJ...465..982D,Wiedenbeck2001,2008ICRC....2..149W}.
However, such calculations were usually made for elemental abundances\footnote{\citet{Wiedenbeck2001,2008ICRC....2..149W} 
and other ACE team publications calculate \emph{isotopic} source abundances.} using the Leaky-Box model or its equivalent.
By current standards, the models and datasets (e.g., semi-empirical cross sections) used in such calculations in the past were not detailed enough,
but reflected the current state of knowledge at that time. The first successful attempt to find the source abundances 
and propagation parameters in a self-consistent way using a proper propagation code \galprop{} was made by \citet{Moskalenko2008}.
The source (injection) abundances were taken first as the solar system abundances, which were then iterated
to achieve an agreement with the propagated abundances as provided by ACE at $\sim$200 MeV/nucleon \citep{Wiedenbeck2001}
assuming a propagation model, such as diffusive reacceleration or plain diffusion. The propagation parameters were then re-adjusted to
reflect the final source abundances. Even though the resulting abundances are fairly close to the previous calculation (Figure~\ref{fig:abd}),
the current paper accomplishes a significantly more challenging task by 
performing a full neural network-assisted scan over the 20 propagation and abundance parameters. Ten more 
nuisance parameters were included into the scan to account for possible systematic errors of different experiments. 
The result is the full set of best-fit values, posterior means and standard deviations.
The latter allows the detailed propagation calculations with meaningful constrains for related areas and for possible signatures of new physics.
 
Thus far, we have only considered the reacceleration model. Other models will be analyzed in forthcoming papers.

%%%%%%%%%%%%%%%%%%%%%%%%%%%%%%%%%%%%%%%%%%%%%%%%%%%%%%%%%%%%%%%%%%%%%%%%%%%
\section{Conclusions} \label{conclusions}
%#########################################################################
We have performed the largest ever -- in terms of number of free parameters, data, resolution and computing time -- study of CR propagation using a fully numerical state-of-the art computer code. By combining \galprop{} with the {\sc BAMBI} package, we were able to perform a full neural network-assisted  scan over the 20 propagation and abundance parameters, as well as 10 nuisance parameters. Two input galdef-files based on the best fits found here will be included in an upcoming update of the publicly available \galprop{} code\textsuperscript{\ref{galprop_link}}.
%  (\url{http://galprop.stanford.edu}).
%

Our results have highlighted two important conclusions. 1) available measurements of the radioactive species $^{10}$Be are not sufficient to significantly remove the degeneracy between the halo height $z_h$ and the diffusion parameter normalization $D_0$; and 2) The propagation parameters 
%that accurately model 
derived from the CR $p$, $\bar p$ and He data
%in the interstellar medium 
are \textit{not} compatible with those found from fitting light elements Be--Si. 
 
We take these results as a probable indication that the interstellar medium properties differ significantly enough over kpc scales to affect propagation of CRs, though we have mentioned other interpretations. This fact has important consequences for CR propagation studies: it is customary to use propagation parameters calibrated to local B/C data to predict fluxes of other CR species including electrons and positrons, 
both locally and as far away from the Earth as the Galactic center or, otherwise, to assume an \emph{ad hoc} functional form for the spatial dependence of the diffusion coefficient.  
Such approaches are particularly misleading in the search for physics beyond the Standard Model, such as signals of dark matter annihilation. 
An excess in antiprotons, positrons or 
$\gamma$-rays could indeed be an indication of a mischaracterized ISM, rather than a need for new physics.
 
\acknowledgements  
This work has been supported by the Royal Society under the International Exchange Scheme, grant number IE120221 as well as by STFC grant ST/N000838/1. 
I.~V.~M., E.~O., T.~A.~P. acknowledge support from NASA~Grant~No.~NNX13AC47G, 
E.~O. additionally acknowledges support from NASA~Grant~Nos.~NNX16AF27G and NNX15AU79G, and T.~A.~P. additionally acknowledges support from NASA~Grant~No.~NNX10AE78G.
R. RdA, is supported by the Ram\'on y Cajal program of the Spanish MICINN and also thanks the support of the Spanish
MICINN's Consolider-Ingenio 2010 Programme under the grant MULTIDARK CSD2209-00064, the \textit{Invisibles} European ITN project (FP7-PEOPLE-2011-ITN,
PITN-GA-2011-289442-INVISIBLES and the ``SOM Sabor y origen de la Materia" (2014-57816) and the ``Fenomenologia y Cosmologia de la Fisica mas
alla del Modelo Estandar e lmplicaciones Experimentales en la era del LHC" (FPA2013-44773) MEC projects, and Severo Ochoa del MINECO:  SEV-2014-0398.
A.~C.~V. was supported by FQRNT (Qu\'ebec) and \textit{Invisibles}. The use of Imperial College High Performance Computing cluster is gratefully acknowledged.

\appendix
\section{Validation of {\sc BAMBI}/{\sc SkyNet} }
\label{validation}
Before launching our high-resolution physics scans, we performed a series of validation scans, with the goal of optimizing the {\sc BAMBI} framework with {\sc MultiNest} as well as determining the reliability of the trained neural nets (NNs). In order to determine the optimal input settings for the network, (i.e., those that maximise speed-up while predicting the likelihood function reliably), several runs were carried out with different values for the two main settings that determine the efficiency and accuracy of the NN training: $n_{\rm hid}$, the number of hidden nodes; and $\sigma$, which sets the desired accuracy for the predicted likelihood value before the network takes over as an interpolator. These tests were carried out with fixed elemental abundances and with a low GALPROP resolution (\texttt{dr} = 1.0, \texttt{dz} = 0.1, \texttt{Ekin} = 2.0, \texttt{starttimestep} = 1.0e9, \texttt{endtimestep} = 1.0e2, \texttt{timestepfactor} = 0.25, \texttt{timesteprepeat} = 20) in order to rapidly obtain trained networks.

\begin{figure}[h]
\centering
\includegraphics[width=0.3\textwidth]{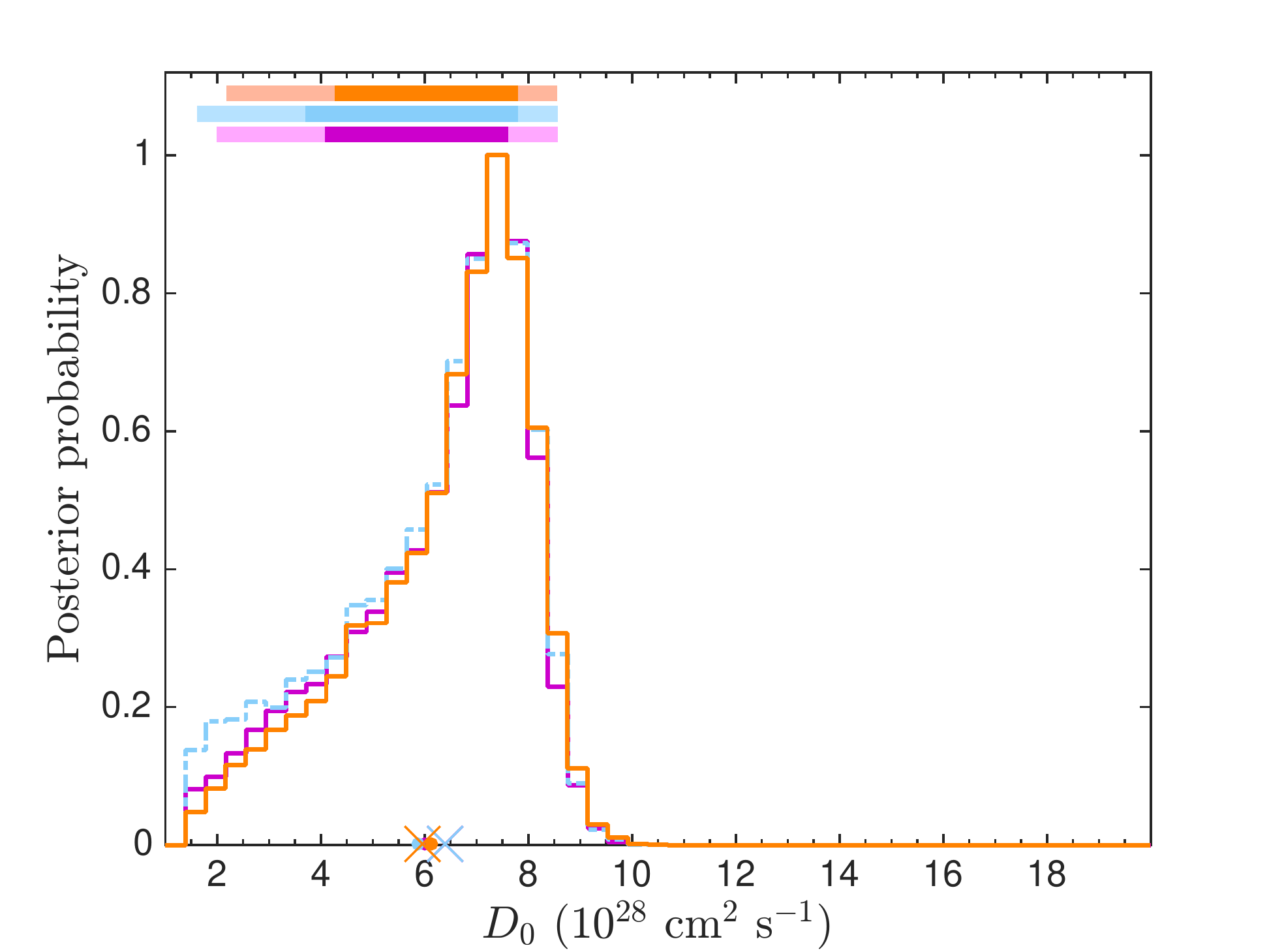}\includegraphics[width=0.3\textwidth]{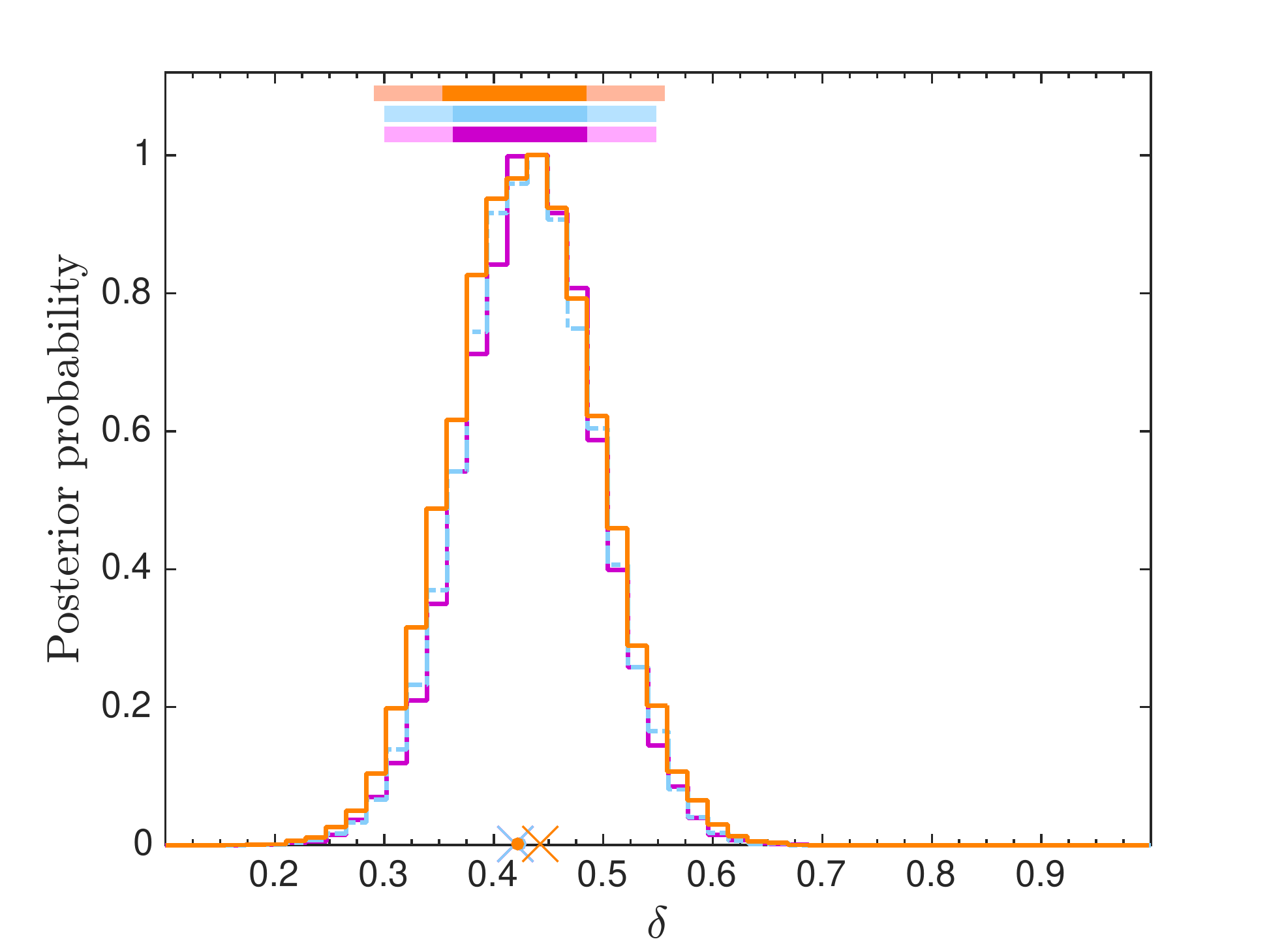}\includegraphics[width=0.3\textwidth]{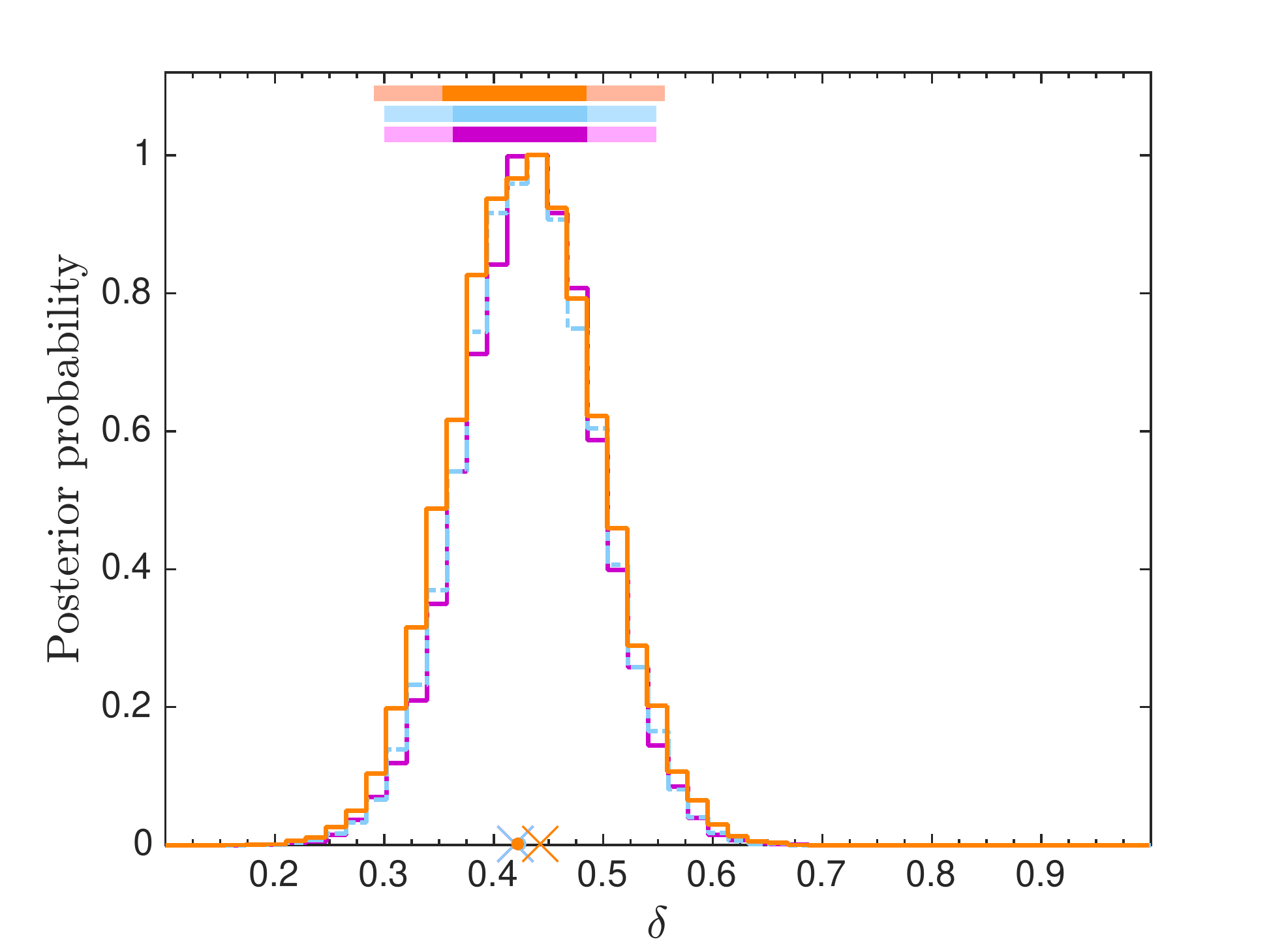}\\
 \includegraphics[width=0.3\textwidth]{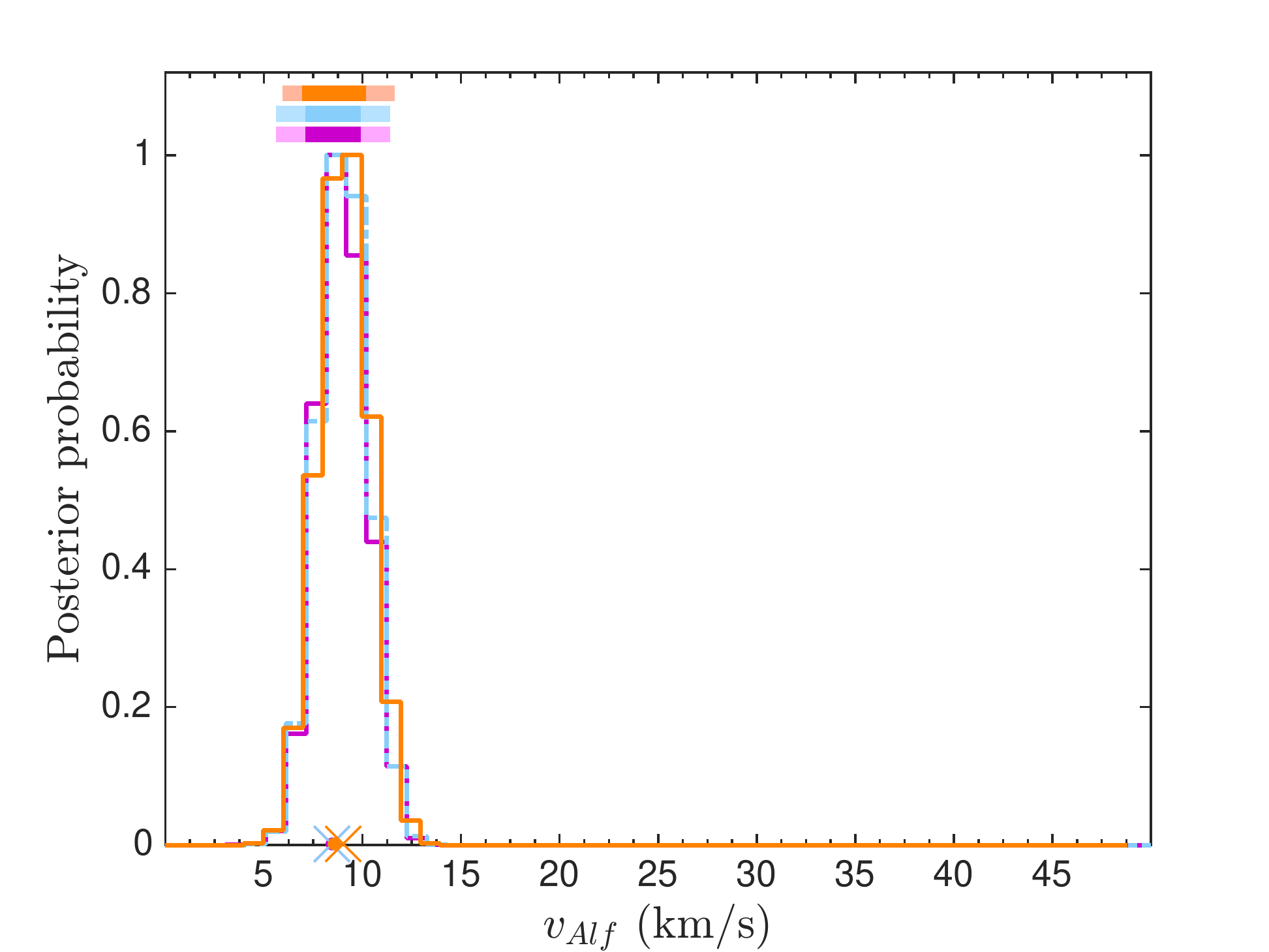}\includegraphics[width=0.3\textwidth]{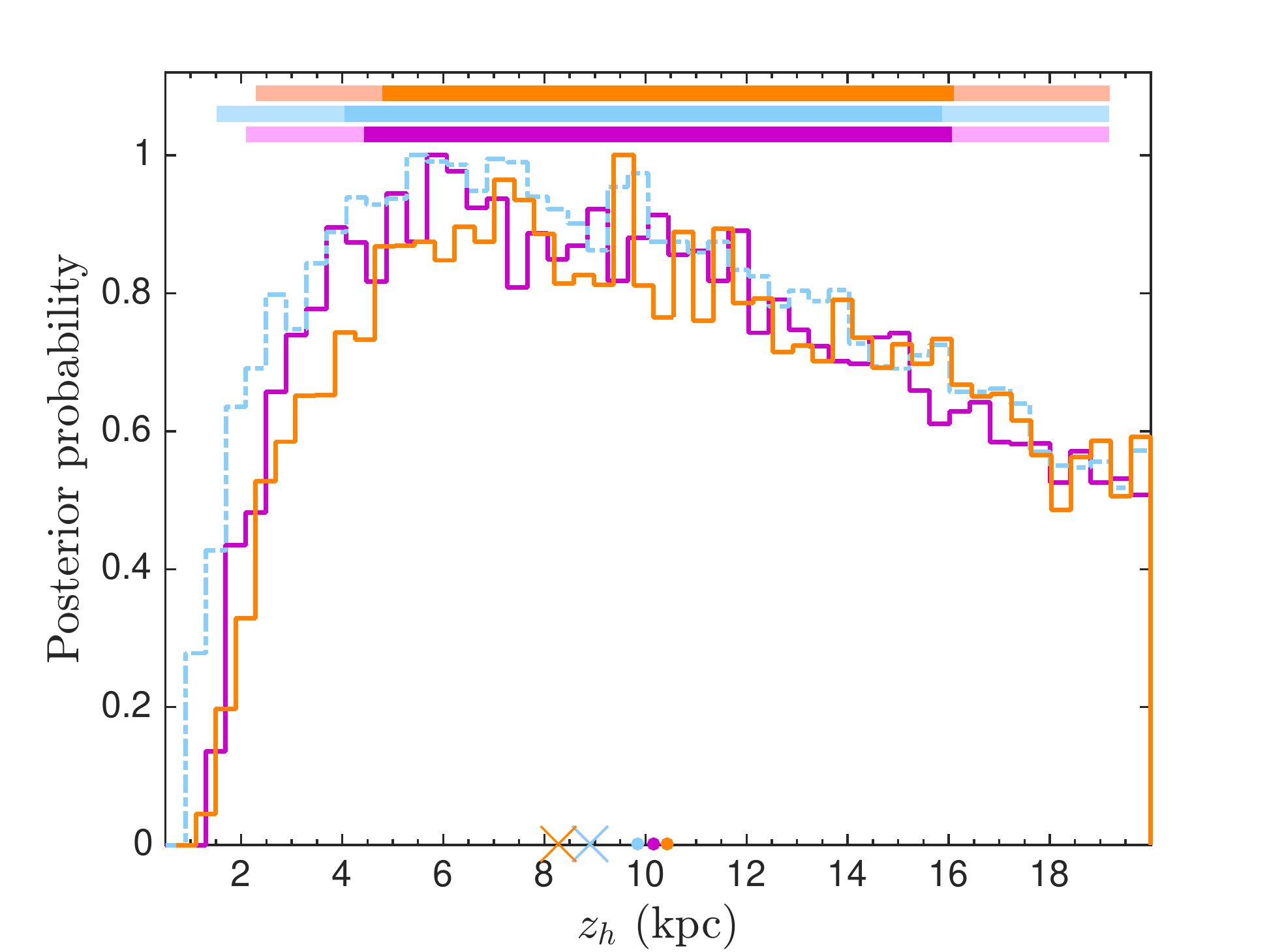}\includegraphics[width=0.3\textwidth]{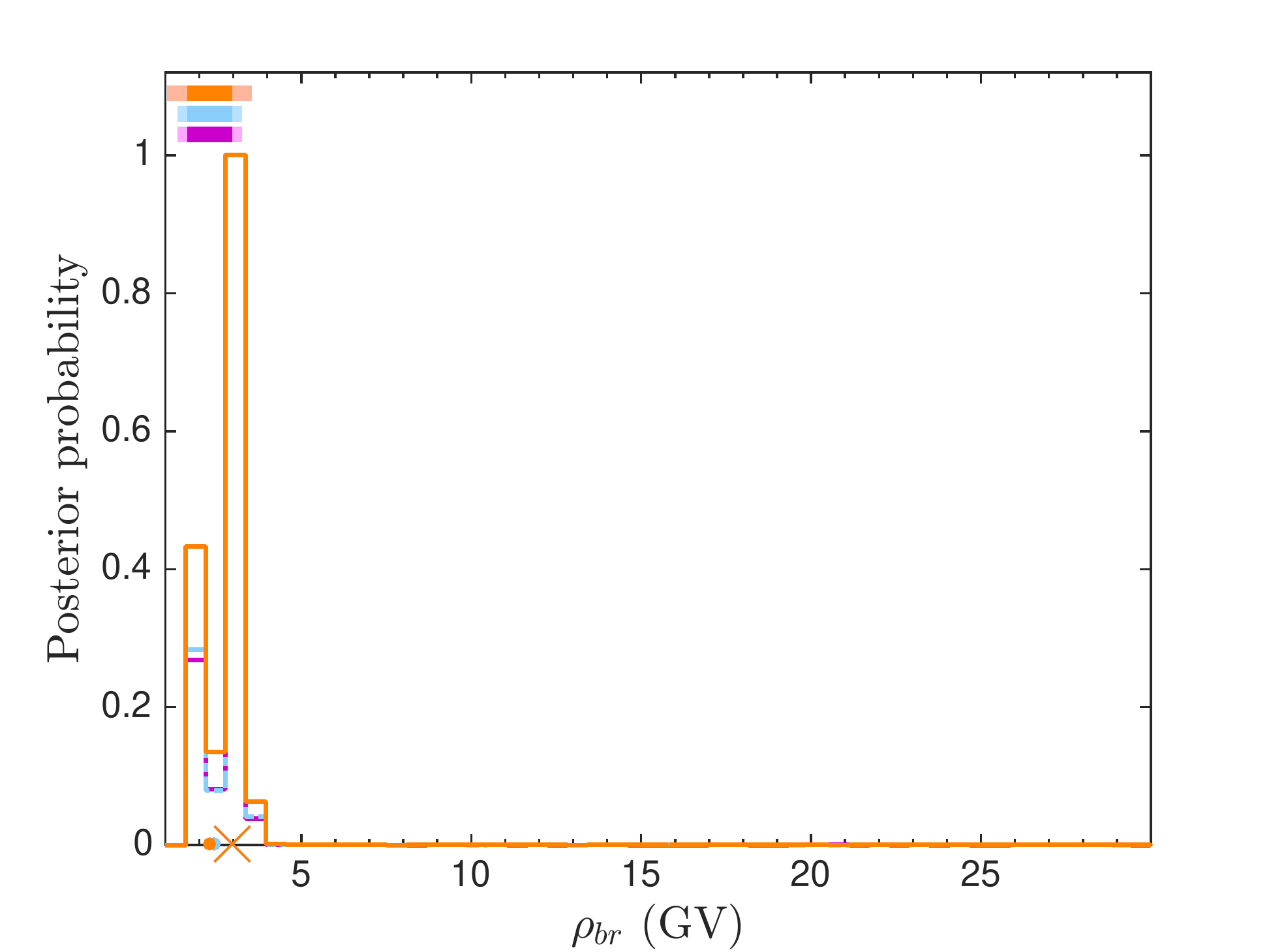} \\
  \includegraphics[width=0.3\textwidth]{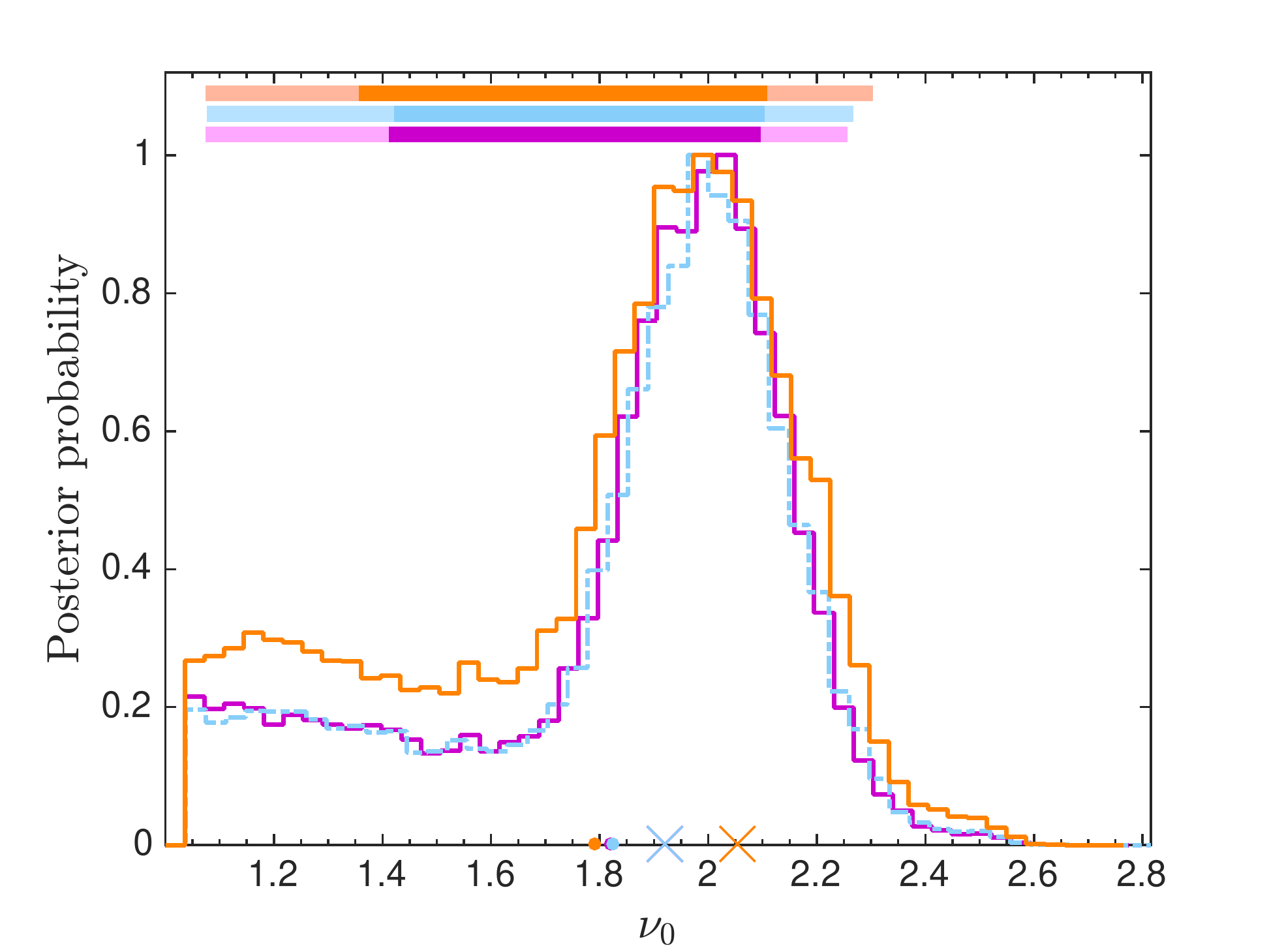}\includegraphics[width=0.3\textwidth]{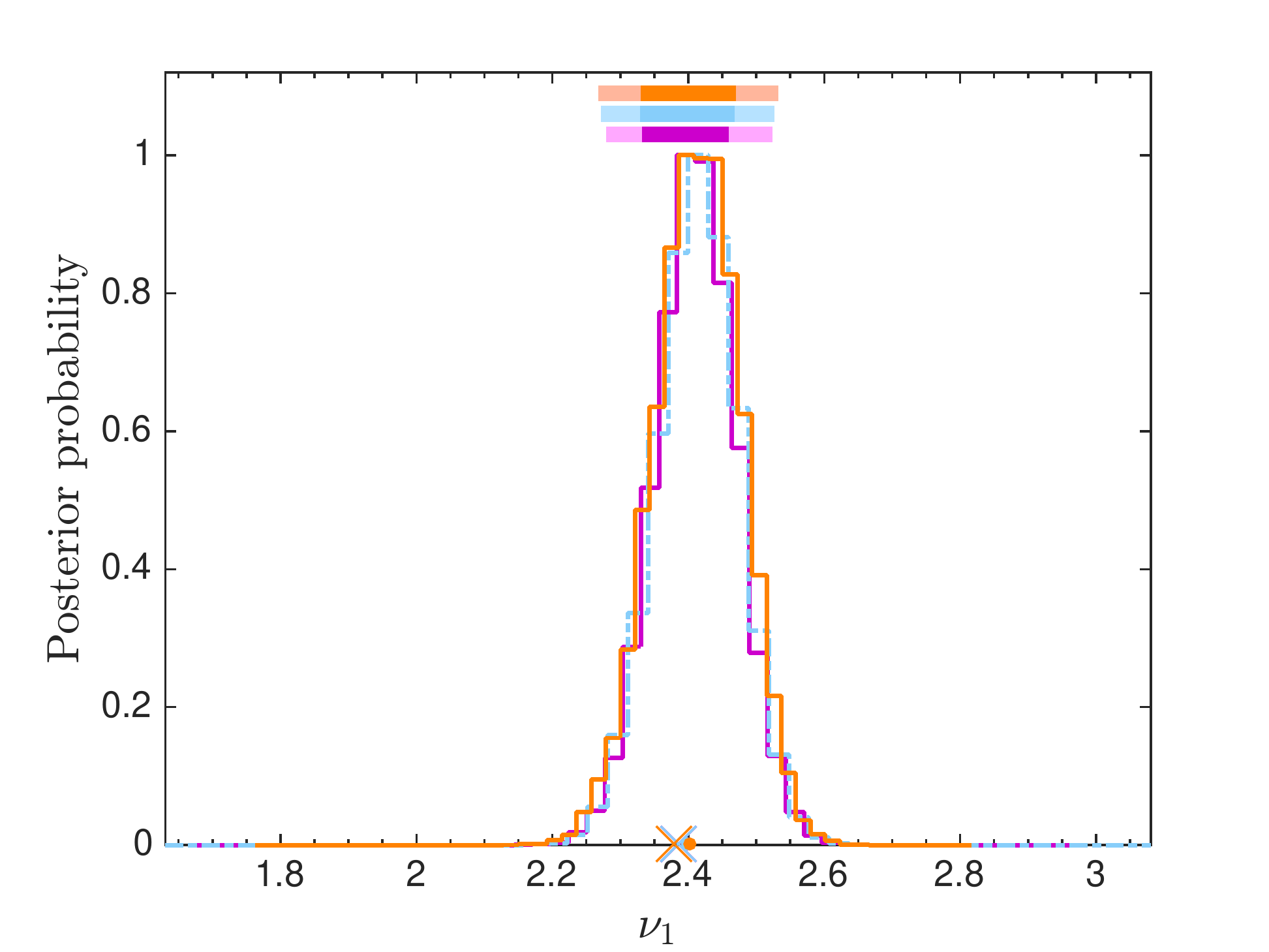}\includegraphics[width=0.3\textwidth]{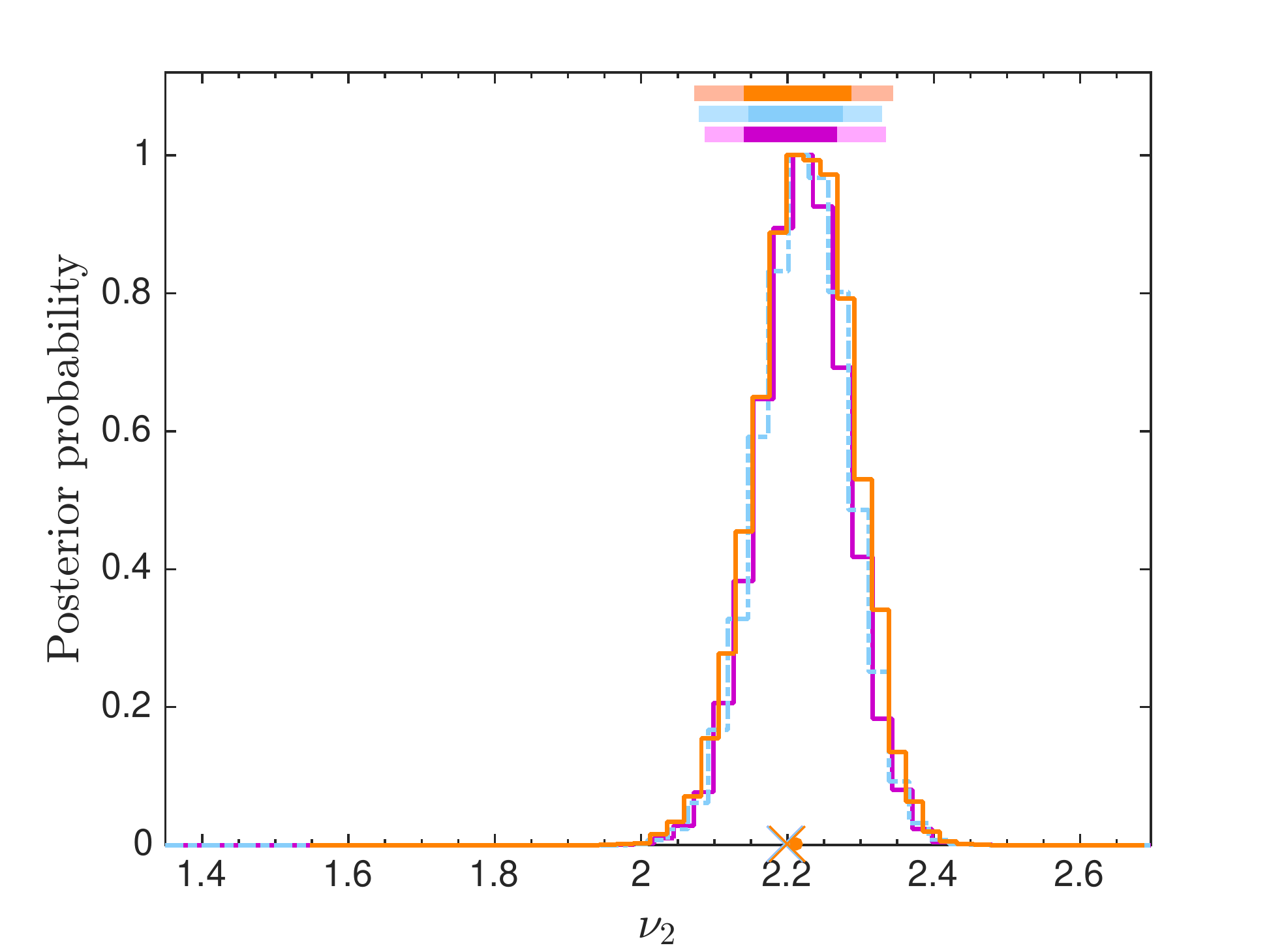} \\
   \includegraphics[width=0.3\textwidth]{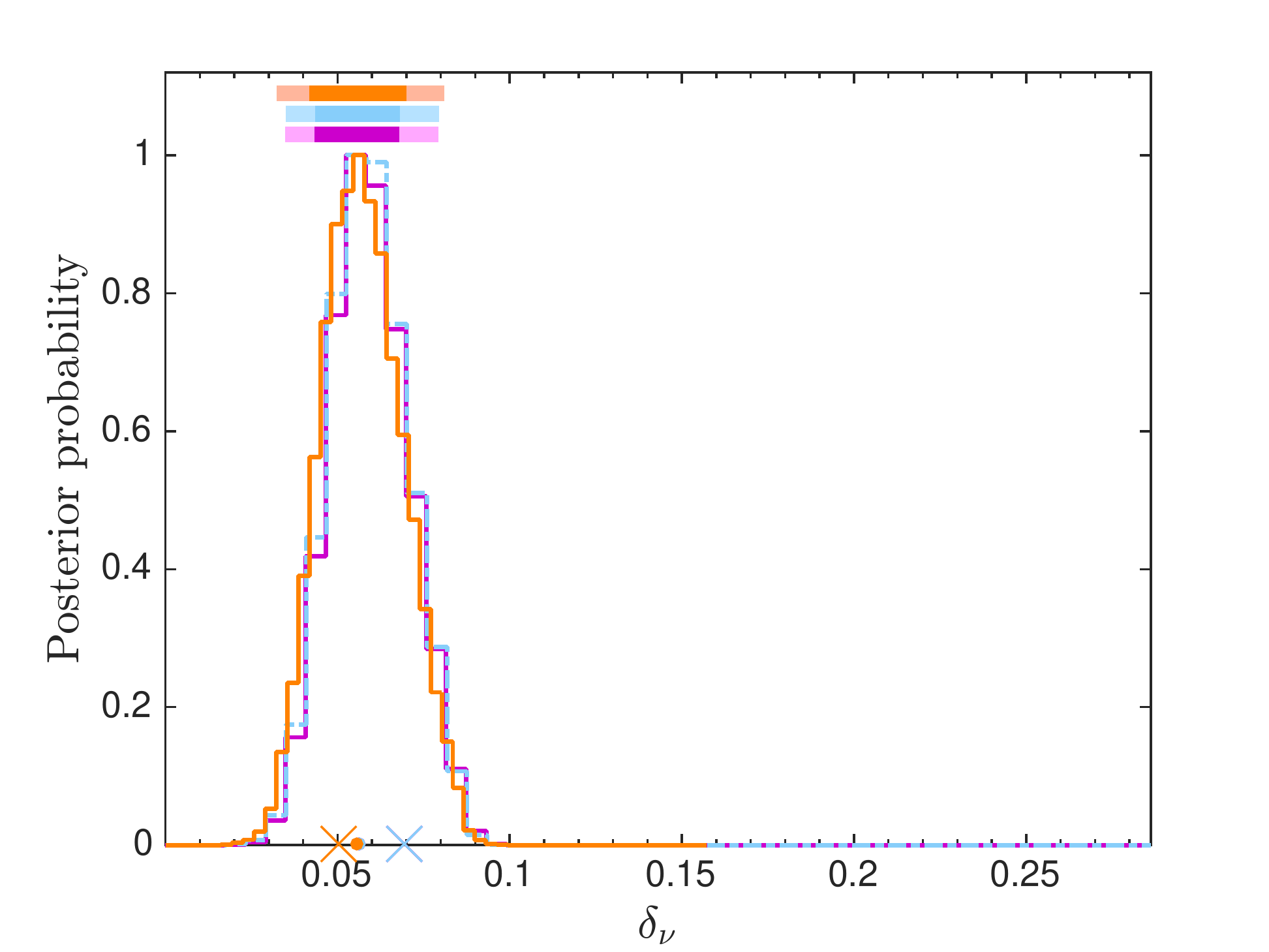}\includegraphics[width=0.3\textwidth]{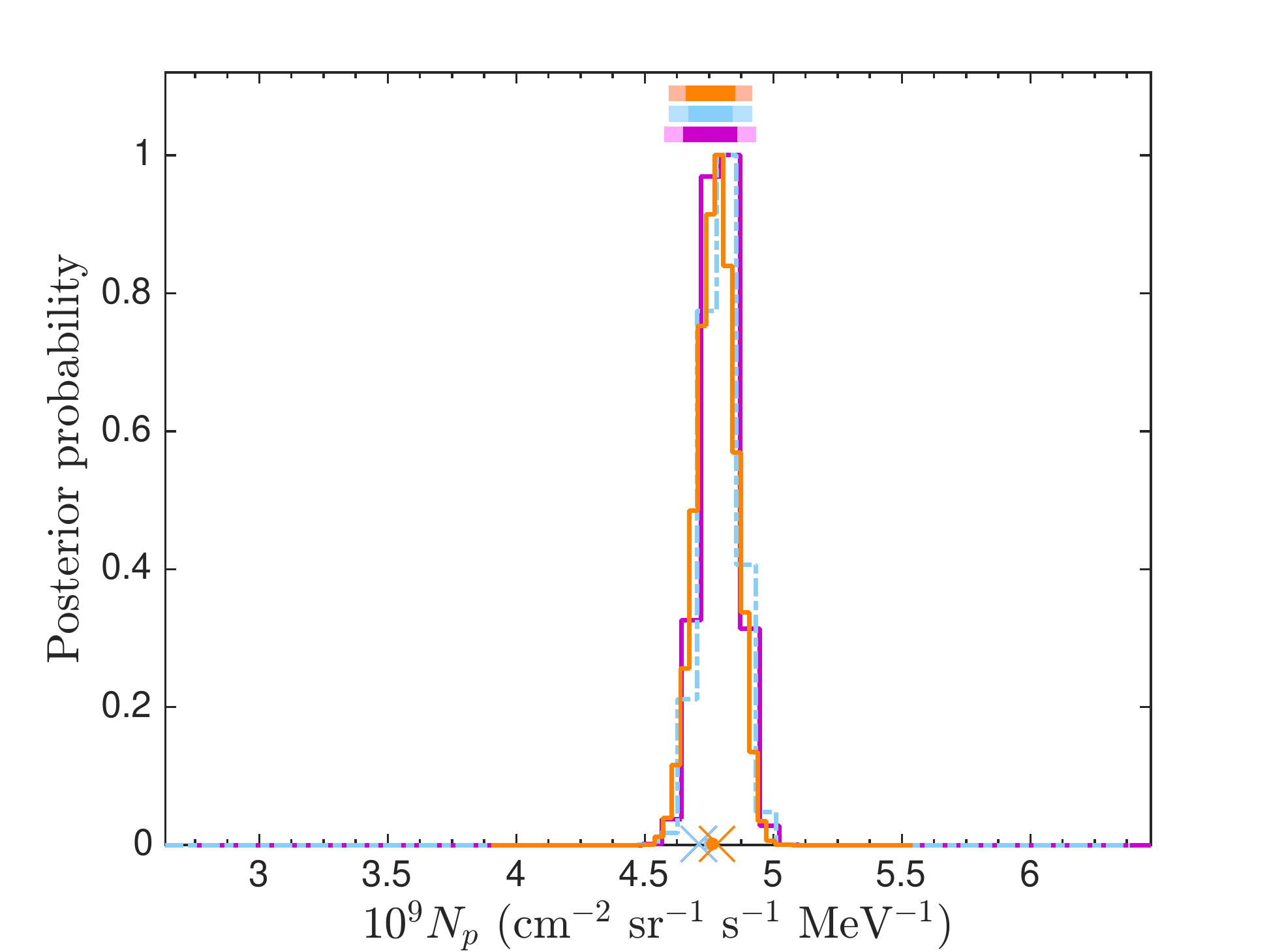}\includegraphics[width=0.3\textwidth]{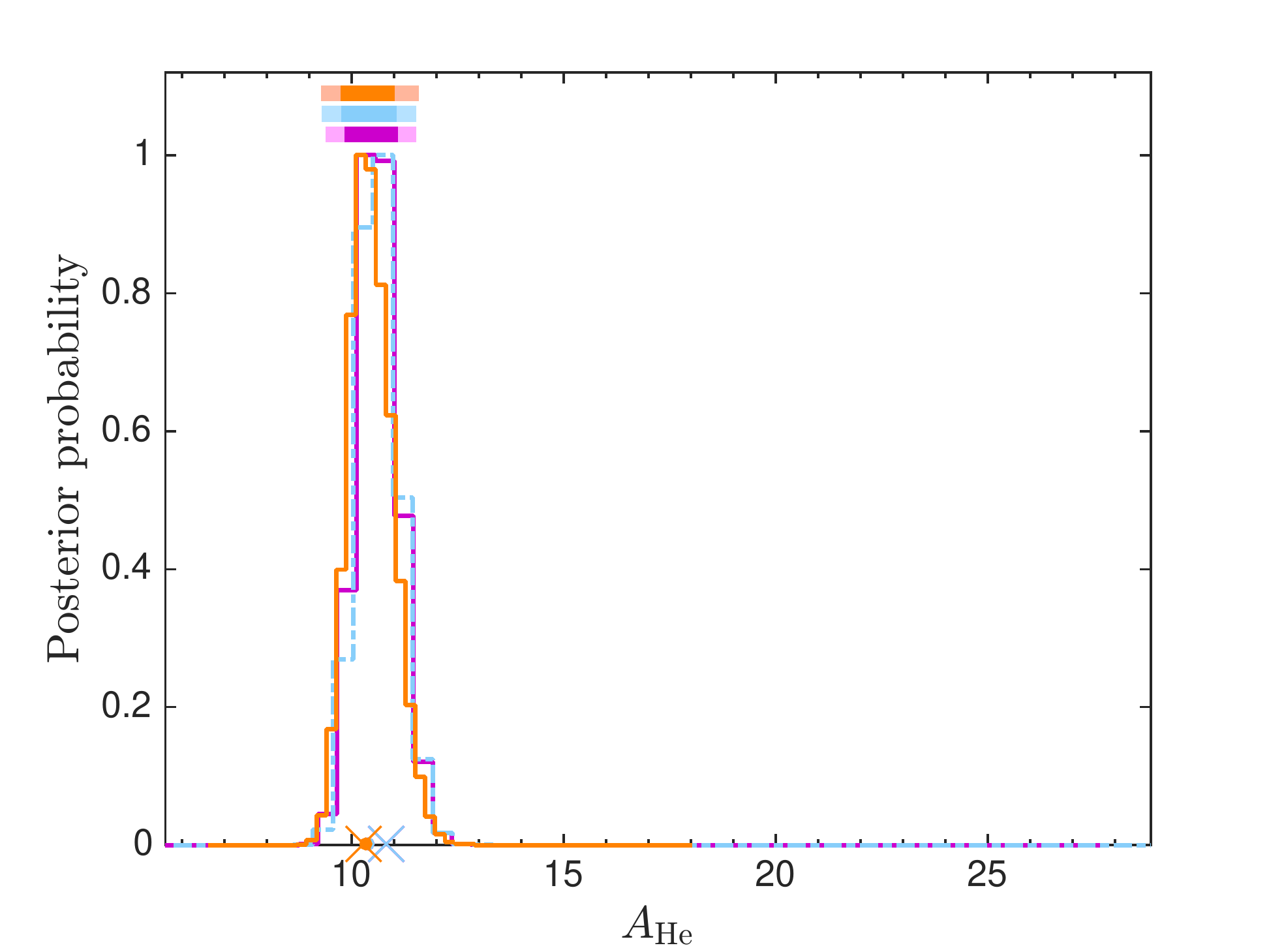} \\
    \includegraphics[width=0.3\textwidth]{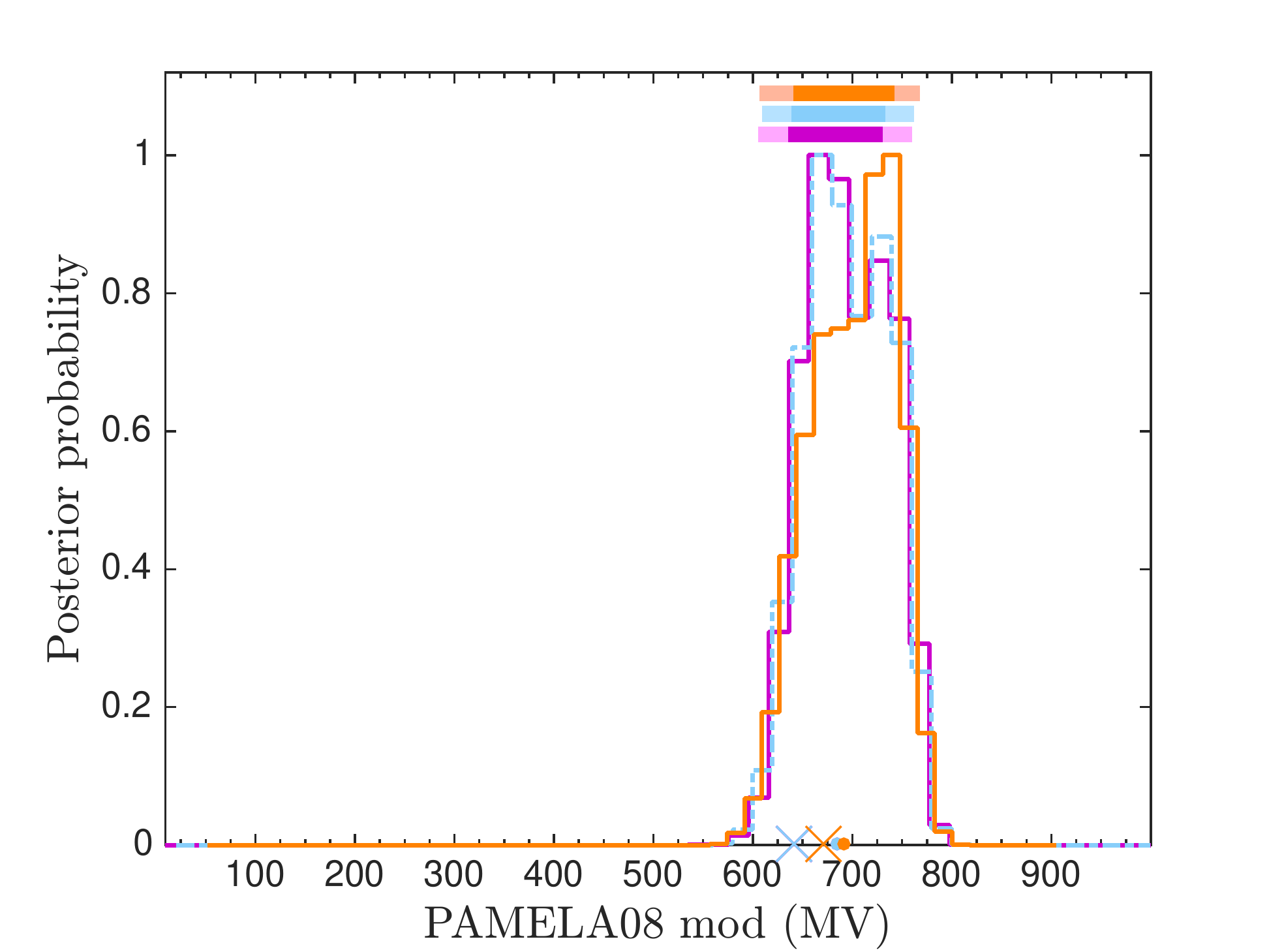}\includegraphics[width=0.3\textwidth]{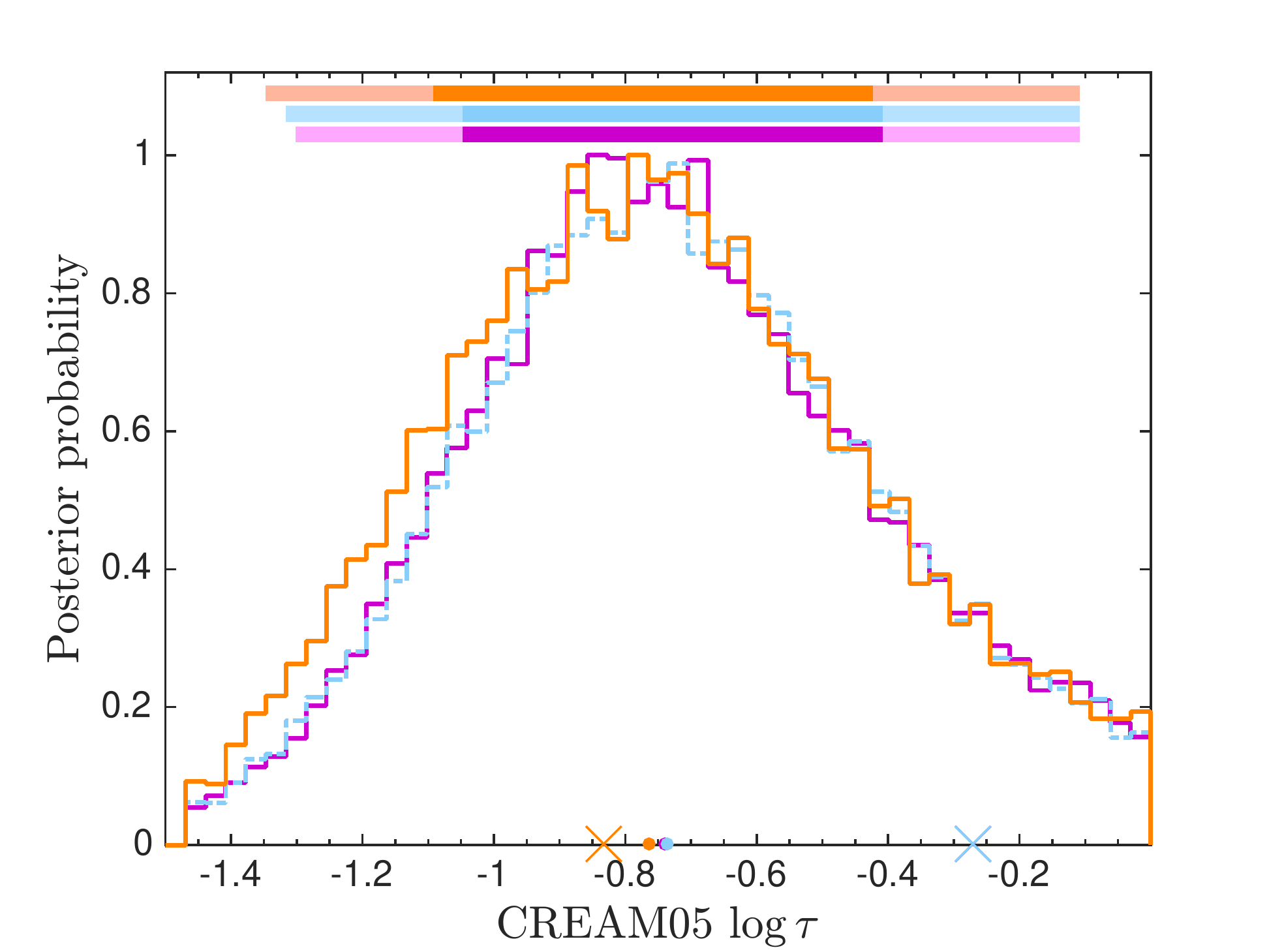}\includegraphics[width=0.3\textwidth]{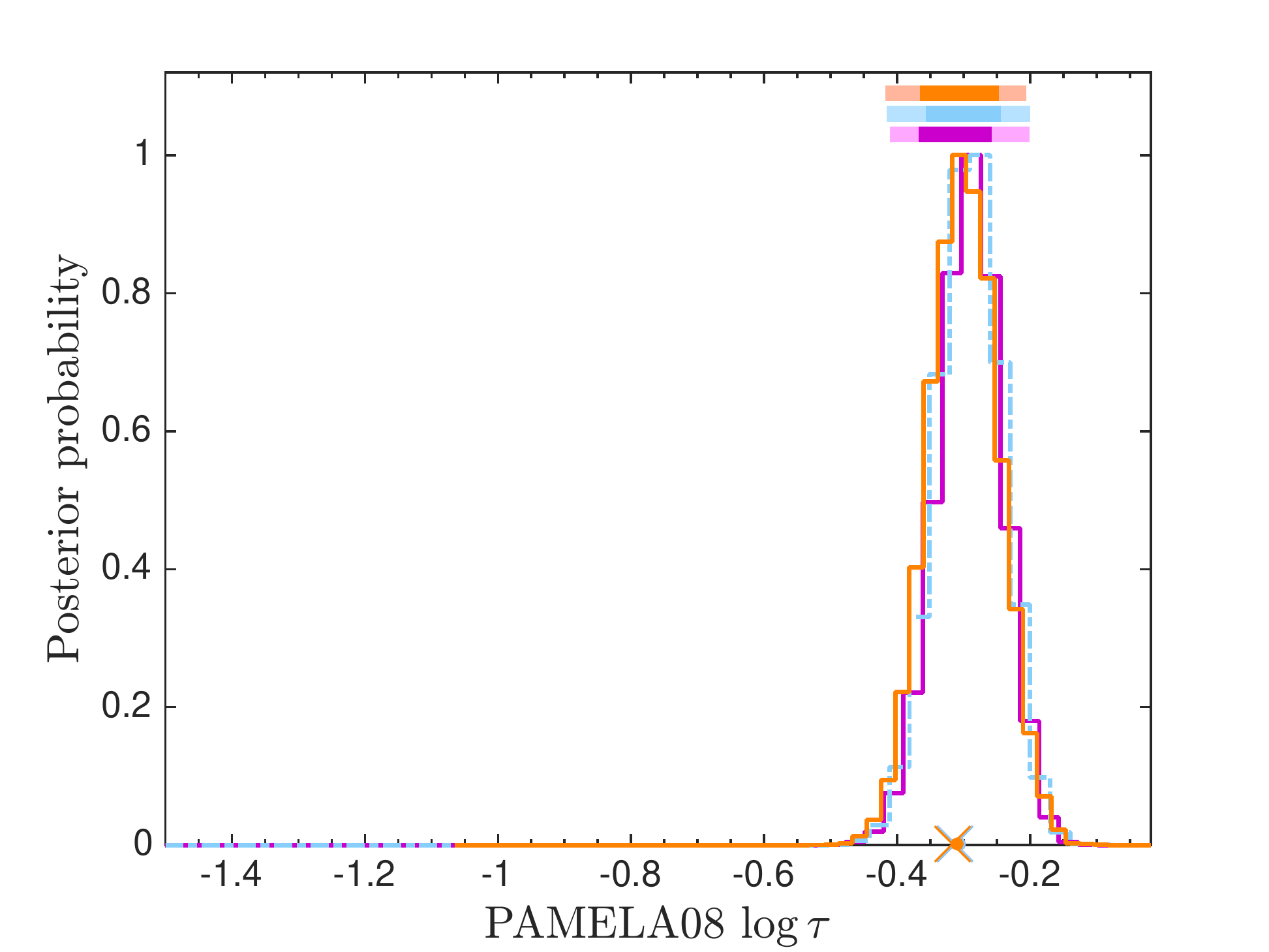} 
\caption{1D posterior distributions (with 68\% and 95\% credible intervals) for the different CR propagation parameters in a low-resolution, \{$p, \bar p$, He\} propagation scenario using {\sc MultiNest} as a sampler (no neural network speed-up, magenta) and from {\sc BAMBI} runs with two different values for the neural network input parameter $\sigma$. Light blue: $\sigma = 0.5$; Orange: $\sigma = 0.8$. All {\sc BAMBI} chains have been post-processed in the same way as in our main paper runs.}
 \label{BAMBI_sigma}
\end{figure}

We found that a training parameter value $\sigma = 0.5$ reproduced accurately the results obtained using {\sc MultiNest} as a sampler (and no BAMBI acceleration). However, in this case only 3\% of the likelihood evaluations were performed by the neural nets, hence with a very minimal speed-up in the computational time. In contrast, $\sigma = 0.8$ led to a good convergence with 21\% of the likelihood evaluations performed by the nets. Since some of the resulting samples gave spurious high-likelihood regions, we further post-processed them to remove any residual inaccuracy. The posterior distributions from these test runs are shown in Fig.~\ref{BAMBI_sigma}, where they are compared with the posterior resulting from a full {\sc MultiNest} run.

The analysis for $\sigma = 0.5$ was carried out for both \texttt{nhid} = 200 and \texttt{nhid} = 300. Both runs led to good parameter inference results, and the number of likelihood evaluations computed using the network was very similar. Based on these results, we decided to fix the input network settings to \texttt{nhid} = 200 and $\sigma = 0.8$, leading to reliable parameter inference with a speed-up of $\sim 20\%$.

\bibliographystyle{apj}

\bibliography{imos,strong,FermiDiffusePaper2,extragalactic,GalBayes2}{}
%#########################################################################

\newpage

\begin{deluxetable}{lccc}
\tabletypesize{\footnotesize}
\tablecaption{\label{tab:params}
Summary of input parameters and prior ranges
}
\tablecolumns{4}
\tablewidth{0pt}
\tablehead{
Quantity & Symbol & Prior range  & Prior type} 
\startdata
%
%\\[-2pt]
\multicolumn{4}{l}{\sc Propagation model parameters $\paramsP$} \smallskip\\
%\hline
\quad Proton normalization ($10^{-9}$ cm$^2$ sr$^{-1}$s$^{-1}$MeV$^{-1}$) & $N_p$ & \range{2}{8} & Uniform \\
\quad Diffusion coefficient\tablenotemark{a} ($10^{28}$ cm$^2$ s$^{-1}$) & $D_0$ & \range{1}{12}  & Uniform\\
%Rigidity break scale ($10^3$ MV) & $\rho_0$ & $4.0$ & \range{3}{4}\\ 
\quad Rigidity power law index & $\delta$  & \range{0.1}{1.0} & Uniform\\ 
%Rigidity power law index 2 & $\delta_2$ &  0.33 & \range{0.33}{0.70}\\ 
\quad Alfv\'en speed (km s$^{-1}$) & $\valf$  & \range{0}{50} & Uniform\\
\quad Diffusion zone height (kpc) & $z_h$  & \range{0.5}{20.0} & Uniform\\ 
\quad Rigidity of first injection break ($10^4$ MV) & $\rho_{br}$ &\range{1}{30} & Uniform \\
\quad Nucleus injection index below $\rho_{br}$ & $\nu_0$ & \range{1.00}{2.50} & Uniform\\
\quad Nucleus injection index above $\rho_{br}$ & $\nu_1$ &  \range{\nu_0}{3.00}& Uniform \\
\quad Nucleus injection index above $220$ GV & $\nu_2$ &  \range{1.5}{\nu_1}& Uniform \\
\quad Difference between $p$ and heavier inj. indices & $\delta_\nu$ &  \range{0.0}{1.0}& Uniform  \medskip\\

\multicolumn{4}{l}{\sc Injection abundance parameters $\paramsA$ \tablenotemark{a}} \smallskip\\
\quad Proton normalization ($10^{-9}$ cm$^2$ sr$^{-1}$s$^{-1}$MeV$^{-1}$) & $N_p$ & \range{2}{8} & Uniform \\
\quad Helium & $X_{\mathrm{He}}$ & \range{0.1}{2}$\times 10^{5}$ & Uniform \\
\quad Carbon & $X_{\mathrm{C}} $& \range{0.1}{6}$\times 10^{3}$ & Uniform \\
\quad Nitrogen & $X_{\mathrm{N}}$ & \range{0.1}{5}$\times 10^{2}$ & Uniform\\
\quad Oxygen & $X_{\mathrm{O}}$ & \range{0.1}{10}$\times 10^{3}$ & Uniform\\
\quad Neon & $X_{\mathrm{Ne}} $& \range{0.0}{1}$\times 10^{3}$ & Uniform \\
\quad Sodium & $X_{\mathrm{Na}}$ & \range{0.0}{5}$\times 10^{2}$ & Uniform\\
\quad Magnesium & $X_{\mathrm{Mg}}$ & \range{0.0}{1.5}$\times 10^{3}$ & Uniform\\
\quad Aluminium & $X_{\mathrm{Al}}$ & \range{0.0}{5}$\times 10^{2}$ & Uniform\\
\quad Silicon & $X_{\mathrm{Si}}$ & \range{0.0}{1.5}$\times 10^{5}$ & Uniform  \\ 
 \medskip \\
%\hline
%5\\[-1pt]
\multicolumn{4}{l}{\sc Experimental nuisance parameters} \smallskip\\

\quad Modulation parameters $\phi$ (MV) 	& 					& 				& Log-normal prior\tablenotemark{b} \\
\qquad HEAO-3 		& $m_\text{HEAO-3}$	&  	& \hspace{0mm} \range{0}{1250}\\ 
\qquad ACE  	& $m_\text{ACE}$  		&   	& \hspace{0mm} \range{0}{1125} \\ 
\qquad CREAM 	& $m_\text{CREAM}$	&  	& Fixed (no modulation) \\ 
\qquad ISOMAX 	& $m_\text{ISOMAX}$	& 	& \hspace{0mm} \range{0}{1075}\\ 
\qquad PAMELA 	& $m_\text{PAMELA}$	&  	& \hspace{0mm} \range{0}{1000}\\ 
\quad Variance rescaling parameters ($j=1,\dots,5$)   & $\log\tau_j$  &  \range{-1.5}{0.0}	& Log-uniform on $\log\tau_j$

\enddata
\tablenotetext{a}{The hydrogen abundance is fixed to $X_{\mathrm{H}} \equiv 1.06 \times 10^{6}$. }
\tablenotetext{b}{We use a log-normal distribution, where $\sigma = 50\%$ of the central value. Quoted limits correspond to $3\sigma$. }
\end{deluxetable}

%
%\begin{minipage}{2.0\textwidth}

\begin{deluxetable}{llr}
\tabletypesize{\footnotesize}
\tablecaption{\label{tab:PropData}
Data used in this analysis
}
\tablecolumns{4}
\tablewidth{0pt}
\tablehead{
Element & Experiment & Energy Range}
\startdata
\multicolumn{3}{l}{\sc Data used in \ppbarHe scan} \smallskip\\
H & PAMELA ('06--'08)\tablenotemark{a}  & 0.44--1000 GeV/n \\
  & CREAM-I ('04--'05)\tablenotemark{b} & 3--200 TeV/n \\
\hline
${\rm \bar H}$     & PAMELA ('06--'08)\tablenotemark{c}  & 0.28--128 GeV/n \\
\hline
He & PAMELA ('06--'08)\tablenotemark{a}  & 0.13--504 GeV/n \\
   & CREAM-I ('04--'05)\tablenotemark{b} & 0.8--50 TeV/n \\
\hline \hline
\multicolumn{3}{l}{\sc Data used in light element scan and abundance scan} \smallskip\\
B/C & ACE-CRIS ('97--'98)\tablenotemark{d} & 72--170 MeV/n \\
    & HEAO3-C2 ('79--'80)\tablenotemark{e} & 0.62--35 GeV/n \\
    & CREAM-I ('04--'05)\tablenotemark{f}  & 1.4--1450 GeV/n \\ 
%   & TRACER                               & 23--2075 GeV/n \\ \hline
\hline
$^{10}$Be/$^{9}$Be & ACE-CRIS ('97-'99)\tablenotemark{g} & 81--132 MeV/n \\
                   & ISOMAX ('98)\tablenotemark{h}       & 0.51--1.51 GeV/n \\ 
\hline
B & HEAO-3 ('79--'80)\tablenotemark{e}   & 0.62--35 GeV/n \\ 
%				&ACE-CRIS('97-'98)		& 60 -- 170 MeV /n\\ \hline
\hline
C & HEAO-3 ('79--'80)\tablenotemark{e}   & 0.62--35 GeV/n \\
  & CREAM-II ('05--'06)\tablenotemark{i} & 86--7415 GeV/n \\
%				& TRACER 			&  26 -- 2300 GeV/n \\
\hline
N & HEAO-3 ('79--'80)\tablenotemark{e}   & 0.62--35 GeV/n \\ 
  & CREAM-II ('05--'06)\tablenotemark{i} & 95--826 GeV/n \\ 
\hline
O & HEAO-3 ('79--'80)\tablenotemark{e}  & 0.62--35 GeV/n \\
  & CREAM-II ('05--'06)\tablenotemark{i} & 64--7287 GeV/n \\
  \hline \hline
%				& TRACER 			& 26--2300 GeV/n \\
\multicolumn{3}{l}{\sc Data used in abundance scan} \smallskip\\
Ne & ACE-CRIS ('97--'98)\tablenotemark{d} & 85--240 MeV/n \\
   & HEAO3-C2 ('79--'80)\tablenotemark{e} & 0.62--35 GeV/n \\ 
   & CREAM-II ('05--'06)\tablenotemark{i} & 47--4150 GeV/n	\\ 
\hline
Na & ACE-CRIS ('97--'98)\tablenotemark{d} & 100--285 MeV/n \\ 
   & HEAO3-C2 ('79--'80)\tablenotemark{e} & 0.8--35 GeV/n \\ 
\hline
Mg & ACE-CRIS ('97--'98)\tablenotemark{d} & 100--285 MeV/n \\
   & HEAO3-C2 ('79--'80)\tablenotemark{e} & 0.8--35 GeV/n \\
   & CREAM-II ('05--'06)\tablenotemark{i} & 27--4215 GeV/n \\ 
\hline
Al & ACE-CRIS ('97--'98)\tablenotemark{d} & 100--285 MeV/n \\
   & HEAO3-C2 ('79--'80)\tablenotemark{e} & 0.8--35 GeV/n \\ 
\hline
Si & ACE-CRIS ('97--'98)\tablenotemark{d} & 120--285 MeV/n \\		
   & HEAO3-C2 ('79--'80)\tablenotemark{e} & 0.8--35 GeV/n \\
   & CREAM-II ('05--'06)\tablenotemark{i} & 27-2418 GeV/n \\
%                    & TRACER 			&  138 -- 2088 GeV/n \\
\enddata
\tablenotetext{a}{\citet{Adriani2011}}
\tablenotetext{b}{\citet{Yoon2011}}
\tablenotetext{c}{\citet{Adriani2010}}
\tablenotetext{d}{\citet{George2009}}
\tablenotetext{e}{\citet{Engelmann1990}}
\tablenotetext{f}{\citet{Ahn2008}}
\tablenotetext{g}{\citet{Yanasak2001}}
\tablenotetext{h}{\citet{Hams2004}}
\tablenotetext{i}{\citet{Ahn2009}}
\end{deluxetable}

\begin{deluxetable*}{lccc | ccc }
\tabletypesize{\footnotesize}
\tablecaption{\label{tab:params_constraints} 
Summary of constraints on all propagation parameters \label{tab:resultstable}
}
\tablecolumns{7}
\tablewidth{0pt}
\tablehead{
& \multicolumn{3}{ c| }{\textbf{\ppbarHe scan}} & \multicolumn{3}{ |c}{\textbf{Light element (B, ..., Si) scan}}   \\ \hline
Quantity & Best fit & Posterior mean and & Posterior  & Best fit & Posterior mean and & Posterior\\
&value & standard deviation & 95\% range&value & standard deviation & 95\% range }\\
\startdata
%
%\\[-2pt]
\multicolumn{3}{l}{\sc Diffusion model parameters $\paramsP$}    \smallskip\\
%\hline
%%
\quad$D_{0}$ ($10^{28}$ cm$^2$ s$^{-1}$)&6.330&6.102$\pm$1.662& [2.138,8.205]&6.188&9.030$\pm$1.610& [5.743,11.256] \\
\quad$\delta$&0.466&0.461$\pm$0.065& [0.343,0.586]&0.375&0.380$\pm$0.018& [0.349,0.412] \\
\quad$v_{Alf}$ (km/s)&8.922&8.970$\pm$1.244& [7.036,11.254]&32.573&30.017$\pm$2.461& [25.484,34.465] \\
\quad$z_h$ (kpc)&9.507&10.358$\pm$4.861& [2.461,19.034]&4.900&10.351$\pm$4.202& [4.544,19.078] \\
\quad$\rho_{br}$ (GV)&2.486&2.345$\pm$0.344& [1.870,2.739]&15.782&16.687$\pm$1.498& [14.051,19.849] \\
\quad$\nu_0$&1.854&1.765$\pm$0.229& [1.230,2.133]&2.012&2.025$\pm$0.073& [1.885,2.155] \\
\quad$\nu_1$&2.352&2.358$\pm$0.063& [2.230,2.468]&2.549&2.548$\pm$0.050& [2.452,2.642] \\
\quad$\nu_2$&2.182&2.186$\pm$0.068& [2.062,2.308]&2.195&2.197$\pm$0.088& [2.042,2.374] \\
\quad$ 10^9 N_p$ (cm$^{-2}$ sr$^{-1}$ s$^{-1}$ MeV$^{-1}$)&4.798&4.791$\pm$0.066& [4.672,4.913]&4.511&4.482$\pm$0.220& [4.055,4.884]  \\
\quad $\delta_\nu$&0.045&0.047$\pm$0.009& [0.030,0.064]&--&--& -- \\
\quad $X_{\mathrm{He}}\times 10^{-4}$&10.261&10.294$\pm$0.505& [9.416,11.240]&--&--& -- \\

\medskip\\
%\hline
%%
%\\[-1pt]
\multicolumn{3}{l}{\sc Experimental nuisance parameters} \smallskip\\

 Modulation parameters $m_j$ \\
\quad PAMELA08 mod (MV) & 637.625&645.740$\pm$26.694& [601.226,696.164]&--&--& -- \\
\quad HEAO80 mod (MV)&--&--& --&622.201&611.039$\pm$93.229& [438.307,789.523] \\
\quad ACECRIS99 mod (MV)&--&--& --&445.975&421.682$\pm$48.797& [330.972,509.777] \\
\quad ISOMAX98 mod (MV)&--&--& --&380.722&492.036$\pm$206.243& [184.184,958.214] \\
  \smallskip\\

\multicolumn{3}{l}{ Variance rescaling parameters $\tau$}\\

\quad PAMELA08 $\log\tau$&-0.237&-0.277$\pm$0.053& [-0.370,-0.181]&--&--& -- \\
\quad HEAO80 $\log\tau$&--&--& --&-0.516&-0.571$\pm$0.089& [-0.740,-0.407] \\
\quad ACECRIS99 $\log\tau$&--&--& --&0.000&-0.263$\pm$0.209& [-0.780,-0.015] \\
\quad CREAM05 $\log\tau$&-0.973&-1.014$\pm$0.260& [-1.440,-0.480]&-0.704&-0.764$\pm$0.140& [-1.053,-0.516] \\
\quad ISOMAX98 $\log\tau$&--&--& --&-0.115&-0.604$\pm$0.378& [-1.380,-0.045] \\
\enddata
%\end{minipage}
\end{deluxetable*}

\begin{deluxetable*}{lccc  }
\tabletypesize{\footnotesize}
\tablecaption{\label{tab:abd_constraints} Summary of constraints on Abundance parameters}
 
\tablecolumns{4}
\tablewidth{0pt}
\tablehead{
\hline
Quantity & Best fit & Posterior mean and & Posterior  \\
&value & standard deviation & 95\% range}\\
\startdata
$ 10^9 N_p$ (cm$^{-2}$ sr$^{-1}$ s$^{-1}$ MeV$^{-1}$)&4.512&4.544$\pm$0.097& [4.369,4.715] \\
$X_{\mathrm{He}} \times 10^{-4}$&9.044&8.975$\pm$0.264& [8.499,9.508] \\
$X_{\mathrm{C}}$&2578.407&2553.666$\pm$66.318& [2442.083,2666.097] \\
$X_{\mathrm{N}}$&210.667&221.389$\pm$12.245& [199.314,246.589] \\
$X_{\mathrm{O}}$&3372.090&3335.543$\pm$82.290& [3184.869,3492.503] \\
$X_{^{20}\mathrm{Ne}}$&304.155&306.029$\pm$26.345& [259.181,357.127] \\
$X_{^{22}\mathrm{Ne}}$&97.767&94.118$\pm$22.321& [50.997,137.982] \\
$X_{\mathrm{Na}}$&33.578&35.931$\pm$2.812& [31.065,41.583] \\
$X_{^{24}\mathrm{Mg}}$&583.254&548.250$\pm$40.044& [472.095,623.988] \\
$X_{^{25}\mathrm{Mg}}$&80.104&87.010$\pm$28.553& [35.980,143.917] \\
$X_{^{26}\mathrm{Mg}}$&85.998&100.340$\pm$23.765& [55.965,147.898] \\
$X_{\mathrm{Al}}$&79.410&78.102$\pm$3.211& [72.186,83.727] \\
$X_{^{28}\mathrm{Si}}$&643.797&629.755$\pm$21.512& [589.202,669.806] \\
$X_{^{29}\mathrm{Si}}$&44.661&47.725$\pm$10.524& [27.996,67.989] \\
$X_{^{30}\mathrm{Si}}$&32.996&38.987$\pm$8.010& [23.997,54.992] \\
\smallskip\\
\multicolumn{3}{l}{\sc Experimental nuisance parameters} \smallskip\\
HEAO80 mod (MV)&593.085&591.606$\pm$11.074& [573.154,610.848] \\
ACECRIS99 mod (MV)&329.543&340.231$\pm$14.137& [315.514,371.142] \\
PAMELA08 mod (MV)&664.817&671.463$\pm$21.223& [630.303,708.612] \\
\smallskip\\
\multicolumn{3}{l}{\quad Variance rescaling parameters $\tau$}\\
HEAO80 $\log\tau$&-0.615&-0.594$\pm$0.062& [-0.721,-0.478] \\
ACECRIS99 $\log\tau$&-1.162&-1.269$\pm$0.120& [-1.465,-1.037] \\
CREAM05 $\log\tau$&-1.039&-1.008$\pm$0.087& [-1.184,-0.853] \\
PAMELA08 $\log\tau$&-1.500&-1.499$\pm$0.001& [-1.499,-1.494] \\
TRACER06 $\log\tau$&-1.712&-1.563$\pm$0.185& [-1.921,-1.228] \\
\enddata
\end{deluxetable*}

\begin{table}[h]
\begin{center}
\caption{\galprop{} resolution parameters used in this study}
\label{restable}
\begin{tabular}{l l l} \hline \hline
\multicolumn{2}{c}{Parameter}  & value \\ \hline
\texttt{dr}& radial spacing (kpc) & 1.0 \\
\texttt{dz} & height spacing (kpc) & 0.2 \\
\texttt{Ekin\_factor} & (log) kinetic energy spacing & 1.2 \\
\texttt{timestep\_factor} & rescaling factor when reducing timesteps & 0.5 \\ 
\texttt{start\_timestep} & size of initial timestep (s) & $10^8$ \\
\texttt{end\_timestep} & size of final timestep (s) & $10^2$\\
\texttt{timestep\_repeat} &  repeats per timestep & 20 \\
\texttt{max\_Z} & number of elements & 14 \\ \hline \hline
\end{tabular}
\end{center}
\end{table}

\end{document}